\documentclass[12pt, a4paper]{cas-sc}
\usepackage{xcolor}
\usepackage{units}
\usepackage{upgreek}
\usepackage{revsymb}
\usepackage{lineno}
\usepackage{overpic}
\usepackage[numbers]{natbib}
\usepackage{multirow}
\def\tsc#1{\csdef{#1}{\textsc{\lowercase{#1}}\xspace}}
\tsc{WGM}
\tsc{QE}
\tsc{EP}
\tsc{PMS}
\tsc{BEC}
\tsc{DE}

\let\oldsim\sim 
\renewcommand{\sim}{{\oldsim}}
\usepackage{subcaption}
\usepackage[title]{appendix}

\begin{document}

\let\WriteBookmarks\relax
\def\floatpagepagefraction{1}
\def\textpagefraction{.001}
\shorttitle{Initial Performance of the E320 Tracker}
\shortauthors{O. Borysov et~al.}
\title[mode = title]{Initial Performance of the E320 Tracker}

\author[1]{Oleksandr {Borysov}}

\author[2,3]{Sébastien {Corde}}

\author[1]{Gal {Evenzur}}

\author[3]{Alexander {Knetsch}}

\author[1]{Alon {Levi}}

\author[4]{Sebastian {Meuren}}

\author[1]{Nathaly {Nofech-Mozes}}

\author[3]{Ivan {Rajkovic}}

\author[2]{Sheldon {Rego}}

\author[3,5]{David A. {Reis}}

\author[1,6,7]{Arka {Santra}}

\author[5]{Tania {Smorodnikova}}

\author[3]{Doug W. {Storey}}

\author[1]{Noam {Tal Hod}}[type=editor,orcid=0000-0001-5241-0544]
\ead{noam.hod@weizmann.ac.il}
\cormark[1]

\author[1]{Roman {Urmanov}}[type=editor,orcid=0009-0001-5686-9050]
\ead{roman.urmanov@weizmann.ac.il}
\cormark[1]

\affiliation[1]{organization={Department of Particle Physics and Astrophysics, Weizmann Institute of Science},
addressline={234 Herzl Street}, 
city={Rehovot},
postcode={7610001}, 
country={Israel}}

\affiliation[2]{organization={Laboratoire d’Optique Appliquée (LOA), CNRS, École polytechnique, ENSTA},
addressline={Institut Polytechnique de Paris}, 
city={Palaiseau},
postcode={91762}, 
country={France}}

\affiliation[3]{organization={SLAC National Accelerator Laboratory},
addressline={2575 Sand Hill Rd}, 
city={Menlo Park},
postcode={94025}, 
state={CA}, 
country={USA}}

\affiliation[4]{organization={LULI, CNRS, CEA, Sorbonne Université, Ecole Polytechnique,},
addressline={Institut Polytechnique}, 
city={Palaiseau},
postcode={91128}, 
country={France}}

\affiliation[5]{organization={Stanford PULSE Institute, SLAC National Accelerator Laboratory},
addressline={2575 Sand Hill Rd}, 
city={Menlo Park},
postcode={94025}, 
state={CA}, 
country={USA}}

\affiliation[6]{organization={Saha Institute of Nuclear Physics},
addressline={1/AF Bidhannagar}, 
city={Kolkata},
postcode={700064}, 
country={India}}

\affiliation[7]{organization={Homi Bhabha National Institute},
addressline={Training School Complex, Anushakti Nagar}, 
city={Mumbai},
postcode={400094}, 
state={Maharashtra}, 
country={India}}








\date{} 

\cortext[cor1]{Corresponding authors}

\begin{abstract}
Our recent study discussed the simulation-based prospects for measuring single positrons produced in electron-laser collisions via the nonlinear Breit-Wheeler deep-tunneling process in the SLAC Experiment 320 at the FACET\textsf{--}II RF LINAC.
In this work, we demonstrate how a tracking detector, that is a scaled-down version of the one discussed in the prospective simulation study, enables the measurement.
This prototype detector, installed in Aug 2024, is built out of five layers of single ALPIDE chips.
The data are taken from several standalone runs completed in Nov 2024 and Feb 2025.
We use positrons generated through conversion of Bremsstrahlung photons as a proxy to the nonlinear Breit-Wheeler process. 
These positrons are produced by the beam electrons in a thin Beryllium foil close to the experiment's interaction point.
The tracking approach used in this initial work is based on a Hough-Transform seeding algorithm followed by a straight line fit confined to the detector volume.
Even with this relatively simple approach, we are able to measure a signal rate of $(1.20\pm 0.06_{\rm stat.}\pm 0.56_{\rm syst.})\times 10^{-1}$ positrons per shot.
This signal rate is comparable to the nonlinear Breit-Wheeler rate expected in the main experiment.
Notably, the measurement is achieved under an extreme, unprecedented background hit density of $\sim 1.7/{\rm mm}^2$, unlike the main experiment, where at least a twice lower density is expected.
This large background is mostly due to secondary particles produced when the large flux of Bremsstrahlung photons interacts with the material of the beamline elements.
When the foil is retracted, the false-positive signal rate is shown to be four orders of magnitude smaller than  the signal rate.
We further show that the high spatial tracking resolution of $\sim5~\mu{\rm m}$ allows to characterize the positrons' spectra.
The results are compared to simulations, which are found to be compatible with the data.
\end{abstract}


\maketitle

\section{Introduction}
\label{sec:intro}
In a recent work~\cite{Borysov:2025ehq} we have outlined concrete prospects for the production and detection of $e^+e^-$ pairs via nonlinear Breit-Wheeler (NBW) mechanism in the strong-field tunneling regime.
This is proposed in the SLAC Experiment 320 (E320), which  collides 10~TW-class laser pulses with the high-quality, 10~GeV electron beam from the FACET\textsf{--}II radio-frequency (RF) linear accelerator (LINAC), where the NBW positrons are detected and characterized with a new tracking detector.
Specifically, we have shown that an estimated yield of $\sim 0.01\textsf{--}0.1$ pairs per collision is expected in the upcoming campaigns of E320, where this small signal rate typically comes along with large backgrounds originating, e.g., from dumping the high-charge primary beam, secondaries induced by the beam halo, as well as photons and low-energy electrons produced in the electron-laser collision itself.
As we argue in~\cite{Borysov:2025ehq}, understanding quantum electrodynamics (QED) in the presence of strong electromagnetic (EM) background fields is a thriving, worldwide experimental effort.
Thanks to the rapid evolution of laser power in the last few decades, with energies already reaching a few $100$~J, strong EM fields with laboratory-frame amplitudes reaching $\mathcal{O}(10^{-5}\textsf{--}10^{-4})$ of the Schwinger QED critical field strength of $\sim 1.3\times 10^{18}$~V/m~\cite{PhysRev.82.664,Sauter} are now achievable~\cite{danson_petawatt_2019}.
While the QED critical field is currently orders of magnitude above any terrestrially producible field strength, the gap can be significantly narrowed and may even be closed in collisions of these lasers with high-energy probe particles (electrons or photons).
The field seen by particles of mass $m$ and energy $E$ is boosted by $\sim\gamma = E/m$\footnote{Here and in the following we will use natural units with $c = \hbar =\epsilon_0 =1$.}.
For example, the boost factor for a $10~{\rm GeV}$ probe electron, is $\gamma \approx 2\times 10^4$.
Multi-GeV electron beams are available in a few RF LINAC facilities (see e.g.~\cite{Yakimenko:IPAC2016-TUOBB02,altarelli2007european}).
In addition, such beams are becoming available also in an increasing number of multi-PW laser wakefield acceleration (LWFA) facilities worldwide (see e.g.~\cite{weber2017p3,gales2018extreme,papadopoulos2016apollon,yoon2021realization}), where the electrons are accelerated by a high-power laser.
Several running and planned experiments worldwide are seeking to study strong-field QED (SF-QED) processes in different setups (see e.g.~\cite{corels_compton_2024,PhysRevLett.132.175002,Matheron:2024hwy,LUXETDR}).
While their goals are coherent, the different approaches are complementary in the prospective precision and the phase-space coverage.
To date, the only measurement of the NBW process was carried out by the seminal SLAC Experiment 144~\cite{Bamber:1999zt} (E144) in the mid 1990s, albeit via the non-sequential absorption of multiple laser photons, i.e., not in the strong-field tunneling regime.
Therefore, a high-priority goal of current and future SF-QED experiments is to measure the NBW process for the first time in the (deep) tunneling regime, via a strong-field tunneling process as discussed in detail in~\cite{Borysov:2025ehq}.
These experiments face extreme challenges owing to the small signal production probability and the large background fluxes associated with the collision and the beam electrons.\\

In this work, we give the first in-depth demonstration with data -- in the context of the running E320 -- using positrons generated through conversion of Bremsstrahlung photons.
The data are collected at FACET\textsf{--}II with a scaled-down prototype version of the exact same technology discussed in~\cite{Borysov:2025ehq}.
The prototype detector, installed in Aug 2024, is built out of five layers of single ALPIDE\footnote{ALPIDE stands for ``ALice PIxel DEtector''}~\cite{Abelevetal:2014dna,AGLIERIRINELLA2017583} chips.
The data are taken from several standalone runs completed in Nov 2024 and Feb 2025, in three configurations: (i) background-only, (ii) signal+background from a 100~$\mu{\rm m}$ Aluminum foil, and (iii) signal+background from a 50~$\mu{\rm m}$ Beryllium foil. 
These (retractable) foils are positioned in proximity to the experiment interaction point and they are used to produce Bremsstrahlung photons that convert to $e^+e^-$ pairs.
Unlike the Kalman-Filter (KF)~\cite{ai2022common} based study in~\cite{Borysov:2025ehq}, the tracking approach used in this initial work is based on a Hough-Transform seeding followed by a straight line fit confined to the detector volume with no attempt of back-propagation to the production point.
We compare the signal positron measured rate and its spectrum with the case where the foils are retracted.
The measured rate and spectrum are also with simulations.
The tracking in this work is done under extremely dense conditions, with a real-particles hit density of $\sim 1.7$~hits/mm$^2$.
This value is extremely challenging.
For reference, we point out that this hit density is roughly double the one expected in the upcoming high-luminosity phase of the LHC (HL-LHC) experiments in their innermost pixel layers (see, e.g., Table~4 in~\cite{Accettura:2023ked}).
To the best of our knowledge it is the largest hit density in which tracking has ever been attempted with real experimental data.
We finally point out that, besides E320, variants of the same prototype detector are already used in a few of the complementary SF-QED experiments in multi-PW facilities like APOLLON~\cite{papadopoulos2016apollon} and ELI-NP~\cite{weber2017p3,gales2018extreme,Matheron:2024hwy}.\\

This document is organized as follows.
In Sec.~\ref{sec:experiment} the experimental setup is described, including a detailed description of the detector and its mode of operation (the power, trigger and data acquisition setup are described in Appendix~\ref{app:daq}).
In Sec.~\ref{sec:datasets} we discuss the specifics of the different datasets used, including the beam and magnets conditions, and the way the data are preprocessed before tracking (the data cleaning procedures are discussed in Appendix~\ref{app:cleaning}).
In Sec.~\ref{sec:houghtransform} and~\ref{sec:selection} we discuss the Hough Transform tracking pipeline and the track selection process, respectively.
In Sec.~\ref{sec:localalignment}
 and~\ref{sec:globalignment}
we describe the local and global alignment algorithms, respectively.
Finally in~Sec.\ref{sec:results} we outline the results for tracking in highly dense background environment and in Sec.~\ref{sec:outlook} we conclude.

\section{Experimental Overview}
\label{sec:experiment}
In analogy to Fig.~1 of~\cite{Borysov:2025ehq}, the main elements along the experimental setup are conceptually highlighted in Fig.~\ref{fig:setup} below, focusing on the last part of the beamline.
The relevant mode of operation in this data campaign is with Bremsstrahlung products induced by the FACET\textsf{--}II primary electron beam in two different thin foils (separately).
The first is a $50~\mu{\rm m}$ thin Beryllium ``window'' placed 84~cm upstream of the E320 vacuum chamber, encapsulating the electron-laser interaction point (IP). 
This window is typically used by plasma experiments (utilizing the same IP chamber) to prevent gases from flowing upstream in the accelerator's vacuumed beamline.
The second is a $100~\mu{\rm m}$ thin Al foil in the IP chamber, about 50~cm downstream the IP itself.
In the electron-laser collision mode of E320 both foils are retracted from the beam path.
The Be window used in this experiment was installed just before this data-taking campaign and hence it does not suffer from drilling or burning effects that one can typically expect after long operation in front of the beam.
\begin{figure}[pos=!ht]
\centering
\begin{overpic}[width=0.90\textwidth]{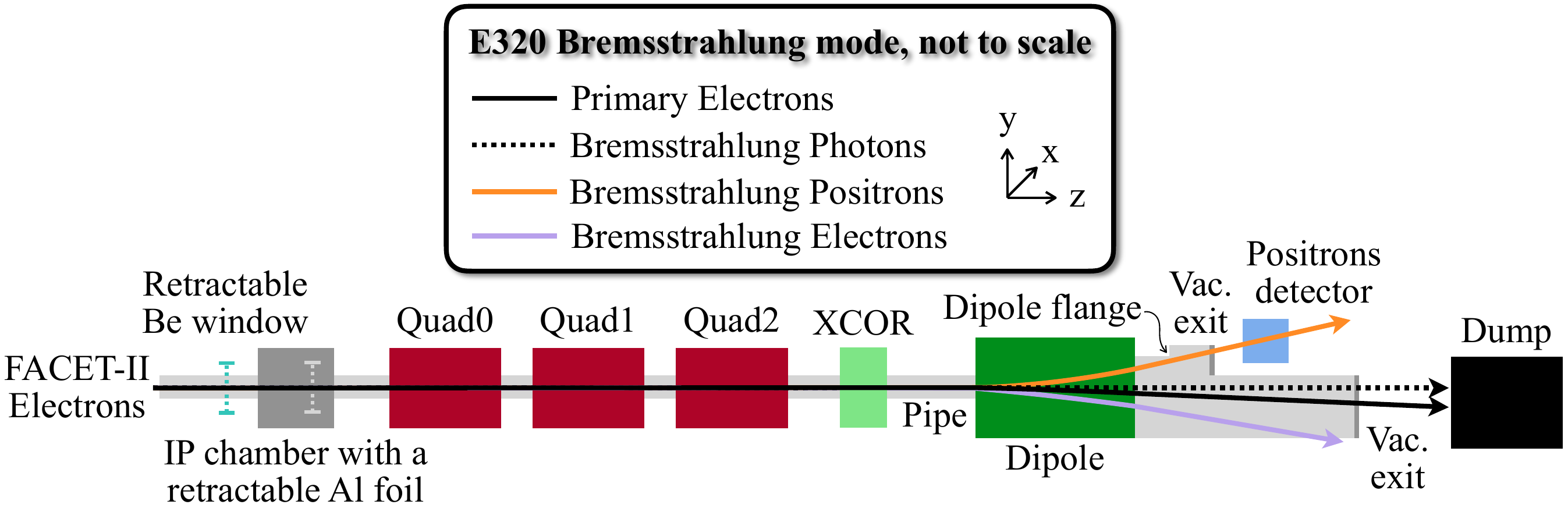}\end{overpic}
\caption{A schematic illustration of the E320 experimental setup in the Bremsstrahlung mode, from the IP chamber to the dump, focusing on the key elements related to the signal positrons detection. The positrons are produced either in the Beryllium (Be) window upstream of the IP chamber or in the Aluminum (Al) foil in it. For comparison with the E320 setup related to electron-laser collision, see Fig.~1 of~\cite{Borysov:2025ehq}.}
\label{fig:setup}
\end{figure}

\subsection{Accelerator}
\label{sec:accelerator}
The FACET\textsf{--}II National User Facility at the SLAC National Accelerator Laboratory provides high-intensity electron beams for studying their interactions with lasers, plasmas, and solids~\cite{Yakimenko:IPAC2016-TUOBB02,PhysRevLett.134.085001,PhysRevAccelBeams.22.101301}.
Electron bunches of up to 2~nC are presently accelerated to 10~GeV (potentially up to 13~GeV if all klystrons are being used) in the $\sim$1~km S-band LINAC with three stages of bunch compression to deliver bunches with a transverse spot size on the order of $10\times 10~\mu{\rm m}^2$, a longitudinal bunch length of $\sim$10~$\mu{\rm m}$, and thus peak currents in excess of $\sim100$~kA to the experimental area.
The multipurpose experimental area is specifically designed to facilitate a variety of experiments, including
plasma wakefield acceleration (PWFA) and probing SF-QED, and is outfitted with a suite of diagnostic devices to measure the incoming and outgoing beam properties.
This includes a magnetic imaging spectrometer to measure the properties of electrons after the IP~\cite{PhysRevAccelBeams.27.051302}.
The machine sends electron bunches to the experimental area at a typical rate of 10~Hz (up to 30~Hz is possible).\\

The three quadrupole and one dipole magnets seen in Fig.~\ref{fig:setup}, which have an active length of $\sim 1$~m each, are typically configured as follows: the upstream and downstream quadrupoles fields are set to $\vec{B}_{\rm Quad0,2}=(yG_{0,2},xG_{0,2},0)$~T/m, while the middle quadrupole is set to $\vec{B}_{\rm Quad1}=(-yG_{1},-xG_{1},0)$~T/m.
The dipole magnet is $\vec{B}_{\rm Dipole}=(B_D,0,0)$~T.
In addition, there is a possibility to correct the horizontal position (along the $x$ axis) of the particles  with an $x\textsf{--}$corrector (XCOR) dipole with $\vec{B}_{\rm XCOR}=(0,B_{\rm XCOR},0)$ that has a length of 0.23~m.
The quadrupole gradients ($G_{i}$), the dipole field strength ($B_D$) and the XCOR field strength ($B_{\rm XCOR}$) can be configured and the values used in this campaign are described in Sec.~\ref{sec:datasets}.
These settings allow to focus the center of the positron distribution at any point along the beamline between the dipole exit and dump.
To make sure that the acceptance is maximal, the focusing settings are chosen such that the object plane is set to the Be window or the Al foil, while the image plane is set to the first detector plane.

\subsection{Tracker Prototype Design}
\label{sec:tracker}
The prototype tracker is a scaled-down version of the full E320 tracker discussed in~\cite{Borysov:2025ehq}.
It is made of the ALPIDE\footnote{ALPIDE stands for ``ALice PIxel DEtector''}~\cite{Abelevetal:2014dna,AGLIERIRINELLA2017583} sensors (chips) that are produced for the upgrade of the ALICE experiment~\cite{Collaboration_2008} at the LHC~\cite{ALICE:2023udb,Ravasenga:2023yqd}.
Since this is the first time that the technology is used in the context of an SFQED experimental environment with extreme background, potential large EM pulses, etc., it was decided to first streamline its design and evaluate its performance.
That is, the purpose of the prototype version is to scout for preliminary hardware-wise, and software-wise (online and offline) experience difficulties before installing the full version of the more complex and expensive tracking detector from~\cite{Borysov:2025ehq}.

The ALPIDE chip has 512 rows and 1024 columns of pixels measuring $27\times 29~\mu{\rm m}^2$ each in one $29.94\times 13.76~{\rm mm}^2$ chip that integrates the sensing volume with the readout circuitry in one all-silicon chip.
The chips have demonstrated very good performance both before and after irradiation~\cite{SENYUKOV2013115,AGLIERIRINELLA2021164859,DANNHEIM2019187,MAGER2016434}, leading to a detection efficiency above 99\%, a fake hit rate much better than $10^{-5}$, a spatial resolution of around $5~\mu{\rm m}$ and a peaking time of around $2~\mu{\rm s}$.

\begin{figure}[pos=!ht]
\centering
\begin{overpic}[width=0.503\textwidth]{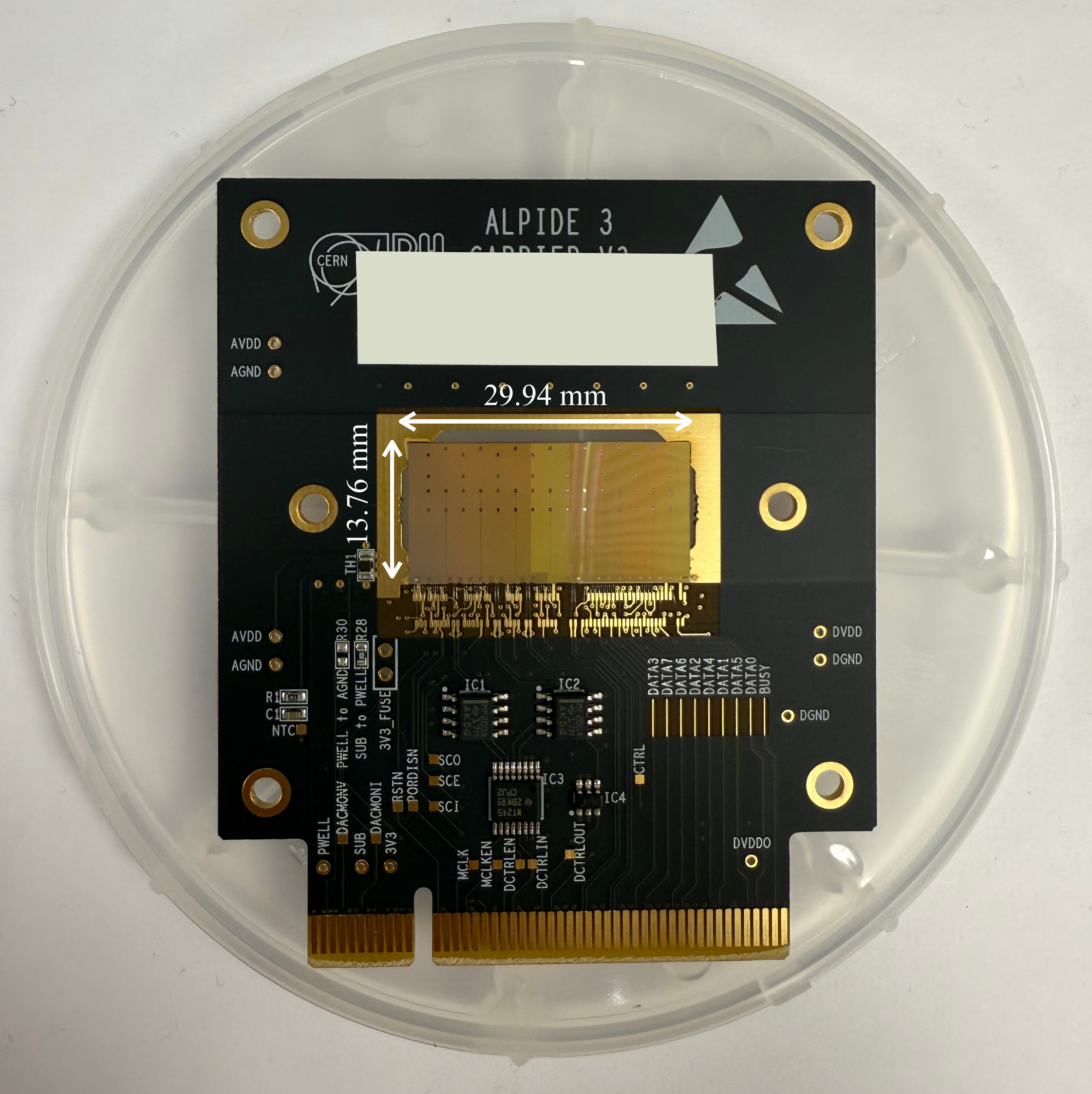}\end{overpic}
\begin{overpic}[width=0.47\textwidth]{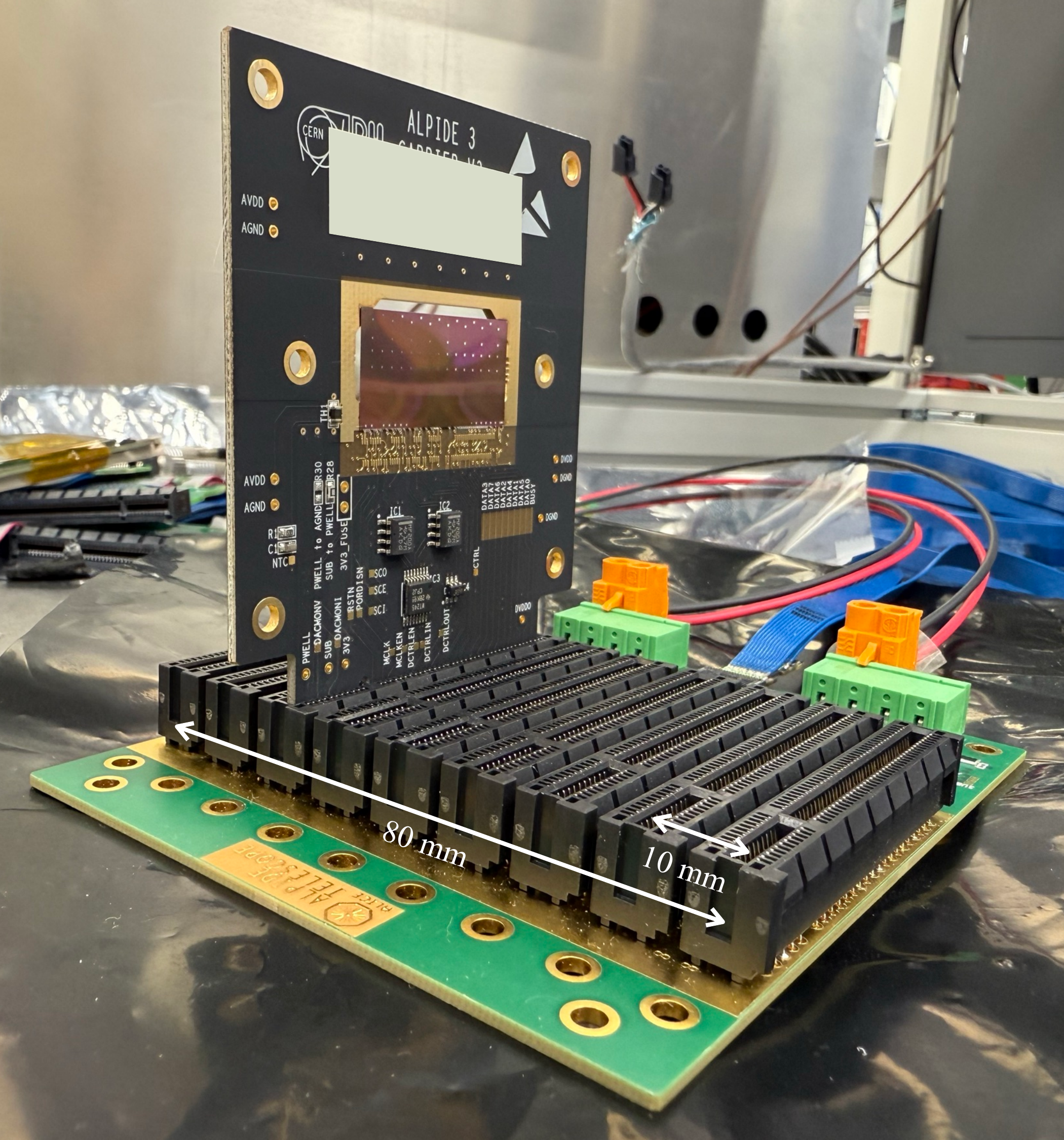}\end{overpic}
\caption{The ALPIDE chip carrier (left) and the 9CA board (right) with one carrier mounted. One can also see the digital and analog power lines as the red and black wires with orange connector, as well as the TwinAx readout and control cable in blue.}
\label{fig:carrier}
\end{figure}

Rather than four layers, each measuring $\sim 270\times 13.76~{\rm mm}^2$ (``stave'' of nine individual chips) as in the full detector version, the prototype has five layers, where each is of the size of a single chip.
The layers of the prototype, labeled as \verb|ALPIDE_0,...,4|, are separated by 20~mm along the $z\textsf{--}$axis, whereas in the full detector the layer separation is 100~mm.
This smaller separation is dictated by the spacing of a nine-chip adapter (9CA) printed-circuit board (PCB) onto which the individual chip carrier PCBs are mounted.
The 9CA can host up to nine carrier PCBs in 10~mm-spaced slots, where in the prototype, we occupy every second slot.
Thus, the 9CA seen in Fig.~\ref{fig:carrier} (right) fixes the chip carriers (left) in space from the PCIe connector side, but it also acts effectively like the ``stave'' object of the full detector in terms of power distribution and readout.

The detector's power distribution, the trigger configuration, the data acquisition setup and the overall integration with the FACET\textsf{--}II accelerator systems is discussed in detail in Appendix~\ref{app:daq}.\\

The 9CA and the five carriers are enclosed in an Aluminum box that serves as a mechanical support as well as a Faraday cage.
Specifically, the chip carriers are fixed from the other edge with a precision Aluminum plate having slots machined with a $<20~\mu{\rm m}$ precision.
We note that the placement of the chips on their carriers themselves is controlled to only about 100~$\mu{\rm m}$ and hence their exact relative position has to be inferred using data as discussed later in Sec.~\ref{sec:localalignment}.
The precision plate of the box is directly mounted on a Newport URS50CPP rotational stage, which is in turn mounted on a Newport UTS100PP translational stage.
The linear stage is fixed with a custom Aluminum bracket to a standard breadboard mounted slightly below the vacuum exit window's bottom edge.
This structure allows us (i) to take the detector $\sim 85$~mm away from the beam axis vertically (along the $y\textsf{--}$axis) when it is not in use during plasma experiments etc., (ii) to scan different positions around the nominal location, as well as to (iii) rotate it by up to approximately 100 degrees from its nominal position in which the chips face the vacuum exit window.
The latter is important for correcting small differences in the detector orientation or, e.g., for rotating the detector so the chips face the sky to measure cosmic muons during accelerator shutdown periods.
Both stages are remotely controlled via the FACET\textsf{--}II systems, where the range of motion is limited by a set of customized electro-mechanical limit-switches.

\begin{figure}[pos=!ht]
\centering
\begin{overpic}[width=0.56\textwidth]{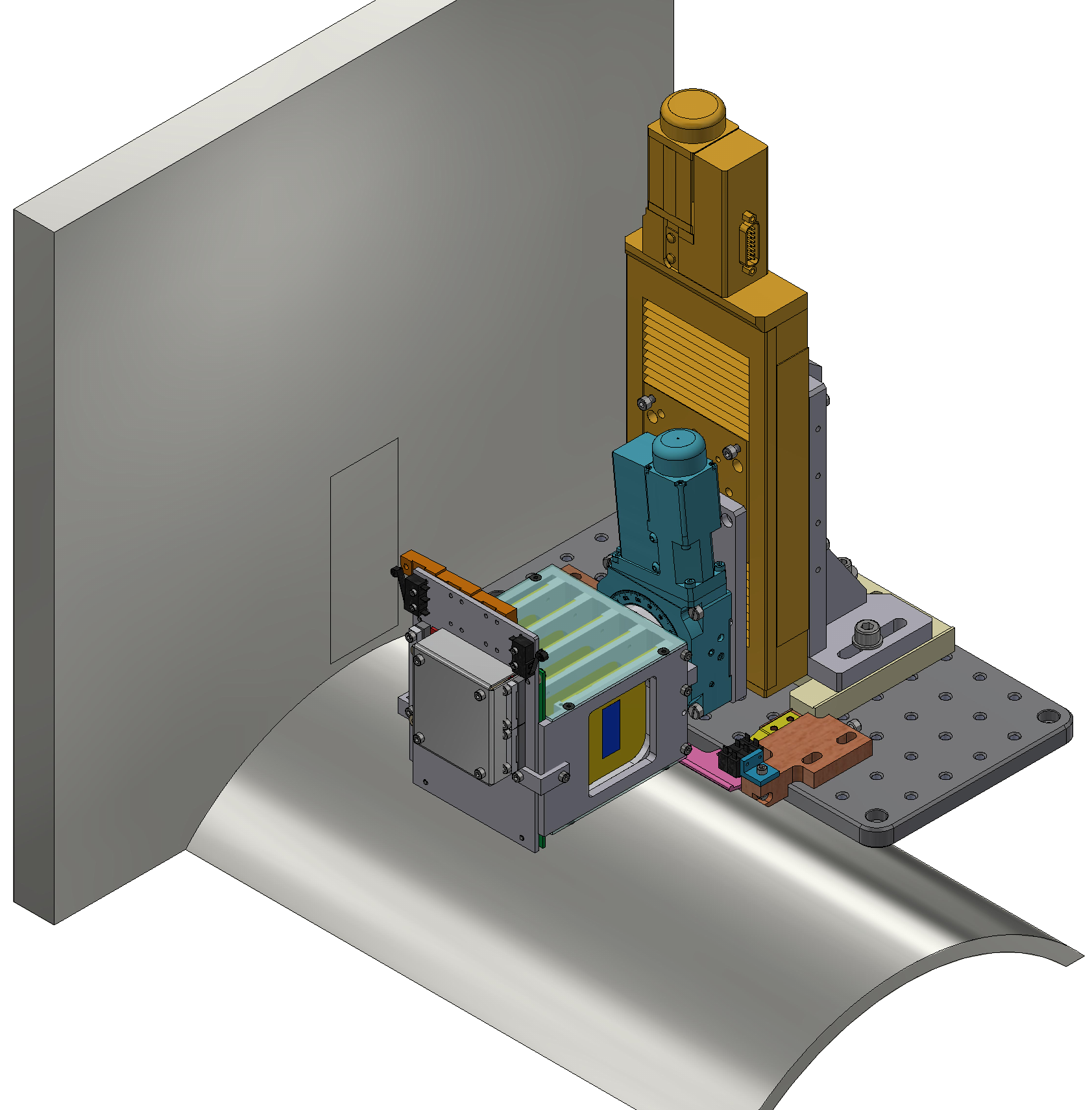}\end{overpic}
\begin{overpic}[width=0.43\textwidth]{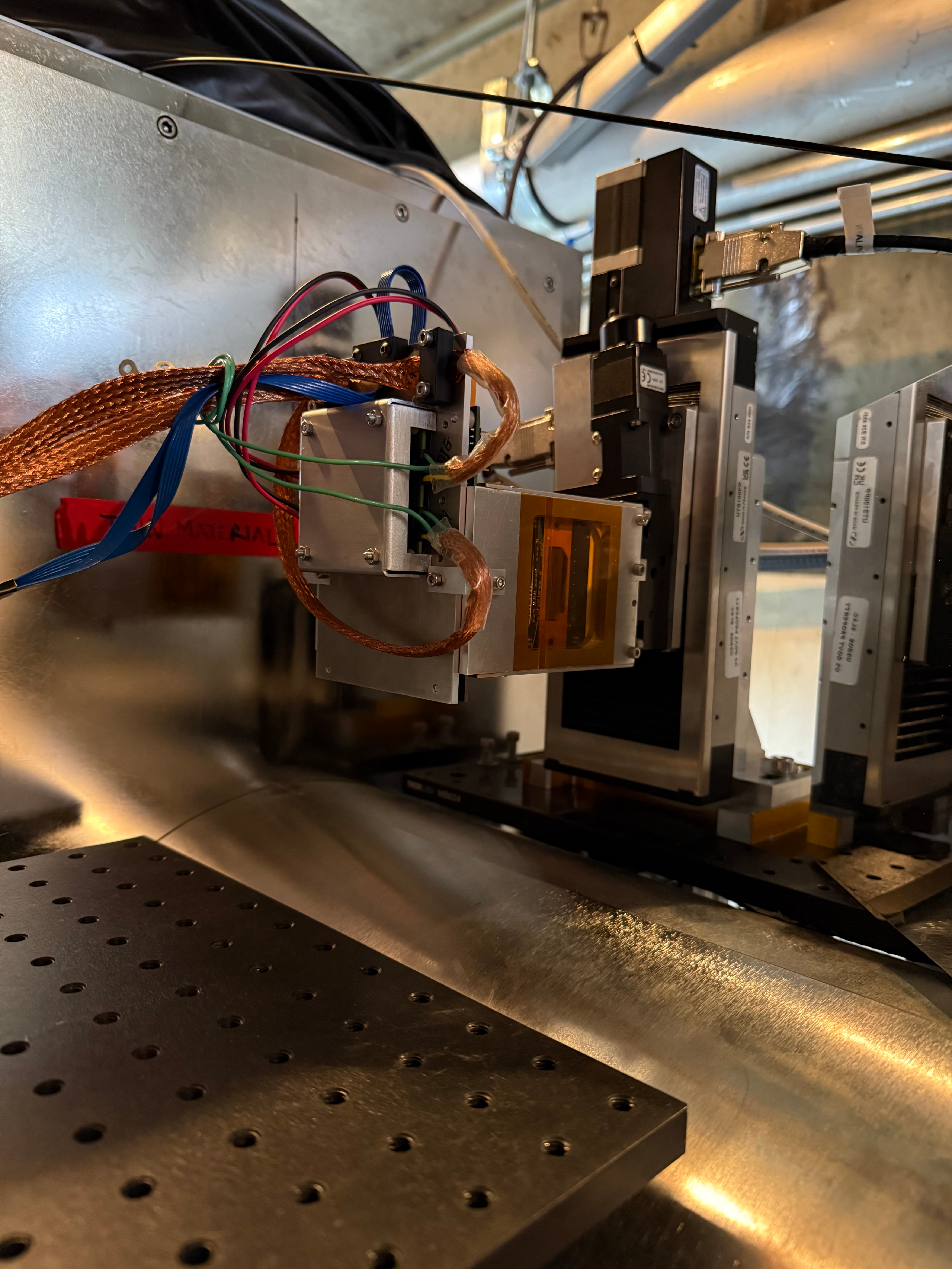}\end{overpic}
\caption{The detector assembly design (left) seen from the side and back with the main elements highlighted: the beampipe, the vacuum exit window (of the positrons) just above it, the breadboard, the bracket, the two stages, the detector box and the range-limiting safety plate below it.
The actual detector installed (right) is pictured from the tunnel aisle side looking upstream approximately. This picture was taken just before installing the safety switching mechanism (seen on the left in brown and pink).
}
\label{fig:detector1}
\end{figure}

\begin{figure}[pos=!ht]
\centering
\begin{overpic}[width=0.623\textwidth]{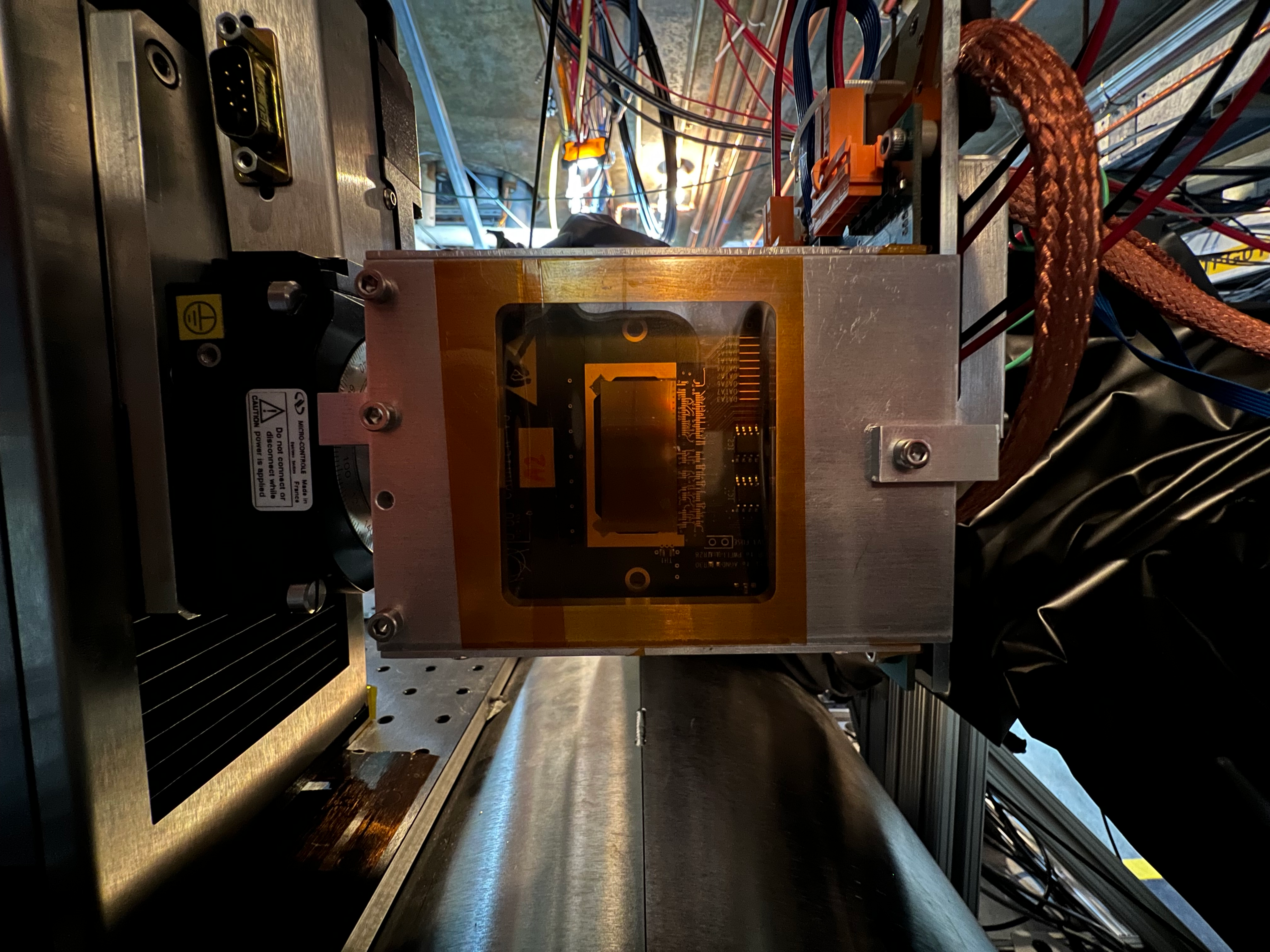}\end{overpic}
\begin{overpic}[width=0.350\textwidth]{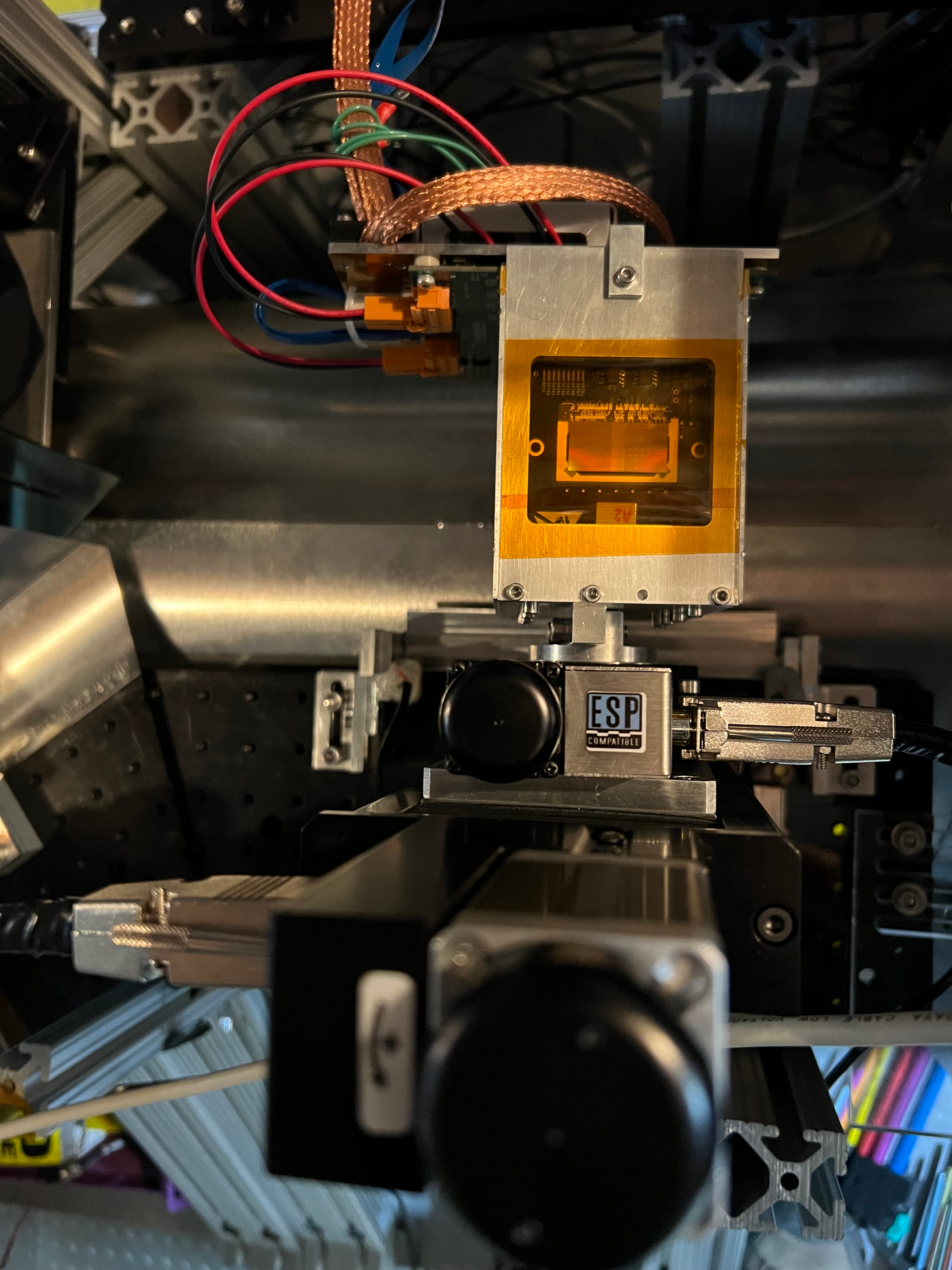}\end{overpic}
\caption{The detector pictured from the vacuum exit window looking downstream (left) and from the top when the detector is rotated to face the sky (right).}
\label{fig:detector2}
\end{figure}

The bracket holding the stages on the breadboard sits on a fixed base-platform having a precision slot that allows the bracket to slide along the $x\textsf{--}$axis.
The bracket is (manually) pushed and fixed horizontally such that the chips are centered at the nominal zero-dispersion axis of the electron beam according to the design.
The detector design and the version installed are shown in Fig.~\ref{fig:detector1} and~\ref{fig:detector2}, respectively.

\begin{figure}[pos=!ht]
\centering
\begin{overpic}[width=0.99\textwidth]{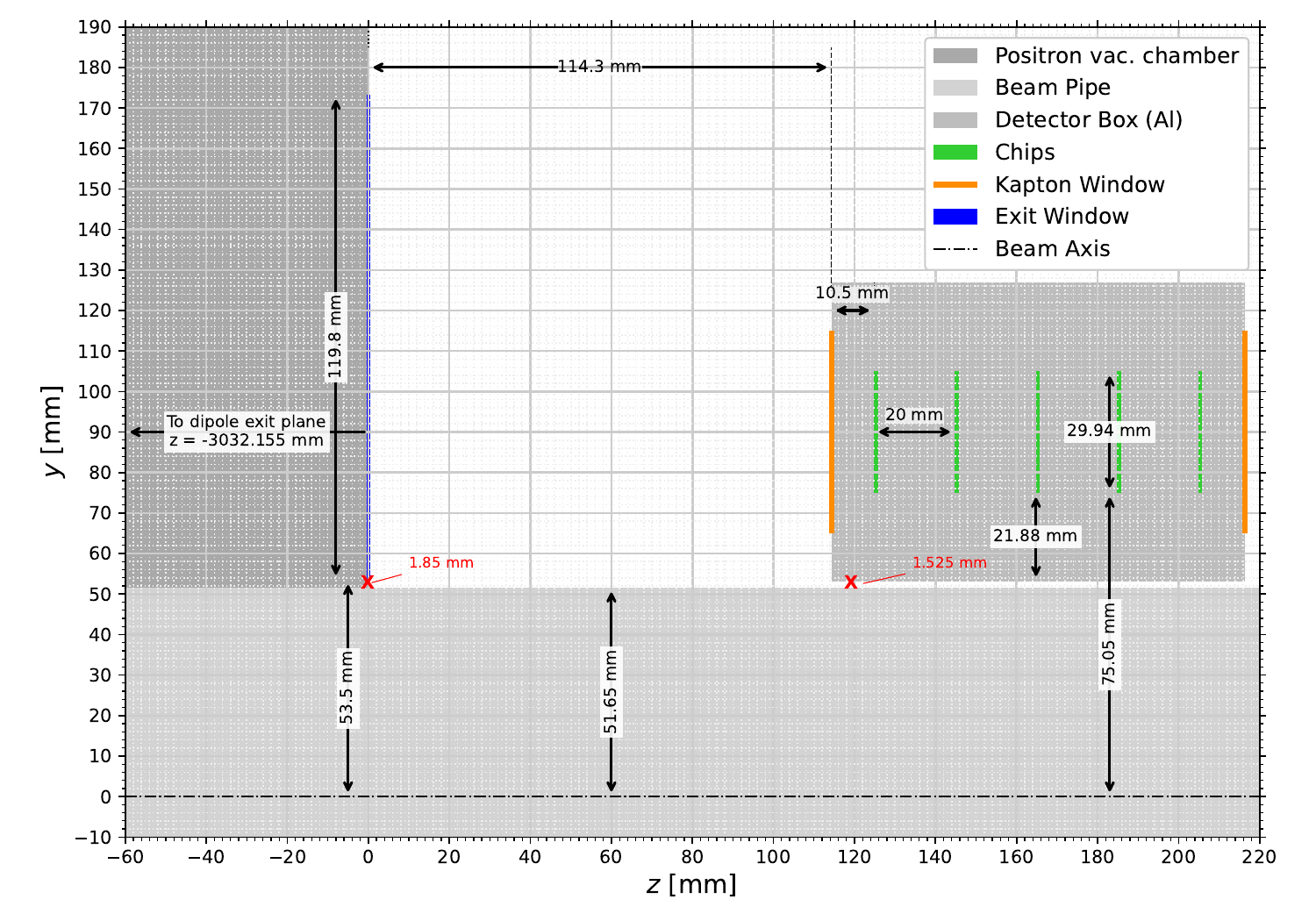}\end{overpic}
\caption{A scaled down illustration showing the side ($y\textsf{--}z$) view of the detector area, focusing on the different distances that are relevant for tracking setup.}
\label{fig:distances}
\end{figure}

The different dimensions, which are relevant for the tracking are marked in Fig.~\ref{fig:distances}.
The distance along the $z\textsf{--}$axis between the detector box wall and the vacuum exit window is 114.3~mm, where the distance of the first chip from the box wall is 10.5~mm.
Thus, the distance along the $z\textsf{--}$axis that the positrons travel in air between the vacuum exit window and the first chip is 124.8~mm.
We note that the two walls of the detector box have a hatch, roughly twice wider than the chip size (to allow clearance for the signal positrons), where these hatches are sealed with a $20~\mu{\rm m}$ thin Kapton tape.
The lowest possible position of the detector along the $y\textsf{--}$axis as allowed by the safety limit-switch is such that the distance between the bottom face of the box and the top of the beampipe is 1.525~mm.
The distance between the bottom face of the box and the middle of the chip along $y$ is 36.85~mm.
The distance along $y$ between the top of the beampipe and the zero-dispersion axis according to the design ($y=0$) is 51.65~mm.
Hence, the lowest point in $y$, where the detector sensitive area starts is 75.05~mm above the beam axis.
For reference, the bottom edge of the vacuum exit window is 53.5~mm above the beam axis and hence there is 21.55~mm (21.88~mm) gap between the bottom edge of the window (top of the beampipe) and the bottom edge of the chips.
This gap is compatible with the angle in the $y\textsf{--}z$ plane at which the positrons exit the window due to the dipole.
We also note that it is possible to control the focusing of the positrons population at the detector ``image'' plane with the quadrupoles. 
Furthermore, their horizontal position along $x$ can be controlled with the XCOR dipole, albeit to a lesser extent.

For completeness, the window height is 119.89~mm, and its width is 50.04~mm.
The dipole exit plane is located 3032.155~mm upstream the vacuum exit window along the $z\textsf{--}$axis, hence the distance between this plane and the first chip is 3156.955~mm.
The dipole aperture in the $x\textsf{--}y$ plane is 44.704~mm in $x$ and 101.854~mm in $y$, where its flange is located 260.04~mm downstream the dipole exit plane and its aperture is slightly smaller than the dipole one along the $y\textsf{--}$axis, at 95.504~mm.

\section{Datasets}
\label{sec:datasets}
Following its installation in Aug 2024 in the FACET\textsf{--}II tunnel, the detector saw its first beam operations in Nov 2024.
A winter shutdown separated this initial run from the Feb 2025 campaign, providing an opportunity to implement a new beam orbit and a few other upgrades.
One important such upgrade between Nov 2024 and Feb 2025 is the switch from a fixed Be window (see Fig.~\ref{fig:setup}) to a retractable system.
It was replaced during the winter shutdown and mounted on a stage, so it can be retractable mostly for the E320 mode of run.
During the replacement, it was discovered that there are severe accumulated impacts from the continuous beam operation in the preceding period.
Namely, a few holes and ``burning'' effects were seen on the surface, where the material was completely destroyed.  

\begin{figure}[pos=!ht]
\centering
\begin{overpic}[width=0.99\textwidth]{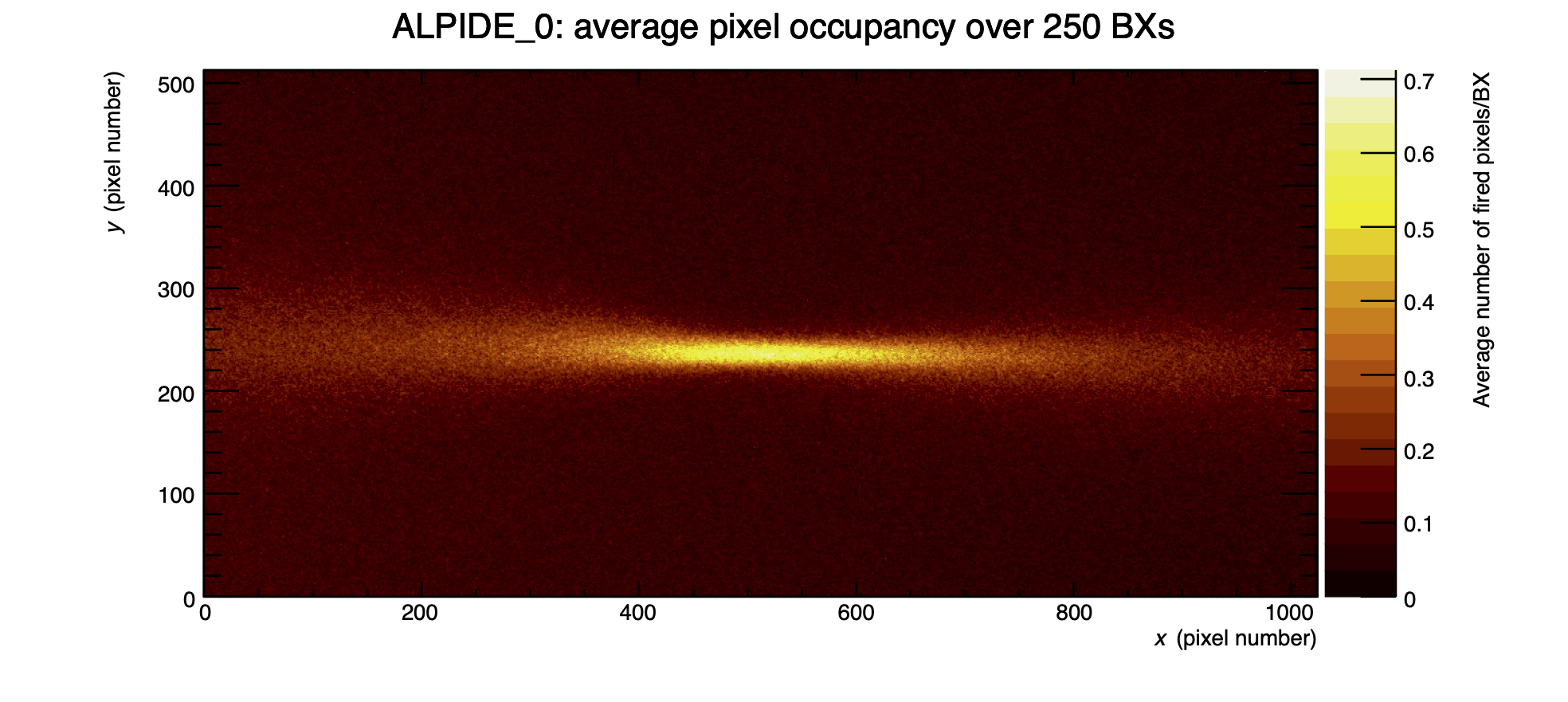}\end{overpic}
\caption{The pixel spatial distribution from a short run (250 BXs) with the Al foil during the Nov 2024 campaign. This picture is similar between all layers of the detector.}
\label{fig:nov_occupancy}
\end{figure}

This retrospective physical observation was supported by the data collected during the first data-taking period of Nov 2024 with the prototype tracker, where extremely large occupancies of $\mathcal{O}(10,000)$ pixels per layer per bunch crossing (BX) were measured (about 2\% of the pixel matrix).
These large occupancies prevented meaningful tracking and prompted effectively the need to mount the Be window on a stage, so it can be retracted during E320 runs.
Nevertheless, we could still observe a very clear positrons distribution, when the Al foil was driven into the beam path.
Fig.~\ref{fig:nov_occupancy} shows the bare pixel occupancy (number of pixels per BX) in the first layer of the prototype tracker during the Al run from Nov 2024.
As can be seen, the positrons spatial distribution is well centered at the chip plane, without needing to physically rotate or move the detector.

Fig.~\ref{fig:nov_occupancy} represents an average situation corresponding to one BX of the beam+Al foil, in a given focusing configuration.
In principle, all ``fired'' pixels (pixels with charge above threshold) are grouped in clusters using the breadth-first search (BFS) algorithm~\cite{cormen2009introduction,moore1959shortest}.
In case of large pixel multiplicities, such as the one seen here, this is significantly faster than the common recursive ``pac-man'' algorithm.
In the specific case seen here, the tight focusing inflates the occupancy at the very center of the distribution such that the clusters in that area are very large, with size exceeding 100's of pixels.
These clusters do not represent well the actual particle multiplicity and this effectively prevents us from running the subsequent tracking algorithms.
Thus, the beam+Al foil runs are mostly good for quickly estimating the position of the positrons at the detector as well as the focusing level.
We only proceed to run the tracking algorithms for a much looser focus.

Finally, following the Nov 2024 experience, and partially also thanks to the preliminary feedback provided by the prototype tracker regarding the large backgrounds in the entire range after the dipole magnet (see Fig.~\ref{fig:setup}), the FACET\textsf{--}II beam orbit has been significantly changed before the 2025 data-taking campaigns started in Feb.
These changes involved improving the quality of the beam as well as its orbit.
With the new orbit, the zero-dispersion axis of the beam has moved such that the spatial distribution of the positrons at the first layer of the prototype tracker has shifted significantly by a few mm's as discussed below.\\

\begin{figure}[pos=!ht]
\centering
\begin{overpic}[width=0.99\textwidth]{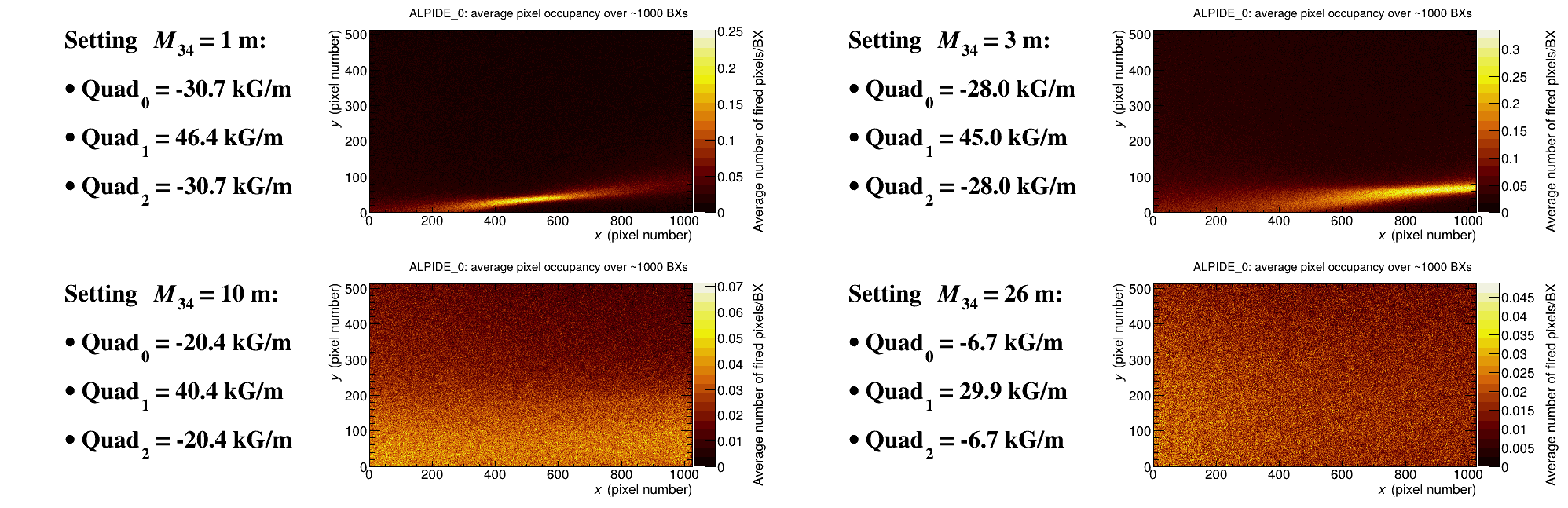}\end{overpic}
\caption{The pixel spatial distribution from a set of short consecutive sub-runs ($\sim 1000$ BXs, each) of Run~490 with the Al foil, while changing the quadrupoles focusing parameters during the Feb 2025 campaign. The beam orbit is different than the one of Nov 2024 seen in Fig.~\ref{fig:nov_occupancy}. This picture is similar across all layers of the detector. The bottom right configuration is similar to the one used in Runs~502 and~503.}
\label{fig:feb_scans}
\end{figure}

The datasets in this study are obtained from four different runs taken during Feb 2025 with the new beam orbit, as listed below.
The run numbers are assigned by the DAQ system - see Appendix~\ref{app:daq}.
The beam conditions and data cleaning algorithm for these runs, based on the different FACET\textsf{--}II monitors, are described in detail in Appendix~\ref{app:cleaning}.
The algorithm removes at most 5\% of all BXs (or triggers) in a run.
\begin{itemize}
\item\textbf{Run~490} with the Aluminum foil in the IP chamber, with $\sim 30$k triggers. The focusing configuration with the three quadrupoles is scanned, where between $\sim 50$k and $\sim 1$k~pixels per layer per BX are recorded. In this run tracking is not attempted.
\item\textbf{Run~502} with a fresh $50~\mu{\rm m}$ thin Beryllium window installed 84~cm upstream the IP chamber. This run is about seven minutes long with $\sim 4$k~triggers, $\sim 2$k~pixels per layer per BX and $\sim 700$~clusters per layer per BX. In this run tracking is attempted. The Focusing used is listed in Table~\ref{tab:settings} and it is similar to the one seen in Fig.~\ref{fig:feb_scans} (bottom-right).
\item\textbf{Run~503} with no material in the beam path for estimating mostly the dump-induced background. This run is about six hours long with $\sim 216$k~triggers, $\sim 100$~pixels per layer per BX and $\sim 20$~clusters per layer per BX.  In this run tracking is attempted.
The Focusing used is listed in Table~\ref{tab:settings} and it is similar to the one seen in Fig.~\ref{fig:feb_scans} (bottom-right).
\item\textbf{Run~510}  parasitic, with the same  Beryllium window as in Run~502. This run is about twelve minutes long with $\sim 7$k~triggers and similar occupancies as in Run~502.
It is only used as a control sample to check the local alignment result obtained with Run~502.
\end{itemize}

Due to the changes in the beam orbit between Nov 2024 and Feb 2025, we start with Run~490 with the purpose of identifying the center of the positrons' spatial distribution at the detector.
This is achieved first with focusing the positrons from the Al foil onto the first tracker layer.
The result is seen in the top left plot of Fig.~\ref{fig:feb_scans} and the quadrupoles' settings are listed on the left.
The rest of the plots in Fig.~\ref{fig:feb_scans} are obtained by defocusing the pattern at the detector through controlling the relevant transport matrix element ($M_{34}$) value~\cite{Brown1982} between 1~m and 30~m.
Comparing Fig.~\ref{fig:nov_occupancy} and~\ref{fig:feb_scans} (top left), it can be seen that the center of the positrons' spatial distribution shifted by a few mm's from the center of the chip (Nov 2024) towards its edge (Feb 2025).
Given that the it was right at the middle of the chip in Nov 2024 (see Fig.~\ref{fig:nov_occupancy}), then by finding the center of the Feb 2025 distribution (Fig.~\ref{fig:feb_scans} top left), we can estimate the effective shift of the beam at the position of the detector.
This can be done by subtracting the two centers positions in terms of pixel numbers and converting to real distance using the known pixel sizes.
Accounting for maxed-out XCOR, this difference corresponds to an effective $\sim -10$~mm shift in $y$ and no shift in $x$ (in the chip frame as seen in Fig.~\ref{fig:feb_scans}).
Notice that, while the XCORR controls the position along the $x\textsf{--}$axis in the lab frame, Fig.~\ref{fig:nov_occupancy} and~\ref{fig:feb_scans} rather show the picture in the chip frame, where the the $x$ and $y$ axes are swapped through a simple rotation by $90^\circ$.
Therefore, the XCORR action in the chip frame is rather along $y$.
The different reference frames are explicitly defined below.
Without the XCOR, the center of the distribution for Feb 2025 was lying almost completely outside of the chip area.
The position with the XCOR is $\sim 1$~mm from the chip edge (or $\sim 5.9$~mm from its center) as seen in Fig.~\ref{fig:feb_scans} (top left).
This means that the XCOR provides a maximum correction of $\sim 3.6$~mm at the detector.
While the physical position of the detector did not change between Nov 2024 and Feb 2025, we can still account for the new beam orbit as an effective shift of the detector box position.
We will show later that the approximate $-10$~mm shift deduced here simply from the center of the distribution on the chip in Run~490 is fully compatible with the same shift found from tracking using Run~502. 

The detector can, in principle, slide along the $x\textsf{--}$axis on the slot seen in Fig.~\ref{fig:detector1} to compensate for the new orbit shift, but it was not done during the Feb 2025 campaign due to tunnel access limitations.
This operation was done later in March 2025, but it is not discussed here further.

As mentioned above, the large focusing used during the first half of  Run~490 (Fig.~\ref{fig:feb_scans} top) has a rather adverse impact on the subsequent tracking step since, combined with the very large positrons multiplicity produced per BX in the Al foil, it leads to a severe pixel-clusters merging problem.
Therefore, the focusing parameters used in Runs~502 and~503 are similar to the ones seen in Fig.~\ref{fig:feb_scans} bottom-right.
These conditions are listed in Table~\ref{tab:settings}.
\begin{table}
    \centering
    \begin{tabular}{lcc|l}
    \toprule
    \midrule
        Setting & Run~502 & Run~503 & Central positions along $z$~[m]\\
    \toprule
        Bunch charge [nC] & 1 & 1 & \\
        Bunch rate [Hz] & 10 & 10 & \\
        Dipole [GeV] & 10 & 10 & 2005.94\\
        Quad$_0$ [kG/m] & -7.64 & -6.66 & 1996.98\\
        Quad$_1$ [kG/m] & 28.55 & 29.86 & 1999.21\\
        Quad$_2$ [kG/m] & -7.64  & -6.66 & 2001.43\\
        XCOR [kG$\times$m] & 0.062 & $10^{-6}$ & 2002.815\\
        Object plane & Be window & Al foil & Be: 1991.99, Al: 1993.27\\
        Image plane & \multicolumn{2}{c|}{First tracker plane} & 2009.5\\
        $E$ image [GeV] & 2.5 & 2.5 & \\
        $M_{34}$ [m] & 26 & 26 & \\
    \midrule
    \bottomrule
    \end{tabular}
    \caption{The settings used in Runs~502 and~503. The $z$ object for Run~502 is the Be window, while for Run~503 it is the Al foil (which was retracted during that run). 
    The quadrupoles, dipole and XCOR magnets' lengths are 0.973~m, 0.914~m and 0.23~m, respectively.
    The $z$ values are given in the FACET-II coordinate system, where for reference, the E320 IP is located at $z=1992.83$~m and the vacuum exit window at $z=2009.37$~m.}
    \label{tab:settings}
\end{table}
In all runs of the Feb 2025 campaign, the electron beam bunch charge was set to 1~nC at 10~Hz repetition.
The image plane was set to the first tracking layer and the imaged energy was set to 2.5~GeV, which is roughly the expected center of the positrons' spatial distribution at the lowest position of the tracker.
The dipole settings are quoted in GeV, where the convention is that 10~GeV electrons are deflected by a fixed 6~mrad at the exit plane of the dipole volume.
With an effective magnet length of 0.914~m, and assuming a perfect uniform field over that volume, we can calculate the field strength at $B_{x}^{\rm Dipole}\simeq 0.22$~T.
The maximal XCOR dipole field (over a length of 0.23~m) is effectively $B_{y}^{\rm XCOR}\simeq 0.026$~T.
This maximal value was used in Run~502 to obtain the top left picture of Fig.~\ref{fig:feb_scans}.
As it can be seen, even the maxed-out field is not enough to center the positrons at the middle of the chip with the new orbit.
It is, however, still important to apply so the distribution is pushed towards the chips' center.
The tilt of the positron spatial distribution in the top left of Fig.~\ref{fig:feb_scans} vs Fig.~\ref{fig:nov_occupancy} is partially due to the XCOR dipole.\\

\begin{figure}[pos=!ht]
\centering
\begin{overpic}[width=0.99\textwidth]{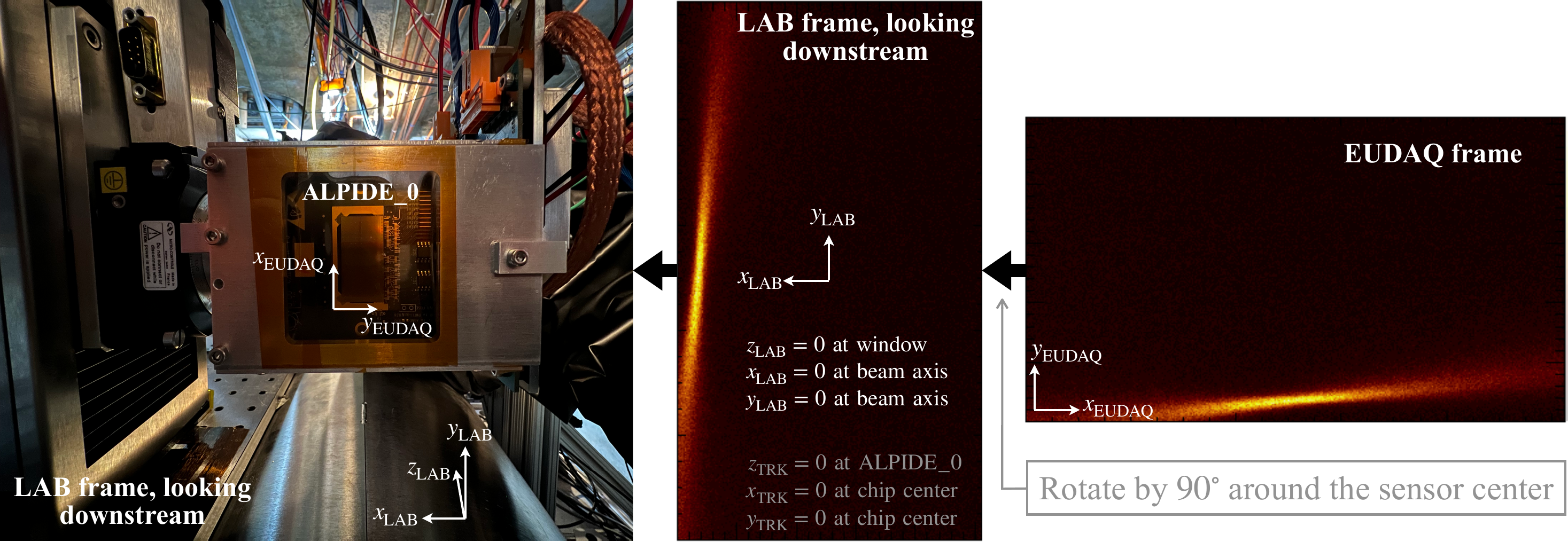}\end{overpic}
\caption{An illustration of the three relevant coordinate systems for this study. The right plot shows the picture in the EUDAQ frame as in Fig.~\ref{fig:feb_scans} top left.
The middle plot shows the picture after the transformation to the LAB frame, looking downstream from the exit window.
The TRK frame is similar to the LAB frame with the exception that the origin is shifted to the middle of the first chip. The left plot shows the physical system with the LAB and EUDAQ coordinate systems illustrated on it.}
\label{fig:coordinates}
\end{figure}

We note the coordinate system seen in Fig.~\ref{fig:nov_occupancy} and~\ref{fig:feb_scans} as the ``EUDAQ frame'' since this is simply showing the pixel number in the two dimensions.
EUDAQ~\cite{EUDAQ2020,Liu_2019} is the data acquisition system used to read the detector out, as discussed in detail in Appendix~\ref{app:daq}.
The EUDAQ frame correlates with the laboratory frame (LAB) by a simple rotation of $\theta=90^\circ$ around the $z\textsf{--}$axis.
Since the EUDAQ frame has no $z$ dimension in principle, we chose to place $z_{\rm LAB}=0$ at the exit window.
The tracking frame (TRK) is similar to the LAB frame with a simple shift from the actual position of the detector such that the origin is right at the center of the first chip (\verb|ALPIDE_0|).
This is illustrated in Fig.~\ref{fig:coordinates}.

\section{Hough Transform Tracking}
\label{sec:houghtransform}
Comparing the baseline rates seen between Run~502 with the Be window and 503 without it (as discussed and seen in Appendix~\ref{app:cleaning}   Fig.~\ref{fig:cleaning_502} and~\ref{fig:cleaning_503}, respectively), it is clear that the large difference in the fired-pixels multiplicity comes from the secondary and far less collimated particles generated in the Be window itself (compared to the primary electron beam).
Pixel clusters obtained in the ALPIDE chip can be almost exclusively attributed to real minimum-ionizing charged particles; mostly secondary electrons and positrons.
These particles are produced mostly by wide-angle photons interacting with the material along the beamline or in the chips themselves (mostly via pair production and Compton scattering).
The contribution of other charged particle species (mostly pions, protons, muons) is negligible as discussed in~\cite{Borysov:2025ehq}.
Hence, we treat these clusters as real particle ``hits'' (see~\cite{Santra:2022sjw} and references therein).
With $\sim 700$ clusters per layer per BX in Run~502,  the tracking has to be done under extremely high hit density of $\sim 1.7$~hits/mm$^2$.\\

In contrast to our KF-based study~\cite{Borysov:2025ehq}, in this work we base our tracking pipeline rather on the a modified Hough-Transform (HT)~\cite{Hough:1959qva,osti_4746348,R_Mankel_2004,Fruhwirth:2020zbo} algorithm.
Conceptually, the pipeline is composed of two steps: the seeding and the track fitting.
The latter step is a maximum-likelihood fit of a straight line in 3D constrained to the volume of the detector only.
That is, without attempting to back-propagate the track through the magnets to its production point in the foils, and without fully accounting for scattering in the material and energy loss, like it is typically done in KF-based algorithms.
The former (seeding) step in this study is rather the heart of the pipeline.
As discussed extensively in~\cite{Borysov:2025ehq}, the seeding processes that initiate any KF-based tracking algorithm require essentially a momentum guess.
Here, we keep the seeding completely momentum-agnostic.
We leave the full KF-based study for a future followup work since it requires a very good knowledge about the actual precise positioning and alignment of all relevant beamline elements as well as the detector itself.
Contrarily, the HT-based seeding is sensitive almost only to the ``local alignment'' of the detector elements with respect to themselves.
Albeit to a lesser extent, it is also sensitive to the ``global alignment'' of all beamline elements (including the detector).
As shown below, a ``momentum-agnostic'' seeding under the extreme hit density conditions mentioned above is very challenging and is demonstrated here for the first time with real data, to the best of our knowledge.\\

The HT is a well established algorithm that has been commonly used for particle tracking in high-energy physics experiments over the last few decades (for a few examples see~\cite{Amram:2008jla,CHESHKOV200635,Storaci:2017850,Fruhwirth:2020zbo}).
In essence, the algorithm transforms between a point representation and a line one, where different points (clusters) ``sitting'' on the same line (track) in the ``real space'', generate lines that intersect at the same location in the ``Hough space''.
By finding these intersections one can group clusters that potentially belong to the same track and attempt to fit the track in the real space.
This group of clusters is called here the ``track-seed'' (or seed for brevity).
In this section, the real space is defined by the TRK reference frame (see Fig.~\ref{fig:coordinates}), where the origin is fixed to the center of the first (\verb|ALPIDE_0|) plane.
In the context of the HT, this choice simplifies the procedure (compared to working in the LAB frame) since the large displacements along $z_{\rm LAB}$ and $y_{\rm LAB}$ lead to unwanted ``condensation'' of the lines in the Hough space, which in turn complicates the search for intersections.
The simplest HT (space) representation is of a straight line, characterized by two unconstrained parameters: a slope $A$ and an intercept $B$.
In this study, we rather use a wave representation since it is also characterized by two parameters, but only one of them is unconstrained: amplitude $\rho$ and an angle $\theta\in [0,\pi]$.
The constrained angle effectively fixes the range, where the intersections have to be searched.
Hence, in the HT we simply write down two wave equations for every cluster in the detector using its $(z,k)$ coordinates, where $k=x_{\rm TRK},y_{\rm TRK}$ and $z=z_{\rm TRK}$:
\begin{equation}
    \rho_k = k\sin\theta_k+z\cos\theta_k.
    \label{eq:wave}
\end{equation}
Hence, there are two waves defined per every cluster in every layer of the detector.
Since seeds can be defined by a straight line equation $k = A\cdot z+B$, the relation between that and the wave representation is simply given by identifying $A \equiv -1/\tan\theta_k$ and $B \equiv \rho_k/\sin\theta_k$.
From Eq.~\ref{eq:wave} it is now possible to see that working in the LAB frame will make the amplitudes of all waves unnecessarily large due to the shifts of the detector in both the vertical and beam directions compared to the TRK frame.
This increases the wave amplitudes, while decreasing the intersection areas.\\

\begin{figure}[pos=!ht]
    \begin{minipage}[c]{0.3\textwidth}
    \centering
    \begin{tabular}{lccc}
    \toprule
    \midrule
    Layer & $x$~[mm] & $y$~[mm] & $z$~[mm]\\
    \toprule
    \verb|ALPIDE_0| & 2.406 & -5.482 & 0 \\
    \verb|ALPIDE_1| & 2.350 & -4.910 & 20 \\
    \verb|ALPIDE_2| & 2.296 & -4.312 & 40 \\
    \verb|ALPIDE_3| & 2.230 & -3.723 & 60 \\
    \verb|ALPIDE_4| & 2.179 & -3.122 & 80 \\
    \midrule
    \bottomrule
    \end{tabular}
    \end{minipage}
    \hfill 
    \begin{minipage}[c]{0.59\textwidth}
    \centering
    \includegraphics[width=\linewidth]{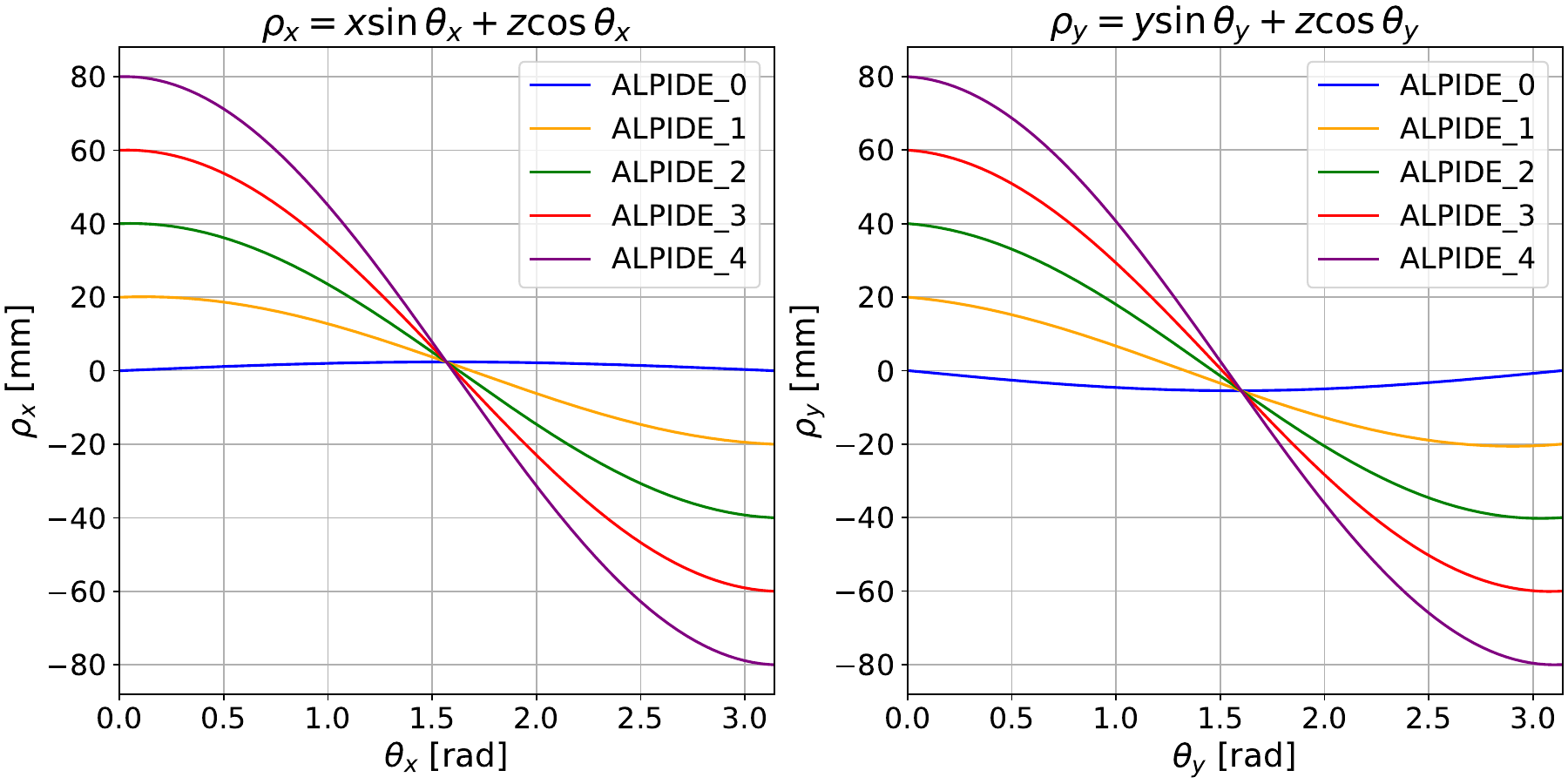}
    \end{minipage}
    \caption{Example of five clusters belonging to the same track in the five detector layers. The $(x,y,z)$ values in the left table indicate the clusters position in the TRK frame. The plots on the middle and right show the corresponding waves in the $z\textsf{--}x$ and $z\textsf{--}y$ Hough-spaces.}
    \label{fig:clusters_example}
\end{figure}

As mentioned, waves of clusters sitting on the same straight line (track) intersect at the same 4D point in the Hough space: $(\theta_x,\rho_x,\theta_y,\rho_y)$.
In reality, different effects (like chips misalignment) essentially force us to work with 4D ``cells'' in the Hough space, rather than with 4D points.
These cells are denoted as $(\theta_x\pm \Delta\theta_x,\rho_x\pm\Delta\rho_x,\theta_y\pm\Delta\theta_y,\rho_y\pm\Delta\rho_y)$, where the determination of the sizes is explained below.
We now need to blindly search for all cells, where pairwise intersections occur to get the seed.
The complexity of this pairwise search in one BX, e.g., in Run 502, is very large due to the large clusters multiplicity per layer.
It can be written as $\mathcal{O}(n_{\rm lyr}^2 m_{\rm cls/lyr}^2)$, where in our case $n_{\rm lyr}=5$ is the number of layers in the detector and $m_{\rm cls/lyr}\sim 700$ is the average number of clusters per layer.
A good seed requires 10+10 intersections in the $\theta_x\textsf{--}\rho_x$ and $\theta_y\textsf{--}\rho_y$ Hough spaces.
In general the number of pairwise intersections is given by $2\times\binom{n_{\rm lyr}}{2}$.
To illustrate the transformation and the intersection concept, we show five real clusters from the Run~502 dataset as an example in Fig~.\ref{fig:clusters_example}.
In this example it is clearly seen that the ten intersections in each of the plots coincide at the same location, indicating that the clusters indeed belong to the same track.
\\

In trying to answer the question of how to efficiently search this 4D Hough space for ``cells'' with $2\times10$ intersections, when there are $\sim 700$ clusters per layer in all five layers of the detector, we first point out that the intersections can be analytically found.
Specifically, an intersection point $(\theta_k^{\rm int},\rho_k^{\rm int})$ of two waves defined by any two clusters $(k_1,z_1)$ and $(k_2,z_2)$ analytically satisfies:
\begin{equation}
\tan\theta_k^{\rm int} = \textstyle{\frac{z_2-z_1}{k_1-k_2}}~~{\rm and}~~\rho_k^{\rm int} = k_1\sin\theta_k^{\rm int}+z_1\cos\theta_k^{\rm int}.
\label{eq:intersections}
\end{equation}
To find these intersections we need to define the Hough space bounds, partition it into cells and query each cell in principle to count the number of intersections in it.
While the $\theta_k$ dimension is constrained as discussed above, the $\rho_k$ dimension of the Hough space is unconstrained, but rather determined form $z$ and $k$.
That is, $-\rho_k^{\rm min}=\rho_k^{\rm max}=\sqrt{k_{\rm max}^2+z_{\rm max}^2}$, where in the TRK frame, the coordinate $z_{\rm max}$ is simply $z(\verb|ALPIDE_4|)$ and $k_{\rm max}$ corresponds to half the sensor dimensions.
Eq.~\ref{eq:wave} further shows that the waves belonging to the same layer originate and end at the same $\rho_k$ value in absolute terms.
That is, we see that the waves start at
$\rho_k(\theta_k=0)= z(\verb|ALPIDE_4|)$ and end at $\rho_k(\theta_k=\pi)= -z(\verb|ALPIDE_4|)$.
Therefore, to avoid counting these points as valid intersections, the bounds of the two Hough spaces are chosen to be smaller than these values:
$\theta_x\in[0.63,2.51]$~rad, 
$\rho_x\in[-61.87,61.86]$~mm, 
$\theta_y\in[0.31,2.83]$~rad and 
$\rho_y\in[-76.66,76.65]$~mm.
Next, we partition these two Hough spaces with the same divider $N_{\theta_x} = N_{\rho_x} = N_{\theta_y} = N_{\rho_y}$~\footnote{The divider does not need to be necessarily the same between all four axes.}.
This divider has to be relatively coarse (a few 100s) before finding the local alignment as discussed in Sec.~\ref{sec:localalignment} below, and fine (a few 1000s) after that.
In this analysis we use 650 and 1700 respectively.
Now, each cell in the 4D space has a unique identifier integer tuple $T = (j_{\theta_x},j_{\rho_x},j_{\theta_y},j_{\rho_y})$, where $j_{\theta_k}\in [0,N_{\theta_k})$ and $j_{\rho_k}\in [0,N_{\rho_k})$ are the cell indices in the  $\theta_k\textsf{--}\rho_k$ reduced and partitioned Hough space.\\

\begin{figure}[pos=!ht]
\centering
\begin{overpic}[width=0.39\textwidth]{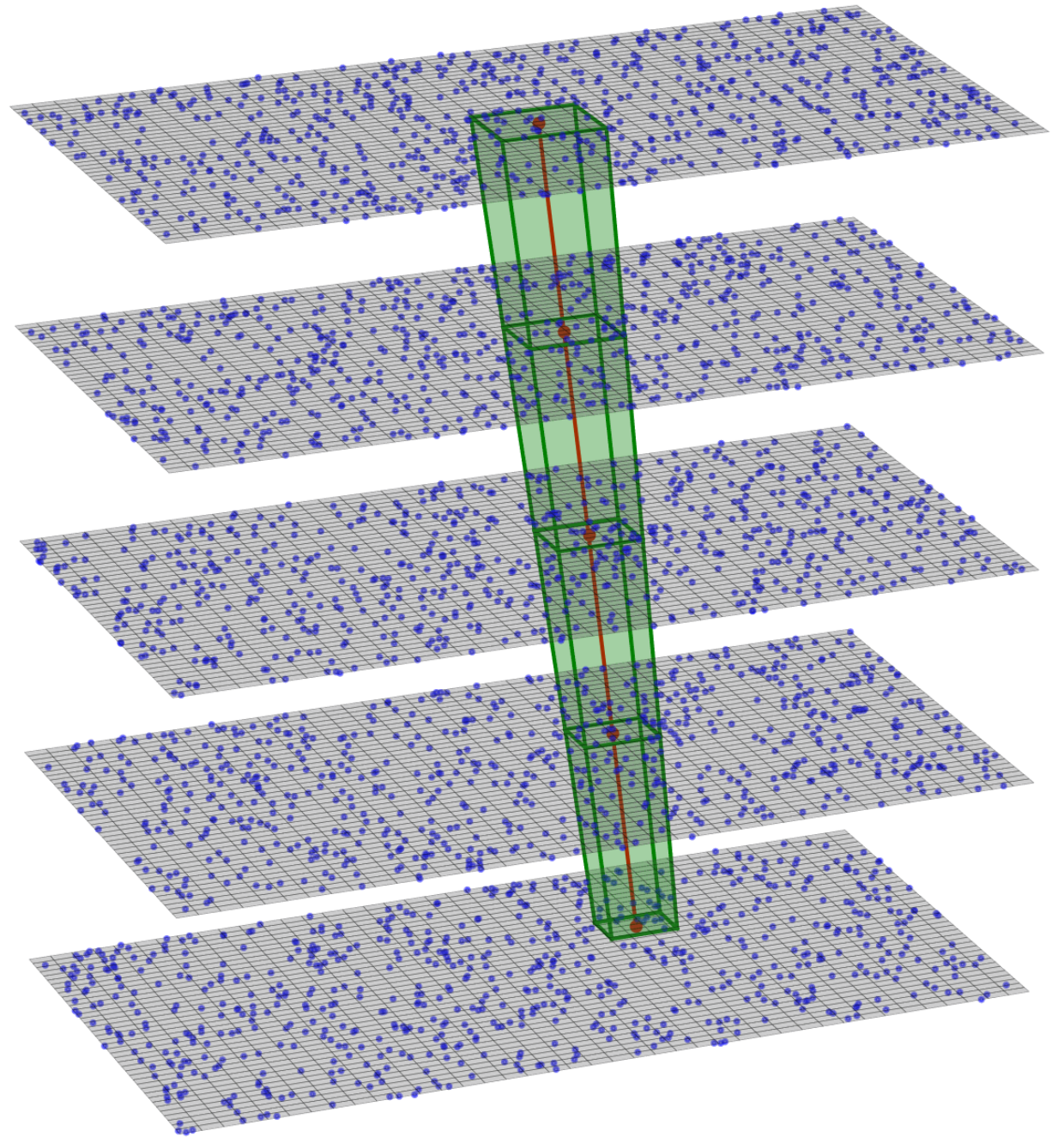}\end{overpic}
\begin{overpic}[width=0.3\textwidth]{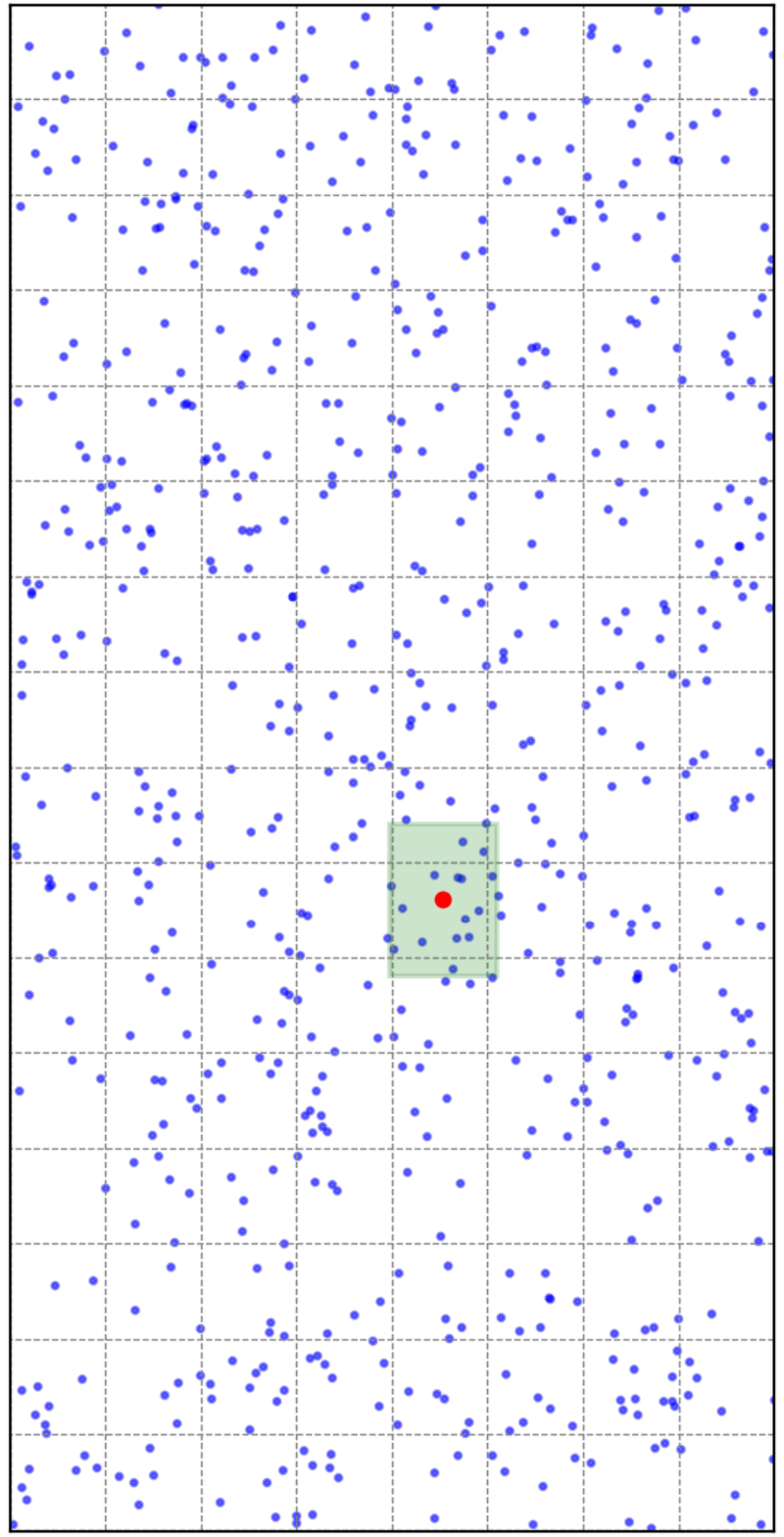}\end{overpic}
\begin{overpic}[width=0.3\textwidth]{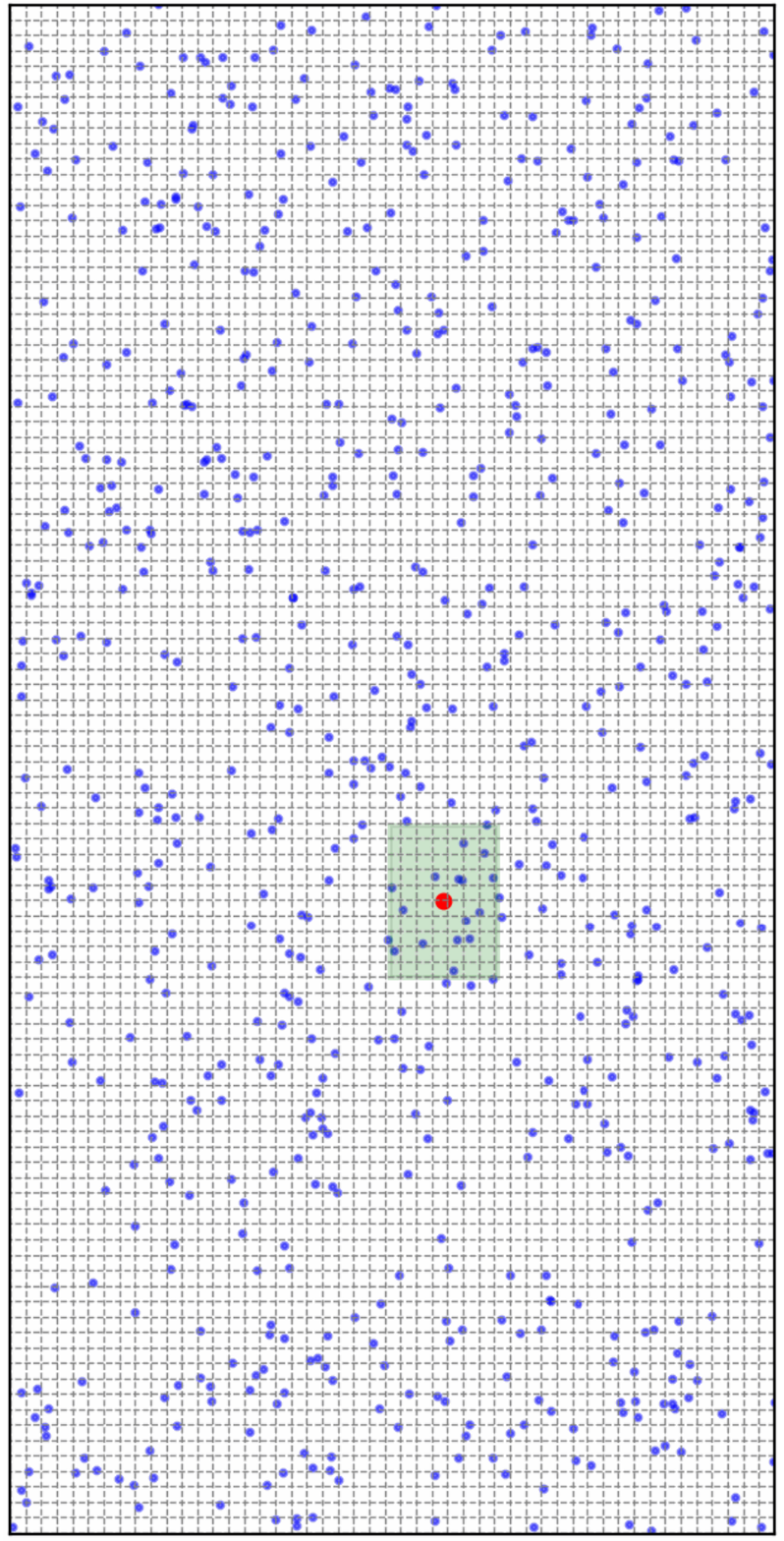}\end{overpic}
\caption{Left: an illustration of the seed-tunnel concept for a typical scenario similar to Run~502, with $\sim 700$ (mostly) random clusters per layer (blue points). The ``good'' clusters (red points) are shown together with the central prediction (red line) and the tunnel (green areas) defined by the five rectangles in the five layers. Middle and right: the impact of the LUT binning on the size of the tunnel. In the coarse case (middle) the list would include all clusters that fall in the bins that overlap with the tunnel in that layer (even if partially) and so it will include too many clusters. In the fine case (right) the list is much smaller.}
\label{fig:seed_tunnel}
\end{figure}

To speed up access and minimize memory usage, we build an ``accumulator'' object that holds all intersection cells.
The accumulator, $\mathcal{A}_{\rm 4D}$ is a dynamically filled, zero-suppressed, hash-table, where the keys are the 4D unique cell identifier tuples $T$, and the values count the intersections in the cells $\mathcal{N}(T)$.
The filling is done once per BX, while checking all possible pairs of clusters in the detector and registering the analytical intersections according to Eq.~\ref{eq:intersections}.
One cell in $\mathcal{A}_{\rm 4D}$ gives a set of five straight lines in the TRK frame using the $(\theta_k,\rho_k)\longleftrightarrow(A_k,B_k)$ relation defined above.
There is one central prediction in the center of the cell and four corners leading to an effective 3D ``seed-tunnel'' (or tunnel for brevity) around the central prediction.
The tunnel is constructed only for those cells that have enough intersections as discussed above.
If a cell has all necessary intersections, $\mathcal{N}(T)\geq 10$, it is accepted immediately.
If a cell has too few intersections, $\mathcal{N}(T)<5$, it is rejected immediately.
If a pivot cell has $5\leq\mathcal{N}(T)<10$ intersections, then to compensate for binning or misalignment effects, we search the nearest-neighbor cells and add all intersections from these.
In this case, the pivot cell is accepted if $\mathcal{N}_{\rm pivot}(T)+\mathcal{N}_{\rm neighbours}(T)\geq 10$.
Once a cell is accepted, its corresponding tunnel is built in the following way.
We translate the cell indices to center and corners (in the Hough space).
We then use these to calculate the slope and intercept parameters at the center $(A_k,B_k)$ and the corners $(A_k^-,A_k^+,B_k^-,B_k^+)$ of the cell.
These five 3D lines allow us to predict (extrapolate) five points per line on the layers in the TRK frame.
These points define a central prediction and a rectangle around it per layer.
Combined together, the five rectangles and central predictions define the seed-tunnel.
The tunnel width is generally not constant since the cells' size in the Hough space depend on the location and orientation of the potential track candidate in the detector.
An illustration of the seed-tunnel concept is shown in Fig.~\ref{fig:seed_tunnel} (left).\\

We now need to check what clusters fall in the seed-tunnel.
This is done with a lookup-table (LUT) object that is filled once per BX.
We fill a list (per layer) with the unique identifiers of all clusters falling in the tunnel.
The tunnel is accepted if there is at least one cluster per layer in it.
We now construct all possible combinations (in the tunnel list) of five clusters (one per layer).
Each such combination is the seed.
The LUT has to cover the entire active chip area and be fine-grained (per layer) such that searching by $(x_{\rm TRK},y_{\rm TRK})$ is immediate.
In principle, fine binning is better, but this has memory consequences.
If the binning is too coarse, there will be too many clusters in the tunnel, leading to too many seeds and hence too many track fit attempts.
The binning we use in this study are $N_x=2000$ and $N_y=4000$.
An illustration of the LUT binning impact with respect to the tunnel size in one layer is shown in Fig.~\ref{fig:seed_tunnel} for the coarse case (middle) and the fine case (right).
Finally, we proceed to the 3D track fit for all accepted seeds.
At the end of the HT seeding algorithm there could be in principle multiple tunnels with multiple seeds per tunnel.
A ``tight'' seeding should ideally lead to $\sim 1$ fit/seed.

\section{Track fit and selection}
\label{sec:selection}
All seeds found by the HT seeding algorithm in a given BX are fitted in the TRK frame as a straight-line fit without imposing any initial momentum or production vertex guess and without attempting to back-propagate the particle to its production point.
The track fitting procedure used in this study employs a maximum likelihood estimation (MLE) to extract the optimal straight-line track parameters $(A_x, B_x, A_y, B_y)$ while rigorously accounting for multiple Coulomb scattering (MPS) only in the detector layers, assuming $X_0({\rm Si})=93.7$~mm, and sensor thickness of $50~\mu{\rm m}$.
This approach incorporates the scattering variance $\theta_{\text{MPS}}^2$ directly into the likelihood model.
The total covariance matrix for the spatial measurements, $V$, is constructed as the sum of the intrinsic detector resolution and the geometrically-dependent scattering uncertainty, yielding $V=V_{\text{meas}}+\theta_{\rm MPS}^2 C_{\rm geom}$.
Here, $V_{\rm meas}$ is a diagonal matrix containing the squared cluster errors, and $C_{\rm geom}$ propagates the angular deviations induced by scattering at each material plane into spatial correlations.
To ensure stable and rapid convergence, the starting guesses for the track intercepts and slopes are derived from a computationally lightweight linear fit to the hit coordinates. Concurrently, the initial value for the scattering variance is estimated using the Highland formula~\cite{ParticleDataGroup:2024cfk} evaluated at a nominal track momentum of 2.5~GeV.
To determine the optimal track parameters, the minimizer optimizes the negative log-likelihood (NLL) function.
This objective function naturally balances the geometrical goodness-of-fit against a penalty for assuming excessively large scattering errors.
Assuming independent measurements in the $x$ and $y$ projections, the quantity to be minimized is given by
\begin{equation}
-2\ln(\mathcal{L})=\mathbf{r}_x^T V_x^{-1} \mathbf{r}_x+\ln(\det(V_x))+\mathbf{r}_y^T V_y^{-1} \mathbf{r}_y+\ln(\det(V_y)),
\label{eq:mle}
\end{equation}
where $\mathbf{r}_x$ and $\mathbf{r}_y$ are the residual vectors between the measured hit coordinates and the predicted track positions in each respective projection.
In this formulation, the total minimized metric is precisely the NLL.
The first term for each axis ($\mathbf{r}^T V^{-1} \mathbf{r}$) acts as an effective $\chi^2$ that minimizes the distance to the hits using the full inverse covariance matrix.
The logarithmic determinant term ($\ln(\det(V))$) acts as a penalty, preventing the (\verb|iminuit|) optimizer from arbitrarily inflating the unconstrained scattering variance $\theta_{\rm MPS}^2$ to trivially reduce the effective $\chi^2$.
The fit yields the four optimized track parameters that define the unperturbed particle trajectory. In addition, it returns the best-fit scattering term $\theta_{\rm MPS}^2$, which represents the variance of the multiple scattering angle acquired by the particle at each layer.
This term can be in principle fixed, e.g., when fitting the local alignment of the detector layers as discussed in Sec.~\ref{sec:localalignment} below.
To retrospectively evaluate the geometrical goodness of fit, the effective $\tilde\chi^2 \equiv \chi^2/N_{\rm DoF}$ metric can be computed, where the number of degrees of freedom ($N_{\rm DoF}$) equals the total number of 1D spatial measurements (clusters) minus the number of freely floating fit parameters.
It yields $2n_{\rm lyr} - 5$, when the scattering variance is fitted, or $2n_{\rm lyr} - 4$ if it is fixed (assuming the track is required to have a cluster in every layer).

The errors on the cluster center estimation at layer $l$, $(\Delta x^l_{\rm trk}$ and $\Delta y^l_{\rm trk})$ are given by the standard error of the mean, calculated as $D_{x(y)}/\sqrt{12 N_{\rm pix}}$, where $D_{x(y)}=0.02924\,(0.02688)~{\rm mm}$ is the size of the ALPIDE pixel in $x$($y$), and $N^{\rm pix}$ is the number of pixels in the cluster.
The average cluster size for MIPs when the ALPIDE is operated with a threshold of $\sim120e$ is $\langle N^{\rm pix}\rangle\sim 2.5$~\cite{Santra:2022sjw}~\footnote{We note that the studies in~\cite{Santra:2022sjw} use $-6$~V bias voltage, while in this data campaign we use zero bias (as in ALICE). However, the bias voltage does not have a critical impact on the average cluster size}.
The expression for the cluster error assumes an independent, uniform charge distribution across the pixels.
While this assumption does not strictly capture the physics of charge sharing, its impact on the final tracking results is negligible since the analysis is restricted to small clusters ($\leq 4$ pixels), which overwhelmingly dominate the dataset.
We note that when deriving the local alignment discussed in Sec.~\ref{sec:localalignment} below, we initially use inflated cluster errors in contrast to the proper errors described above.
These initially inflated errors correspond to the size of the cluster in $x$ and $y$: $D_{x(y)}\times N^{\rm pix}_{x(y)}$, where $N^{\rm pix}_{x(y)}$ is the size of the cluster along $x(y)$ measured in pixels.
This distinction is important only in the very first iteration of the local alignment process, when relative displacements between the chips of the five layers can be expected at the level of 100s of $\mu{\rm m}$ and the proper errors may inflate the $\tilde\chi^2$ estimation unnecessarily.
We denote the goodness of fit metric in the inflated error case as $\tilde\chi^2_{\rm inf}$.\\

The MLE fit result is used to characterize the track properties, where we apply several selection requirements on these properties as discussed below.
Although it is not strictly the metric used during minimization, the $\tilde\chi^2$ defined above can still be retrospectively defined per track as a measure of the ``goodness of fit''.
Besides the $\tilde\chi^2$ (or $\tilde\chi^2_{\rm inf}$), the tracks are characterized by the size of the five clusters, the set of two slopes and two interceptions, and the $\theta_{\rm RMS}$ parameter coming from the MLE fit.
The track parameters can now be transformed to the LAB frame.
Assuming straight lines, the slopes allow us to naively extrapolate the track trajectory to different planes along $z_{\rm LAB}$ as long as these positions are outside of any magnetic field volume and as long as there is no substantial amount of solid material in the path of the particle.
We ignore here the effects of MPS and energy loss (E-loss) in the thin vacuum exit window material and the air.
Nevertheless, we show below that even with these simplistic assumptions we can still reconstruct and select good positron-like tracks from the sample.
The planes at which we examine the extrapolated tracks in the LAB frame are therefore the vacuum exit window ($z_{\rm window}$), the dipole flange ($z_{\rm flange}$), and the dipole exit ($z_{\rm dipole}$).
Contrarily, the KF-based algorithm discussed in~\cite{Borysov:2025ehq} can essentially back-propagate the track all the way to its production point, while accounting for the effects of energy loss and multiple scattering as well as motion in magnetic fields.
However, in this initial work we limit ourselves to the HT seeding and MLE track fitting.
Hence, the discussion is bound to the simple extrapolation down to the dipole exit plane only.
We use this extrapolation to compare the spatial distribution of the fitted tracks at the dipole exit plane to a reference ``spot''.
To define this spot, we simulate the positrons in the target and forward-propagate them through the setup all the way to the detector.
We then filter the simulated positrons that are fully within the detector acceptance and check their spatial distribution at the dipole exit plane.
The simulation is discussed in Sec.~\ref{sec:globalignment}\\

To conclude the track selection discussion, we address the shared-clusters cut applied to the reconstructed tracks.
As the HT seeding allows for cluster-sharing between different seeds, the final population of the reconstructed tracks may include a significant fraction of tracks composed of clusters produced by different particles.
Tracks composed of clusters produced only by a single particle in the detector, will typically have smaller $\tilde\chi^2$ compared to those that contain a ``mixture'' of clusters.
For example, a track can have four clusters from a real signal positron and one cluster due to a random background particle.
Thus, this cut is tasked with removing the ``mixed'' tracks from the sample, leaving only those with the smallest $\tilde\chi^2$.
This procedure is not 100\% efficient, and at the end of the process we are left with tracks that represent ``mixed'' cluster combinations.
The event-wise and track-wise selection criteria are listed in table~\ref{tab:selection} based on the definitions of track characteristics given above.
The event-level selection is kept trivially loose.
\begin{table}[pos=!ht]
\centering
{\footnotesize
\begin{tabular}{p{0.22\textwidth} p{0.68\textwidth}}
\toprule
\midrule
\textbf{Criteria}  & \multicolumn{1}{c}{\textbf{Description of selection requirements}}\\
\toprule
\multicolumn{2}{c}{\textbf{Event level}} \\
\midrule
BeamQC & Beam cleaning as discussed in Appendix~\ref{app:cleaning}\\
0Err   & Event has zero hardware errors as discussed in Appendix~\ref{app:daq}.\\
$N_{\rm hits/det}>0$ & At least one fired pixel per layer.\\
$N_{\rm cls/det}>0$ & At least one cluster per layer.\\
$N_{\rm tunnels}>0$ & At least one HT tunnel as discussed in Sec.~\ref{sec:houghtransform}\\
$N_{\rm seeds}>0$ & At least one HT seed as discussed in Sec.~\ref{sec:houghtransform}\\
$N_{\rm tracks}>0$ & At least one MLE track fitted.\\
\midrule
\multicolumn{2}{c}{\textbf{Track level - baseline}}         \\
\midrule
$\tilde\chi^2_{\rm inf}\leq 50$ & Loose cut on $\tilde\chi^2_{\rm inf}$ (defined with the clusters' inflated errors). This cut is relevant only for the local alignment procedure discussed in Sec.~\ref{sec:localalignment}.\\
Cluster size is $N^{\rm pix}\leq 4$  & Remove the non-MIPs tail of the cluster size distribution (for all five clusters).\\
$y\textsf{--}z$ slope is positive & Remove tracks originating from the top hemisphere in the LAB frame.\\
$\theta_{\rm RMS}^2<10^{-8}$ & Remove tracks with too large MPS prediction.\\
Extrapolated to $z_{\rm window}$ & The track points to the vacuum exit window aperture.\\
Extrapolated to $z_{\rm flange}$ & The track points to the flange exit aperture.\\
Extrapolated to $z_{\rm dipole}$ & The track points to the dipole exit aperture.\\
Reject shared clusters & If two tracks have a shared cluster, the one with the best-$\tilde\chi^2_{\rm inf}$ (or $\tilde\chi^2$) is kept.\\
\midrule
\multicolumn{2}{c}{\textbf{Track level - tight}}         \\
\midrule
$\tilde\chi^2\leq 3$ & Nominal cut on $\tilde\chi^2$ defined with the clusters' proper errors.\\
Spot cut at the dipole & The track points to a small spot at the dipole exit plane, consistent with the expected exit area of the positrons.\\
\midrule
\bottomrule
\end{tabular}
}
\caption{The event-level, the baseline track and the tight track selection criteria.}
\label{tab:selection}
\end{table}

\section{Local alignment algorithm}
\label{sec:localalignment}
Since the placement of the ALPIDE chips on the carrier boards is controlled only to  $\lesssim 100~\mu{\rm m}$ (in $x\textsf{--}y$), and since the mechanical structure that fixes the carrier boards in space is machined with tolerances of $\lesssim 20~\mu{\rm m}$ (in $x\textsf{--}y$ and $z$), we need an algorithm to determine the actual relative alignment of the chips and correct it as necessary.
To reduce the number of free parameters, this can be done with respect to some fixed reference plane, e.g., the first chip.
Typically, the alignment procedure can be performed, e.g., with long cosmic muon runs during accelerator shutdown (as discussed in Appendix~\ref{app:daq}).
However, in this study this is rather done by using data from Run~502, which we also use to estimate the positron-like tracks rates.
We have checked that the alignment result works equally well for a statistically independent data of Run~510 (similar to Run~502).
This means that the result obtained from Run~502 is unbiased and can be used for estimating different quantities using the same data.
The alignment result may still slightly change with time even if the detector is untouched.
The source of these changes include, e.g., temperature and humidity differences in the (uncontrolled) FACET\textsf{--}II tunnel during different periods of the year.

As outlined below, the algorithm (in the TRK frame) starts by tightly distilling the tracks sample.
It then proceeds to fit the misalignment parameters using this sample.
We note that the distilling can be arbitrarily tight and rather empirical.
Hence, this step should not be seen as generally applicable for other problems a priori, while the rest of the steps are generic.\\

\noindent\textbf{Step 1, finding the positrons population:}
We apply the baseline selection with the inflated clusters' errors.
We then use a very loose seeding as discussed in Sec.~\ref{sec:houghtransform} with the 4D HT partition set to be coarse with only $N_{\theta_x}\times N_{\rho_x}\times N_{\theta_y}\times N_{\rho_y} = 650^4$ cells.
The effective tunnel widths for the five layers are relatively wide and range in $190-420~\mu{\rm m}$ along $x_{\rm TRK}$ and $220-580~\mu{\rm m}$ along $y_{\rm TRK}$.
In both cases the width grows gradually, when going from \verb|ALPIDE_0| to \verb|ALPIDE_4|.
The difference between the width in $x_{\rm TRK}$ and the width in $y_{\rm TRK}$ is expected due to the inclination of the positron-like tracks.
These wide tunnels, which correspond to $\sim 2\%$ of the chip's long dimension, allow to compensate the misalignment and therefore ``catch'' the real positrons at the expense of longer runtime due to the large number of possible combinations falling in the relatively wide tunnel.
For reference, the average execution time of one BX on a 2021 Apple MacBook Pro with M1 Max processor is 6.4~seconds, with averages of $\sim 2200$ pixels per layer (five layers), $\sim 700$ clusters per layer, $\sim 28$ tunnels, $\sim 75$ fitted track seeds and $\sim 5$ selected tracks.
Most of this time is taken by (i) the BFS clustering step, and (ii) the pairwise search in the HT space to construct the tunnels.
We then apply the shared clusters rejection and identify the positrons-like track candidates as a peak in the $\tilde\chi_{\rm inf}^2$ distribution.
We note that when using the proper cluster errors at this step, before the local alignment is found, there are no distinct populations emerging from the proper $\tilde\chi^2$ distribution and hence $\tilde\chi^2_{\rm inf}$ is used.\\

\begin{figure}[pos=!ht]
\centering
\begin{overpic}[width=0.99\textwidth]{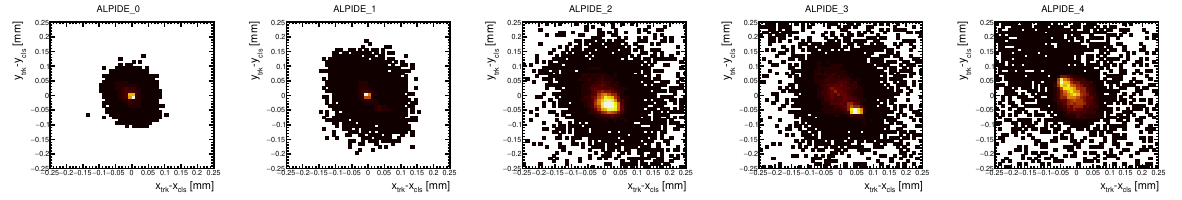}\end{overpic}
\caption{The distribution of the tracks' 2D residuals during step 2 of the alignment process, before applying the respective cut as well as the $\chi^2_{\rm inf}$ and spot cuts.}
\label{fig:residuals_2D_during_alignment}
\end{figure}

\begin{figure}[pos=!ht]
\centering
\begin{overpic}[width=0.7\textwidth]{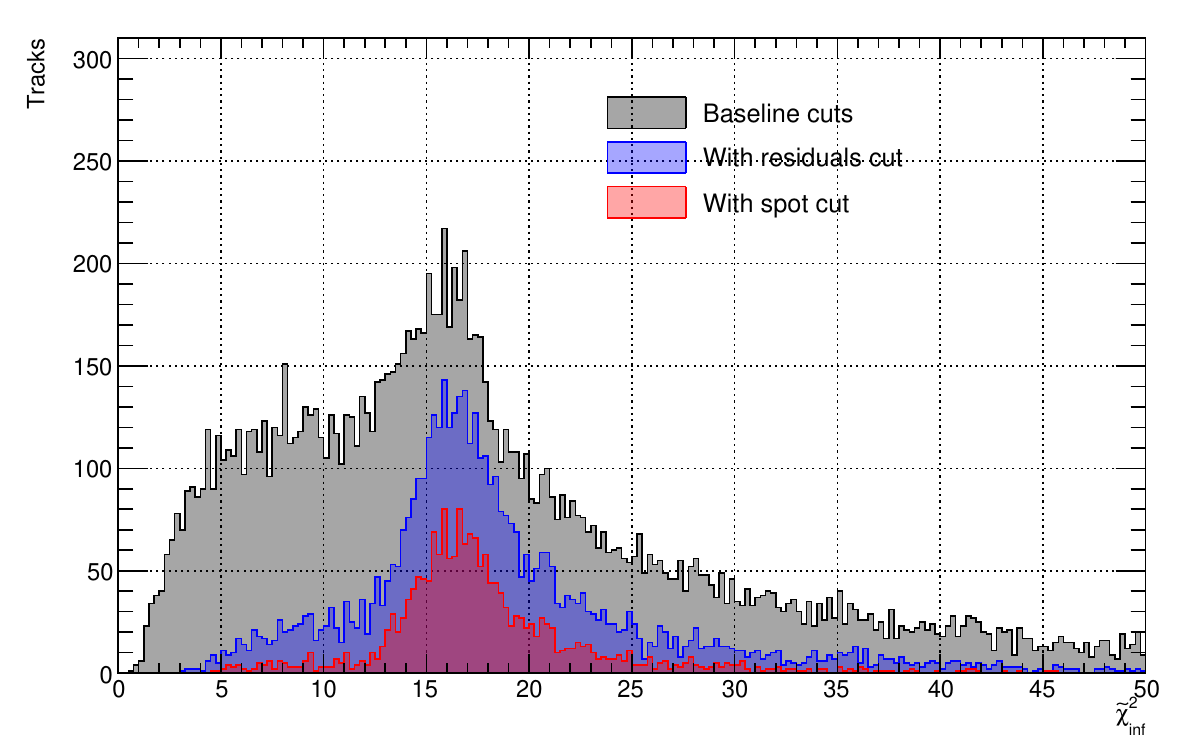}\end{overpic}
\caption{The evolution of the tracks' $\tilde\chi^2_{\rm inf}$ distribution (defined using the inflated clusters errors) after step 2 of the alignment process, excluding the $13<\tilde\chi^2_{\rm inf}<20$ cut.}
\label{fig:chi2_evolution}
\end{figure}

\noindent\textbf{Step 2, clean bad tracks:}
This step aims at cleaning as much as possible the collection of tracks that participate in the alignment fit. 
The cleaning procedure can thus be arbitrarily tight here, where all cuts mentioned below are removed after the alignment fit is completed. 
We start by plotting the 2D residual distributions (per layer) of the surviving tracks from Step 1: $y_{\rm TRK}^{\rm trk}-y_{\rm TRK}^{\rm cls}$ versus $x_{\rm TRK}^{\rm trk}-x_{\rm TRK}^{\rm cls}$, where superscript trk (cls) signify the track (cluster) position in the layer.
This is shown in Fig.~\ref{fig:residuals_2D_during_alignment}.
We now look for potentially clear displacements appearing as localized peak in the residuals distribution of one or more layers.
We isolate this peak by applying a rectangular cut around it, so the tails are removed.
In our case this cut corresponds to $0.02<x_{\rm TRK}^{\rm trk}-x_{\rm TRK}^{\rm cls}<0.08$~mm and $-0.07<y_{\rm TRK}^{\rm trk}-y_{\rm TRK}^{\rm cls}<-0.03$~mm in \verb|ALPIDE_3| only.
With this cut, the tails in the $\tilde\chi^2_{\rm inf}$ distribution are now reduced significantly as seen in Fig.~\ref{fig:chi2_evolution} (blue vs gray).
Finally, tracks extrapolated back to the dipole exit plane point very clearly to a localized spot at $(x_{\rm LAB}^{\rm spot},y_{\rm LAB}^{\rm spot}) = (8,-2)$~mm, with a size of $\lesssim 5 \times 5~{\rm mm}^2$.
Although this spot is positioned well within the dipole aperture it does not appear at the expected location since the setup is not yet aligned globally (this is discussed in later Sec.~\ref{sec:globalignment}). 
Nevertheless, we can still a priori attribute this (mispositioned) spot to real positrons and cut a $5\times 5~{\rm mm}^2$ square around its peak to remove outlier tracks that do not conform with the expected small origin there.
The respective $\tilde{\chi}^2_{\rm inf}$ distribution is shown in Fig.~\ref{fig:chi2_evolution} (red).
Finally, we cut at $13<\tilde\chi_{\rm inf}^2< 20$ to remove the residual tails.
The number of surviving tracks is 1,348.
Only this clean sample of tracks proceed to the subsequent fit of the misalignment parameters as explained below.\\

\noindent\textbf{Step 3, fit the misalignment:}
We now perform a multidimensional fit of the misalignment using \verb|ALPIDE_0| as a fixed reference layer, where the other layers' position are allowed to move with respect to that.
The metric to be minimized is defined as
\begin{equation}
\mathcal{M} = \frac{1}{\mathcal{N}}
\sum_{i=1}^{N_{\rm trgs}}
\sum_{j=1}^{N_{\rm trks}}
{\rm NLL}_{ij},~~~~\mathcal{N}=
\sum_{i=1}^{N_{\rm trgs}}
\sum_{j=1}^{N_{\rm trks}} 1,
\label{eq:metric}
\end{equation}
where the sums run over the number of BXs (or triggers) in the sample ($N_{\rm trgs}$), the number of fitted tracks per trigger ($N_{\rm trks}$).
The tracks NLL values are obtained as explained in Sec.~\ref{sec:selection}, where the MPS parameter is fixed to avoid potential compensation between the ``inner'' (track) and ``outer'' (alignment) fits.
An unconstrained ``inner'' track fit directly competes with the outer geometry optimization; if left floating, the scattering variance $\theta_{\rm MPS}^2$ artificially inflates to absorb spatial residuals of $\mathcal{O}(100~\mu{\rm m})$ caused by misaligned planes, effectively flattening the likelihood gradient.
By fixing $\theta_{\rm MPS}^2$ to the expectation for the nominal momentum of 2.5~GeV, the track is constrained, ensuring that true geometric misalignment manifest as steep NLL penalties that reliably drive the convergence of the ``outer'' alignment minimizer.
The minimization fit of $\mathcal{M}$ has twelve free parameters.
That is, four fitted layers times three parameters per layer, $(\delta x, \delta y, \theta_z)$, where $\delta x~(\delta y)$ is the shift of the layer along $x_{\rm TRK}~(y_{\rm TRK})$ and $\theta_z$ is its rotation angle around the $z_{\rm TRK}$ axis.
All values are initialized to 0 (the perfect alignment case), where $\delta x$ and $\delta y$ are allowed to vary in $\pm 500~\mu{\rm m}$, and where $\theta_z$ is allowed to vary in $\pm 50$~mrad.
We use the \verb|scipy.optimize.minimize| fitter with the `Powell' method, which is a derivative-free optimizer that uses geometry (simplexes) to find the minimum.
If the refitted tracks data within each iteration of the minimizer is noisy, as in this case, the derivative of, e.g., a $\chi^2$ minimization can be unstable.
Derivative-free methods may be slower, but robust against noisy cost functions.
To bridge the derivative-free optimization with rigorous uncertainty estimation, the final spatial parameters obtained from the \verb|scipy| Powell minimizer are passed as the initial coordinates to an \verb|iminuit| object, where the HESSE algorithm is subsequently invoked to extract the covariance matrix.
This allows to extract the statistical uncertainties and correlations for the fitted alignment parameters.
The maximum iterations is set to 2000, while typically the fit converges after less than 10 iterations.
We test the fit stability against variations of the cuts from step 2 and the fit settings themselves.
For all variations, including truncating the fitter maximum iterations parameter, the same result is obtained within a reasonable range of $\pm 10~\mu{\rm m}$ for ($\delta x,\,\delta y$) and $\pm 0.1$~mrad for $\theta_z$.
We can therefore conclude that the result does not correspond to a local unstable minimum and the potential impact of ``weak modes'' (see~\cite{blobel2006software} for a review) is small, if present.
The fit result points to (absolute value) shifts at the level of $\sim 21\textsf{--}64~\mu{\rm m}$ and small (absolute value) rotations at the level of $<1.6$~mrad.\\
\begin{table}[pos=!ht]
\centering
{\footnotesize
\begin{tabular}{lccc}
\toprule
\midrule
Layer  & $\delta x~[\mu{\rm m}]$ & $\delta y~[\mu{\rm m}]$ & $\theta_z~[{\rm mrad}]$\\
\toprule
\verb|ALPIDE_0| (fixed) & 0 & 0 & 0\\
\midrule
\verb|ALPIDE_1| & $-21.0\pm1.3$  & $+37.2\pm1.7$  & $-1.6\pm0.1$\\
\verb|ALPIDE_2| & $+25.6\pm2.5$  & $-48.5\pm3.3$ & $+1.3\pm0.2$ \\
\verb|ALPIDE_3| & $+63.8\pm3.8$  & $-63.4\pm5.2$ & $-1.4\pm0.3$\\
\verb|ALPIDE_4| & $-23.7\pm4.9$ & $+26.4\pm6.6$  & $+0.5\pm0.4$ \\
\midrule
\bottomrule
\end{tabular}
}
\caption{The initial alignment fit result, where the first layer is fixed and the others are floating. The errors quoted are statistical only (Hessian).}
\label{tab:initil_alignment}
\end{table}

\noindent\textbf{Step 4, re-fit the misalignment:}
We use the initial alignment fit results obtained in step 3, and apply the shifts and rotations for all layers (other than \verb|ALPIDE_0|) in the geometry description of the tracking software.
In practice, this means that all clusters (in the TRK frame) are transformed according to the two linear translations and the single rotation per layer as seen in Table~\ref{tab:initil_alignment}.
Then, the process described in Step 3 is repeated iteratively, where each iteration starts from the alignment fit result of the previous one.
Since the initial alignment is found, we can now start the process from a much tighter seeding, use the proper cluster errors, cut more tightly on the (proper) $\tilde\chi^2$ and relax the residuals and spot cuts.
The 4D HT partition is now set to be tight with  $N_{\theta_x}\times N_{\rho_x}\times N_{\theta_y}\times N_{\rho_y} = 1700^4$ cells.
The effective tunnel widths for the five layers are now much narrower and range in $70-160~\mu{\rm m}$ along $x_{\rm TRK}$ and $80-200~\mu{\rm m}$ along $y_{\rm TRK}$.
These tight tunnels are roughly three times smaller in each dimension compared to the loose seeding of step 1.
The choice of this specific partition is motivated by two aspects: (i) it positions us approximately at the accuracy limit achievable in this problem (the $\sim 100~\mu{\rm m}$ mentioned above), and (ii) it is the largest number for which we get $\mathcal{O}(1)$ tunnels and seeds per BX.
More specifically, we now see averages of $\sim 1.7$ tunnels, $\sim 2.1$ fitted track seeds and $\sim 1.2$ selected tracks per BX.
After four iterations, we see very small differences with respect to Table~\ref{tab:initil_alignment} and the final alignment fit results are summarized in Table~\ref{tab:final_alignment}.
\begin{table}[pos=!ht]
\centering
{\footnotesize
\begin{tabular}{lccc}
\toprule
\midrule
Layer  & $\delta x~[\mu{\rm m}]$ & $\delta y~[\mu{\rm m}]$ & $\theta_z~[{\rm mrad}]$\\
\toprule
\verb|ALPIDE_0| (fixed) & 0 & 0 & 0\\
\midrule
\verb|ALPIDE_1| & $-23.5\pm4.6$ & $+39.3\pm2.6$  & $-1.8\pm0.2$ \\
\verb|ALPIDE_2| & $+28.8\pm3.2$  & $-53.9\pm5.4$ & $+1.6\pm0.3$  \\
\verb|ALPIDE_3| & $+64.9\pm3.8$  & $-64.2\pm5.2$ & $-1.5\pm0.3$ \\
\verb|ALPIDE_4| & $-24.6\pm4.9$ & $+27.5\pm6.6$  & $+0.5\pm0.4$  \\
\midrule
\bottomrule
\end{tabular}
}
\caption{The final alignment fit result, after additional four iterations beyond the initial one from Table~\ref{tab:initil_alignment}, where the first layer is fixed and the others are floating. The errors are taken as the maximum between the difference between all five iterations (including the initial one) and the statistical (Hessian) errors from the initial iteration.}
\label{tab:final_alignment}
\end{table}
The final fit result is very similar to the initial one and points to slightly larger (absolute value) linear shifts at the level of $\sim 23\textsf{--}65~\mu{\rm m}$ and slightly larger (absolute value) rotations at the level of $<1.8$~mrad.
The errors are estimated as the maximum of the difference between all five iterations (including the initial one) and the statistical (Hesse) errors from the initial iteration.
The errors are at the level of a few microns in $\delta x$ and $\delta y$, and $<0.5$~mrad in $\theta_z$.\\

\begin{figure}[pos=!ht]
\centering
\begin{overpic}[width=0.7\textwidth]{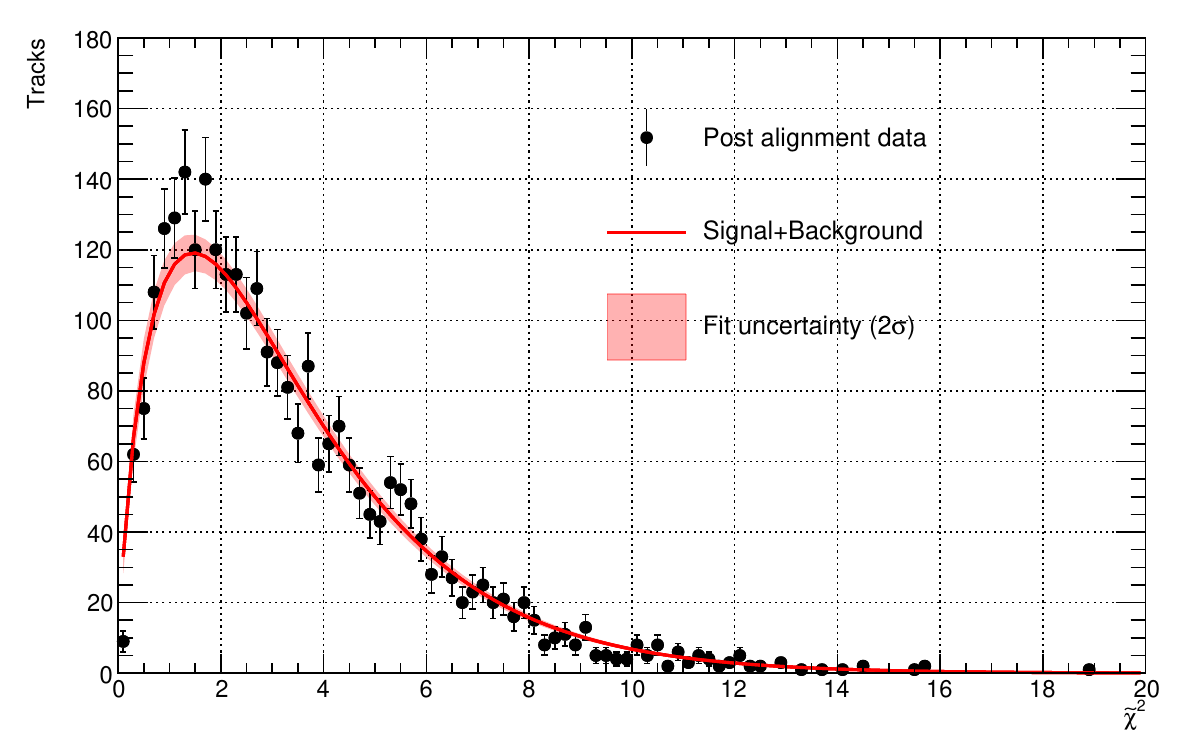}\end{overpic}
\caption{The distribution of the tracks' $\tilde\chi^2$, defined using the proper clusters errors, at the end of the alignment process, excluding the tight $\tilde\chi^2$ cut. The fit is using a $\chi^2$ probability distribution function.}
\label{fig:chi2}
\end{figure}

\begin{figure}[pos=!ht]
\centering
\begin{overpic}[width=0.7\textwidth]{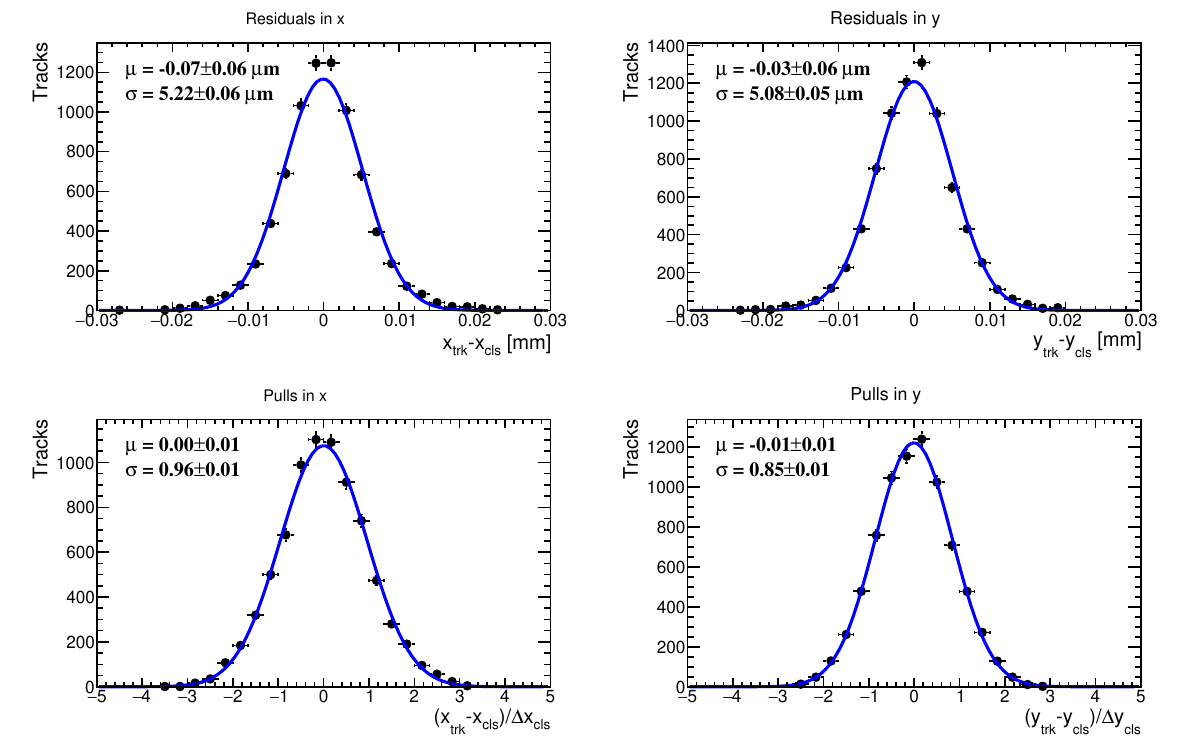}\end{overpic}
\caption{The inclusive residuals distribution in $x_{\rm TRK}$ (top-left) and $y_{\rm TRK}$ (top-right), and the pulls distribution in $x_{\rm TRK}$ (bottom-left) and $y_{\rm TRK}$ (bottom-right) of the five layers at the end of the alignment process, including the $\tilde\chi^2<3$ cut.}
\label{fig:residuals_pulls}
\end{figure}

\noindent\textbf{Step 5, implement the final alignment:}
Starting from the final alignment result of Table~\ref{tab:final_alignment}, we finally proceed to look at the $\tilde\chi^2$, residuals and pulls distributions using the tight seeding and the proper cluster errors as described in step~4.
The $\tilde\chi^2$ distribution is shown in Fig.~\ref{fig:chi2} along with a fit to a $\chi^2$ probability distribution function.
Recalling that the sample still contains some background, the distribution peaks around unity and behaves as expected from the signal tracks.
The fitted MPS angle is $\theta_{\rm MPS}<1~\mu{\rm rad}$.
The 1D residuals ($x_{\rm TRK}^{\rm trk} - x_{\rm TRK}^{\rm cls}$ and likewise for $y$), and 1D pulls ($(x_{\rm TRK}^{\rm trk} - x_{\rm TRK}^{\rm cls})/\Delta x_{\rm TRK}^{\rm cls}$ and likewise for $y$) distributions are shown in Fig.~\ref{fig:residuals_pulls} after applying a $\tilde\chi^2<3$ cut.
A fit of the 1D residuals to a Gaussian returns means of $|\mu|\lesssim 0.07~\mu{\rm m}$ and widths of $\sigma\simeq 5~\mu{\rm m}$ in both dimensions.
Moreover, we see that fitting the pulls to a Gaussian returns means of $|\mu|\lesssim 0.01$ and widths of 0.96 along $x_{\rm TRK}$ and $0.85$ along $y_{\rm TRK}$.
While it is desirable to have means of exactly zero (for both the residuals and pulls), as well as widths of exactly $5~\mu{\rm m}$ (residuals) or unity (pulls), this result is already compatible with the ALPIDE chips' quoted characteristics~\cite{SENYUKOV2013115,AGLIERIRINELLA2021164859,DANNHEIM2019187,MAGER2016434}.
Hence, we accept the fitted alignment at this point.
The individual (per chip) residuals and pulls distributions are examined (not shown here) and fitted in the same way to check for biases.
We find residuals' means of $|\mu|\lesssim 0.2~\mu{\rm m}$ and widths of $4\lesssim\sigma \lesssim 7~\mu{\rm m}$ in both dimensions, where the pulls' means are $|\mu|\lesssim 0.07$ and their widths are $0.7\lesssim\sigma \lesssim 1.2$.


\noindent\textbf{Step 6, estimate systematic uncertainties on the track angles:}
We start by considering a hypothetical extreme case of a systematic $y\textsf{--}z$ angle leading to a wrong subsequent momentum inference.
In this case, e.g., \verb|ALPIDE_0| is shifted down by $\delta y=-100~\mu{\rm m}$ and \verb|ALPIDE_4| is shifted up by $\delta y=+100~\mu{\rm m}$ a priori at the same time, while the middle layers are gradually arranged between these two extremes to create a systematic inclination.
Recalling that the distance between the first and last layers is 80~mm, we can in principle get a maximum local error on the track angle at the level of $\tan^{-1}(200~\mu{\rm m}/80~{\rm mm})\sim 2.5$~mrad.
However, this would be a considerable overestimate since we see that there is no systematic behavior returned by the alignment fit, where furthermore, the errors in Table~\ref{tab:final_alignment} for the largest displacements are at the level of $\sim 10\%$.
Instead, we conservatively use the maximum (absolute) value shifts returned by the fit $\lesssim 65\,(64)~\mu{\rm m}$ in $x\,(y)$, as the potential source of a wrong inclination.
That is, conservatively taking 100\% error on the alignment and translate that into angular error over the 80~mm long detector.
The uncertainties on the tracks inclinations in $x_{\rm LAB}\textsf{--}z_{\rm LAB}$ and $y_{\rm LAB}\textsf{--}z_{\rm LAB}$ are therefore both 0.8~mrad.
The uncertainty due to alignment error on $\theta_z$ is again obtained assuming 100\% error on the maximum value obtained from the fit ($\theta_z^{\rm max}=1.8$~mrad).
This leads to additional maximum effective deviations of $\lesssim 42~\mu{\rm m}$.
In the same way discussed above, this leads to an uncertainty of 0.5~mrad on both inclinations.
Combined together in quadrature, we get a track inclination errors of 1~mrad in both the $x_{\rm LAB}\textsf{--}z_{\rm LAB}$ and $y_{\rm LAB}\textsf{--}z_{\rm LAB}$ planes.\\

With this, the subsequent analysis uses the alignment fit result, the proper clusters' errors and respective $\tilde\chi^2$.
A breakdown of the different ensembles is given below after the local alignment, before the global alignment and with the (last) spot cut from Table~\ref{tab:selection} relaxed.
Starting with 3,655 good BXs in Run~502 (out of a total of 3,826, see Appendix~\ref{app:cleaning} for details), where the average pixel (cluster) occupancy is 2,198 (690) per BX per layer, we are left with 1,559 positron-like tracks that pass the selection modulo the last (spot) cut from Table~\ref{tab:selection}.
This ensemble can now be examined further to extract the global alignment of the detector in the LAB frame as discussed below.

\section{Global alignment algorithm}
\label{sec:globalignment}
In this section we work in the LAB frame.
As discussed in Sec.~\ref{sec:datasets}, between Nov 2024 and Feb 2025 there was a sizable difference in the beam orbit leading to a large shift of the beam in $|x_{\rm LAB}|\sim 10$~mm at the tracker area.
We therefore need a way to find the actual $x_{\rm LAB}$ and $y_{\rm LAB}$ positions and tilts (if any) of the detector box with respect to the new orbit.
While the global alignment is less/not important if one focuses only on estimating the signal and background rates, it allows us to relate the angular distribution of the tracks to their longitudinal momentum spectrum, $p_z$.
In principle, the global alignment can be inferred from the collection of available BPMs along the full FACET\textsf{--}II beamline, but in this campaign the full collection was not saved.
Hence, we need a different way to extract the global alignment of the detector.
We do this by looking at the points, where the back-extrapolated tracks intercept the dipole exit plane and examining their spatial distribution there between data and simulation.
The simulation methods are explained below.
If shifts are observed in data with respect to the simulation, they can be viewed either as the actual physical shifts of the beamline with respect to the detector (since its location hasn't changed from the Nov 2024 campaign), or as conceptual shifts of the detector itself with respect to the actual new beam orbit.
In the discussion below we follow the latter approach to iteratively find the global alignment.
The tracking discussed so far is confined to the detector volume.
Hence, the information that can be extracted about the particles' $p_z$ is limited.
Nevertheless, rough estimates can still be extracted using the field-angle-momentum relation from the naive formula of charged particle motion in a uniform field approximation,
\begin{equation}
p~[{\rm GeV}]=0.3\frac{B~[{\rm T}] \cdot L_B~[{\rm m}]}{\sin\phi},
\label{eq:momentum_naive}    
\end{equation}
where $\phi$ is the track's exit angle from the dipole in the $y_{\rm LAB}\textsf{--}z_{\rm LAB}$ plane, $L_B=0.914$~m is the dipole volume length and $B = 0.22$~T is the dipole field.\\

As a proxy for the ideal configuration, we use toy Monte Carlo (MC) generated with the Xsuite~\cite{Iadarola:2023fuk} software to determine the shifts using all five magnets (three quadrupoles, one XCOR and one spectrometer dipole).
Xsuite uses the common transport matrices formalism and hence the simulation of large statistics samples can be very fast. 
We have checked that the Xsuite results are effectively identical to those obtained with an independent (much slower) code that solves the differential equations describing the particle propagation with numerical integrations.
We operate under a few simplified assumptions, which all have small impact on the actual results:
\begin{itemize}
\item the initial geometry is assumed to be such that all beamline elements, including the detector, are perfectly aligned to their nominally designed positions and orientations,
\item the primary Gaussian electron beam is assumed to be perfectly aligned with the $z_{\rm LAB}$, with emittance $\epsilon_x=\epsilon_y=0.05/\gamma$~mm-rad, a width of $\sigma_x = \sigma_y = 50~\mu{\rm m}$ and length of $\sigma_z = 150~\mu{\rm m}$,
\item Xsuite is used to shoot the primary beam electrons and it is assumed they are all converted to positrons at the target (see discussion below),
\item all magnets' fields are assumed to be perfectly uniform and exist only in the specified volumes,
\item the impact of multiple scattering and of energy loss in both the vacuum exit window and the air medium between the window and the detector is assumed to be negligible.
\end{itemize}
If $z_{\rm LAB}=0$ is set to the IP, the Gaussian beam is shot from $z_{\rm LAB}=-2$~m and propagated to the $50~\mu{\rm m}$ Beryllium foil ($z_{\rm LAB}=-0.84$~m) or the $100~\mu{\rm m}$ Aluminum foil ($z_{\rm LAB}\simeq 0.5$~m) with a constraint that it is focused at the IP.
The mono-energetic beam electrons are then manually ``converted'' to positrons at the foil targets.
First, their charge is flipped.
Second, their energy is re-sampled randomly form a spectrum resulting from a convolution of the Bethe-Heitler approximation and a differential distribution of the $e^+e^-$ pair production in the target foil of thickness $t$.
The Beryllium (Aluminum) radiation length is $X_0=35.3~(8.897)$~cm.
For the Bethe-Heitler approximation for a photon with energy $k$ we use 
$\frac{{\rm d}N_\gamma}{{\rm d}k}\propto \frac{t}{X_0}\frac{1}{k}(\frac{4}{3} - \frac{4}{3}\frac{k}{E_{\rm max}} + \frac{k^2}{E_{\rm max}^2})$, where $E_{\rm max}=10$~GeV is the primary electron beam energy and hence also the Bremsstrahlung photon maximum energy.
The positron production approximation we use is
$\frac{{\rm d}N_{e^+}}{{\rm d}E}\propto \int_{E}^{E_{\rm max}} \frac{{\rm d}N_\gamma}{{\rm d}k} \frac{1}{k}\left[1-\frac{4}{3}\frac{E}{k}(1 - \frac{E}{k})\right] dk$, where $E$ is the positron energy and where the Bethe–Heitler distribution for the positron's energy fraction, $v\equiv E/k$, is $\frac{{\rm d}N_{e^+}}{{\rm d}v}\propto 1-\frac{4}{3}v(1-v)$~\cite{ParticleDataGroup:2024cfk}.
Finally, after resampling the particles' energy, their transverse momenta  components are Gaussian-smeared slightly with $\sigma_{\rm T}^{\rm smear}=100$~keV, accounting for effects like multiple scattering (MPS).
We see no large impact even if this value is inflated e.g. to 800~keV.
The resampled positrons are transported by Xsuite from the point of production through the setup all the way to the last detector layer.
If any of the particles collide with the boundaries of the beamline elements anywhere upstream the detector, the particle is removed.\\

\begin{figure}[pos=!ht]
\centering
\begin{overpic}[width=0.49\textwidth]{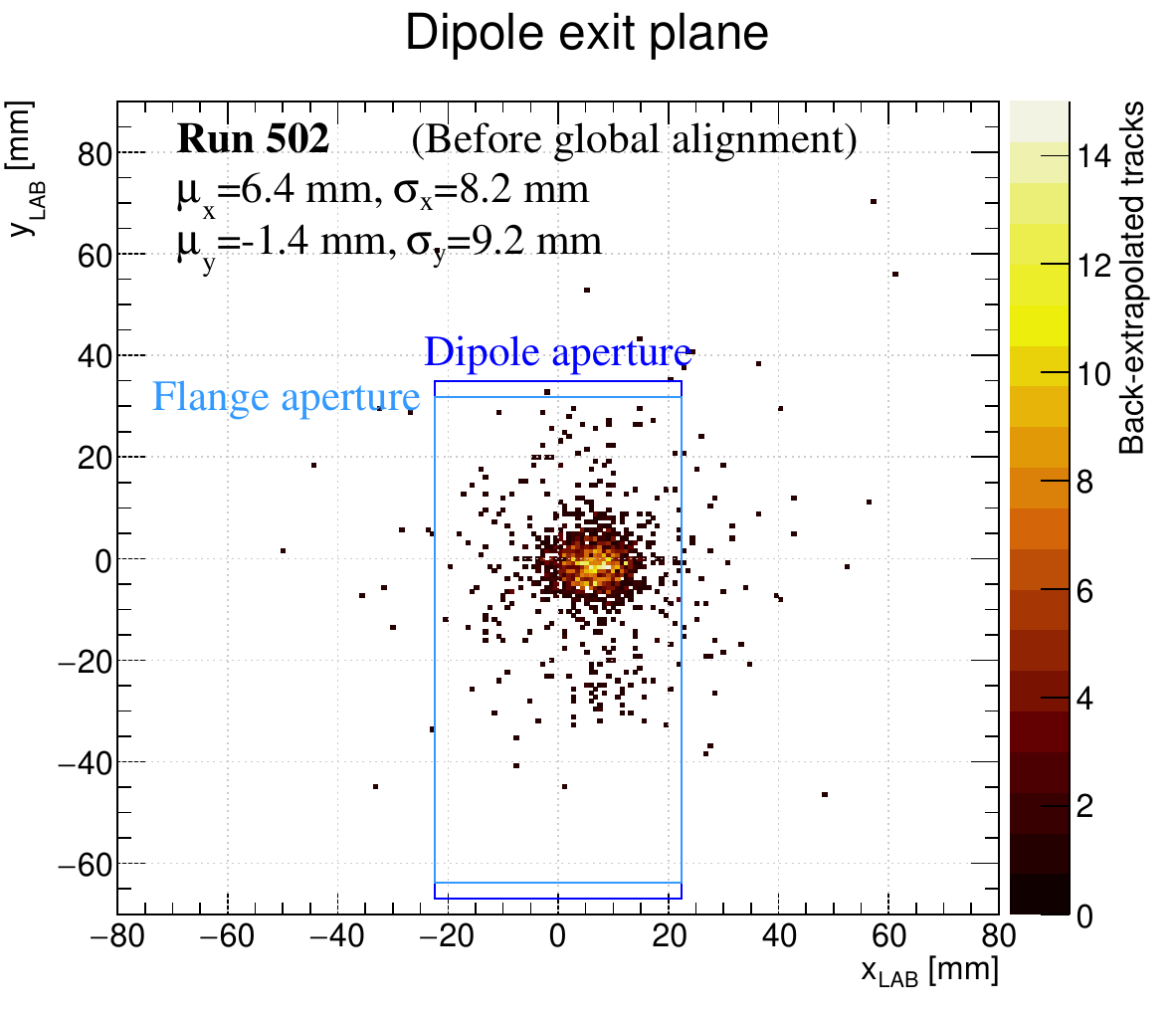}\end{overpic}
\begin{overpic}[width=0.49\textwidth]{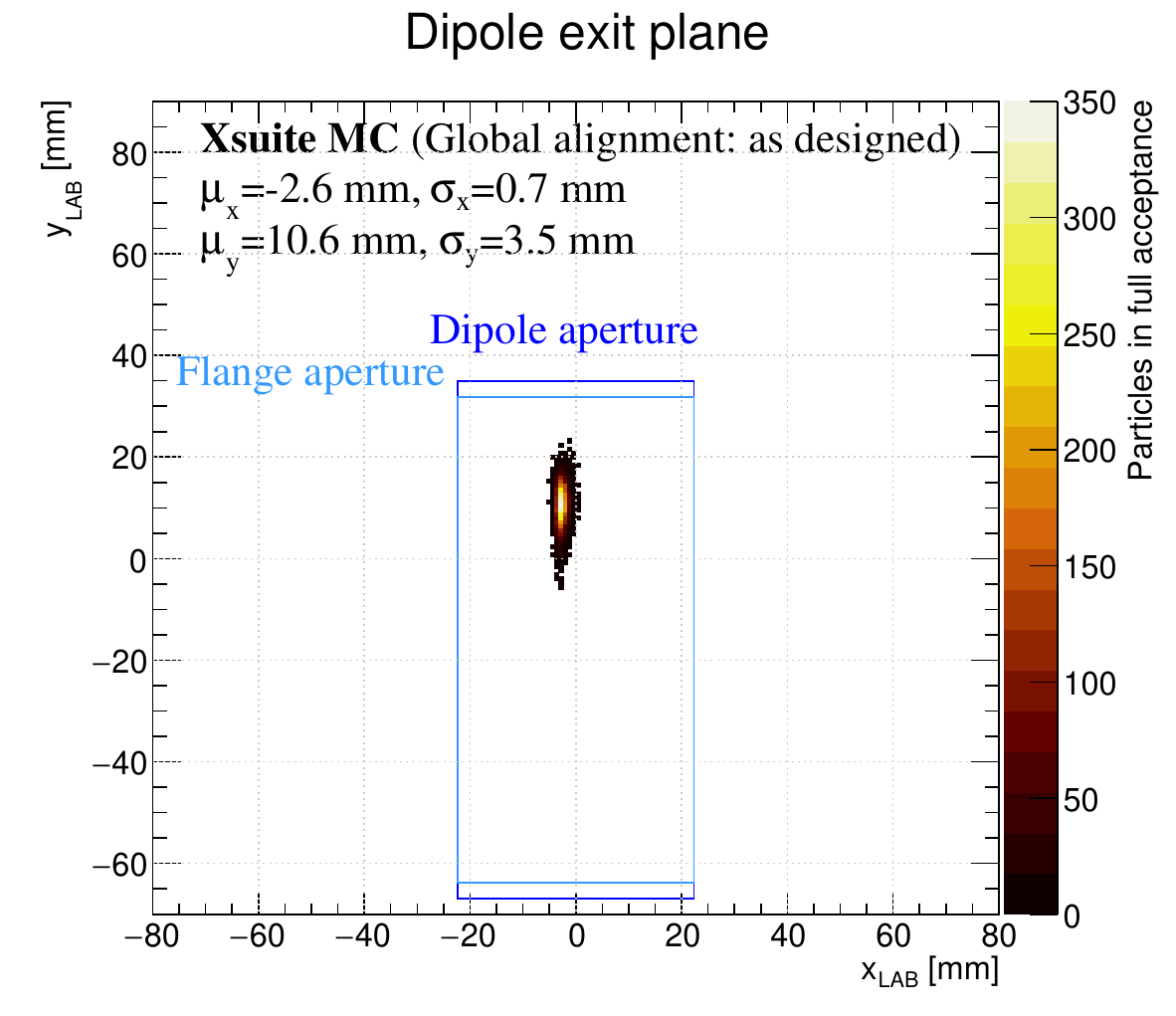}\end{overpic}
\caption{Left: the positron-like tracks in Run~502, before global alignment and without the apertures and spot cuts. Right: the result of the toy MC (using Xsuite) with 250,000 initial particles generated as discussed in the text, assuming the ``as designed'' scenario, for particles traversing exactly five layers.}
\label{fig:dipole0}
\end{figure}

We generate 250,000 particles and compare the distribution of those that have left five ``hits'' in the detector with the reconstructed tracks from Run~502.
We first focus on the distribution at the dipole exit plane in Run~502. 
To do that, we relax all spatial cuts at the dipole exit and dipole flange planes (see Table~\ref{tab:selection}).
Fig.~\ref{fig:dipole0} (right) shows the back-extrapolated tracks to the dipole exit plane in Run~502, before applying the global alignment and requiring further cuts.
This is compared with the the toy MC generated using Xsuite as discussed above for the ``as designed'' case.
It can be seen that even without adjusting the global alignment for the new beam orbit, the bulk of the positron-like tracks are indeed pointing to the dipole aperture (left plot in Fig.~\ref{fig:dipole0}), albeit not at the expected spot for the ``as designed'' case (right plot in Fig.~\ref{fig:dipole0}).
While the spot position in the toy MC case is shifted from zero to $x_{\rm LAB}=-2.6$~mm only due to the XCOR impact, the spot position after tracking is rather shifted to the other direction, $x_{\rm LAB}=6.4$~mm.
Likewise, the shifts in $y_{\rm LAB}$ are $10.6$~mm in the toy MC versus $-1.4$~mm in Run~502.
As mentioned above these shifts can be viewed as conceptual shifts of the detector with respect to the actual beam axis.
The much larger spread of the spot in $x_{\rm LAB}$ for Run~502 (8.2~mm as in the left plot of Fig.~\ref{fig:dipole0} or 3.4~mm with a square $7\times 7~{\rm mm}^2$ outliers removal cut around the spot center) compared to the toy MC (0.7~mm as in the right plot of Fig.~\ref{fig:dipole0}), can be attributed to two main causes:
\begin{enumerate}
\item the large background, recalling that we operate at an extreme hit density of $\sim 1.7$~hits/${\rm mm}^2$, leading to a presence of ``mixed'' tracks,
\item the fact that in the toy MC the effect of energy loss and multiple scattering is not taken into account, where this is expected to be prominent specifically for the vacuum exit window.
\end{enumerate}

\begin{figure}[pos=!ht]
\centering
\begin{overpic}[width=0.99\textwidth]{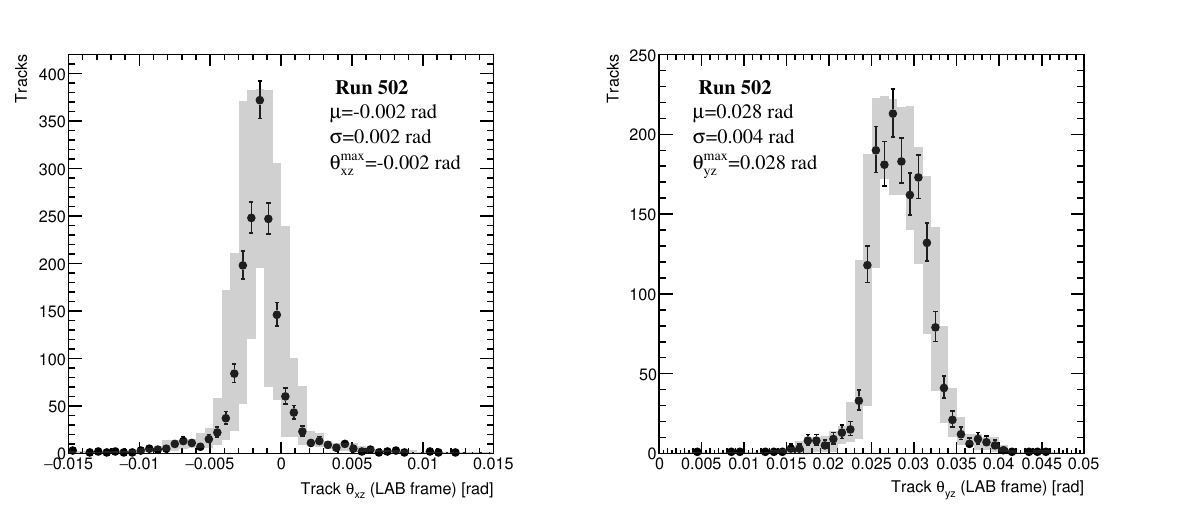}\end{overpic}
\caption{The angle in the $x_{\rm LAB}\textsf{--}z_{\rm LAB}$ plane (left) and in the $y_{\rm LAB}\textsf{--}z_{\rm LAB}$ plane (right) of the positron-like tracks in Run~502, before global alignment from all tracks, without the spot cut. The shaded band represents the uncertainty due to the local alignment.}
\label{fig:angles0}
\end{figure}

Before determining the actual shift in $x_{\rm LAB}$ and $y_{\rm LAB}$, we need to look at another useful quantities that can be used, independently from the global alignment.
That is, the tracks inclination in the $x\textsf{--}z$ and $y\textsf{--}z$ planes.
The angles shown in Fig.~\ref{fig:angles0} are obtained in the volume confined to the detector and hence they are independent of the LAB frame assumptions.
The peak of $\theta_{xz}^{\rm max}$ is at $-2$~mrad, where this is fully consistent with the mean shift of the particles in $x_{\rm LAB}$ due to the XCOR dipole, as obtained in the Xsuite simulation.
Specifically, we see that $\langle x_{\rm LAB}\rangle =-4.9\pm0.9$~mm at the first detector layer, where over a distance of $\Delta z_{\rm LAB}=3156.955$~mm to the dipole exit plane (see Fig.~\ref{fig:distances}), this translates to a typical angle of $-1.5\pm0.3$~mrad considering only the error in $\langle x_{\rm LAB}\rangle$.
This suggests that the detector is effectively not tilted in the $x_{\rm LAB}\textsf{--}z_{\rm LAB}$ plane.
Contrarily, the peak of $\theta_{yz}^{\rm max}$ is at $28$~mrad, where this is not consistent with the Xsuite simulation at the dipole exit plane, for example as seen in Fig.~\ref{fig:dipole0}, suggesting that a tilt in the $y_{\rm LAB}\textsf{--}z_{\rm LAB}$ plane must be introduced.
This tilt introduces an additional uncertainty on the momentum inference as discussed below.

The shaded band in Fig.~\ref{fig:angles0} represents the uncertainty due to the local alignment as discussed in Sec.~\ref{sec:localalignment}.
We obtain this band from 1,000 toys, where in each toy the nominal $\theta_{xz}$ and $\theta_{yz}$ values of all tracks are simultaneously varied using a random (truncated) Gaussian shift with a mean of zero and a width $\sigma$.
Each toy is used to construct a toy histogram similar to the nominal one and the envelope from all toys around the nominal is taken as the band.
There are two systematic uncertainty sources that determine the Gaussian widths size.
The first source is due to the systematic uncertainty values obtained in the local alignment procedure as discussed in Step 6 of Sec.~\ref{sec:localalignment}.
These amount to $\sigma=1$~mrad in both $x_{\rm LAB}\textsf{--}z_{\rm LAB}$ and $y_{\rm LAB}\textsf{--}z_{\rm LAB}$.
The respective Gaussian shifts are truncated at these $\sigma$ values, since these represent already conservative assumptions and letting the shifts float beyond that would lead to an overestimated uncertainty.
Finally, we note that the choice of taking the toys envelope is also leading to a conservative uncertainty band (compared to, e.g., an RMS choice).
The bands seen in Fig.~\ref{fig:angles0} include only this source of uncertainty.\\

To determine the angle between the detector and the beamline in the $y\textsf{--}z$ plane, we first rewrite Eq.~\ref{eq:momentum_naive} explicitly such that $\phi = \theta_{yz}^{\rm max}-\theta_{yz}^{\rm beam-det}$.
We then invert the momentum-angle relation from Eq.~\ref{eq:momentum_naive} and use the facts that the focused energy is fixed to 2.5~GeV (see Table~\ref{tab:settings}), and that the peak of the $\theta_{yz}$ distribution is at $\theta_{yz}^{\rm max}=28$~mrad (see Fig.~\ref{fig:angles0}).
It is then easy to see that the relative angle between the detector and the beamline is $\theta_{yz}^{\rm beam-det}=4$~mrad.
The uncertainty on that angle is estimated from the bin width of the $\theta_{yz}$ distribution, i.e., from the uncertainty on $\theta_{yz}^{\rm max}$ as an additional 1~mrad error.
Finally, when Eq.~\ref{eq:momentum_naive} is rewritten using $\theta_{yz}^{\rm max}$ and $\theta_{yz}^{\rm beam-det}$, it is also easy to see that the two 1~mrad uncertainties on these angles can lead to a sizable uncertainty on the naive $p_z$ estimation.
This impact is discussed in the next section.
In line with the simplistic assumptions listed above in the context of the Xsuite simulation, we neglect the uncertainties due to the error on the actual magnetic field map knowledge, as well as the error on the focused energy value.\\

\begin{figure}[pos=!ht]
\centering
\begin{overpic}[width=0.49\textwidth]{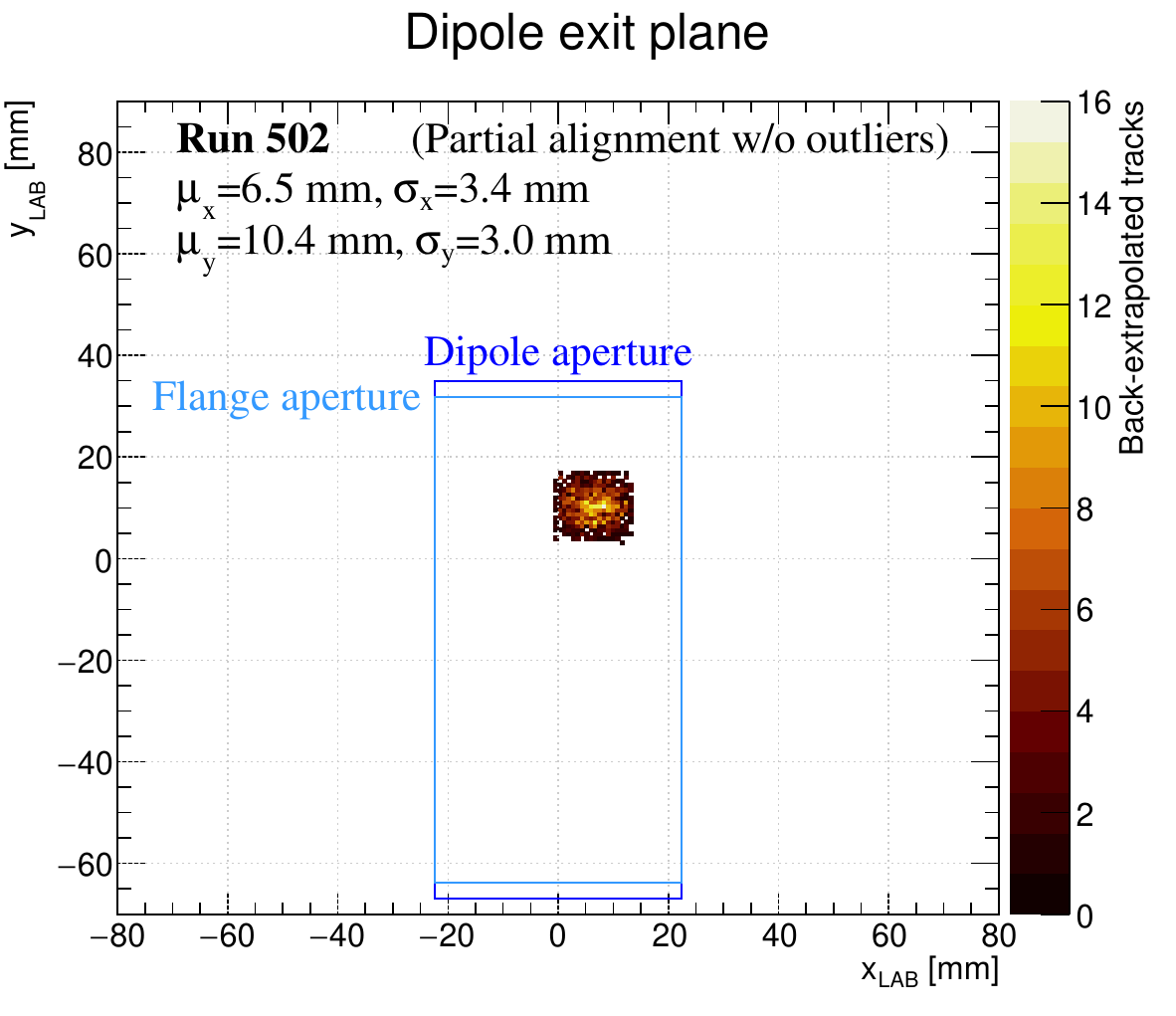}\end{overpic}
\begin{overpic}[width=0.49\textwidth]{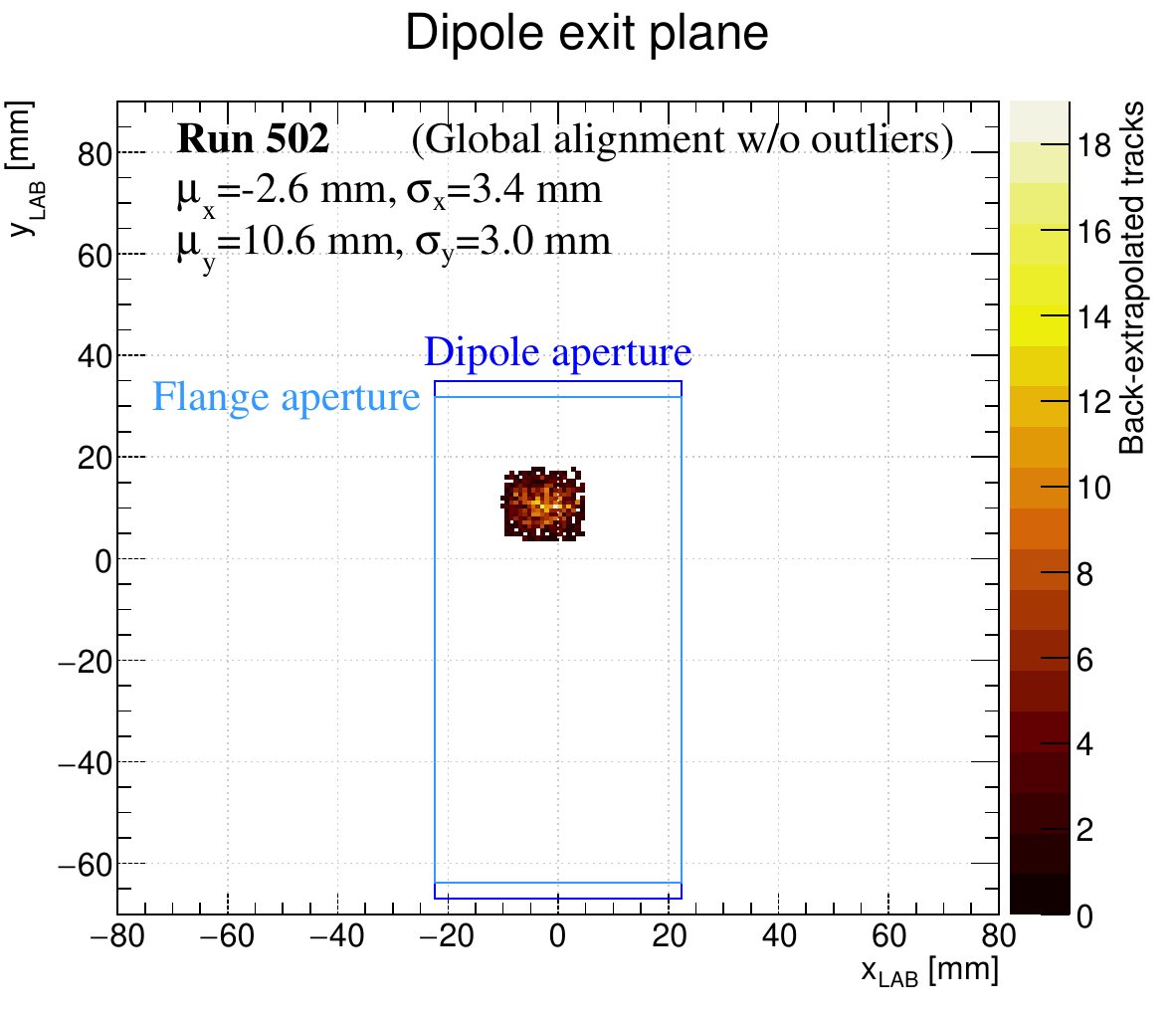}\end{overpic}
\caption{The transverse positions of the positron-like tracks in Run~502 at the dipole exit plane. Left: after partial global alignment, with only the $y\textsf{--}z$ tilt. Right: after full global alignment, including the tilt and two shifts. All cuts are applied, excluding the spot cut, which is replaced by a looser square outliers cut.}
\label{fig:dipole1}
\end{figure}

We first implement a correction to the tilt in $y_{\rm LAB}\textsf{--}z_{\rm LAB}$ (rotation of 4~mrad around the $x_{\rm LAB}$ axis) as obtained from Fig.~\ref{fig:angles0} and the discussion above.
The application of the tilt shifts the center of the spot at the dipole exit to $(6.4,10.3)$~mm.
To obtain the unbiased center of the spot position, this estimate is repeated with a square $7\times 7~{\rm mm}^2$  outliers removal cut around the spot center.
The unbiased spot location is now $(6.5,10.4)$~mm, as shown in Fig.~\ref{fig:dipole1} (left).
That is, the difference between the spot position in the toy MC and Run~502 now stands at $-9.1$~mm in $x_{\rm LAB}$ and $0.2$~mm in $y_{\rm LAB}$.
The shift in $x_{\rm LAB}$ is compatible with the shift seen for the center of the positrons spatial distribution in Run~490 (see the top-left plot in Fig.~\ref{fig:feb_scans}.) compared to Fig.~\ref{fig:nov_occupancy}.
The shift in $y_{\rm LAB}$ represents the actual distance between the bottom of the detector box and the top of the beampipe.
The situation after implementing these two final shifts is seen in Fig.~\ref{fig:dipole1} (right) including the same outliers removal cut used earlier.
The spot is now centered exactly as expected from the toy MC.
To conclude, after applying all shifts and tilts as discussed above, we can see that the mean in $x_{\rm LAB}$ and $y_{\rm LAB}$ are positioned as in the ``as designed'' case seen using the toy MC.
The values used are:
no tilt in $x_{\rm LAB}\textsf{--}z_{\rm LAB}$  (around $y_{\rm LAB}$), tilt in $y_{\rm LAB}\textsf{--}z_{\rm LAB}$  (around $x_{\rm LAB}$) of 4~mrad, shift in $x_{\rm LAB}$ of $-9.1$~mm, and shift in $y_{\rm LAB}$ of $+0.2$~mm.
Clearly, these four simple corrections do not represent all phenomena and fine structures of the new orbit.
For example, the small angle of the positrons spatial distribution seen in Fig.~\ref{fig:feb_scans} (top left) cannot be accounted for by the analysis discussed above.
However, the global alignment procedure described above is sufficient for the purpose of this preliminary study.

\section{Results: Tracking in a dense background environment}
\label{sec:results}
\begin{figure}[pos=!ht]
\centering
\begin{overpic}[width=0.49\textwidth]{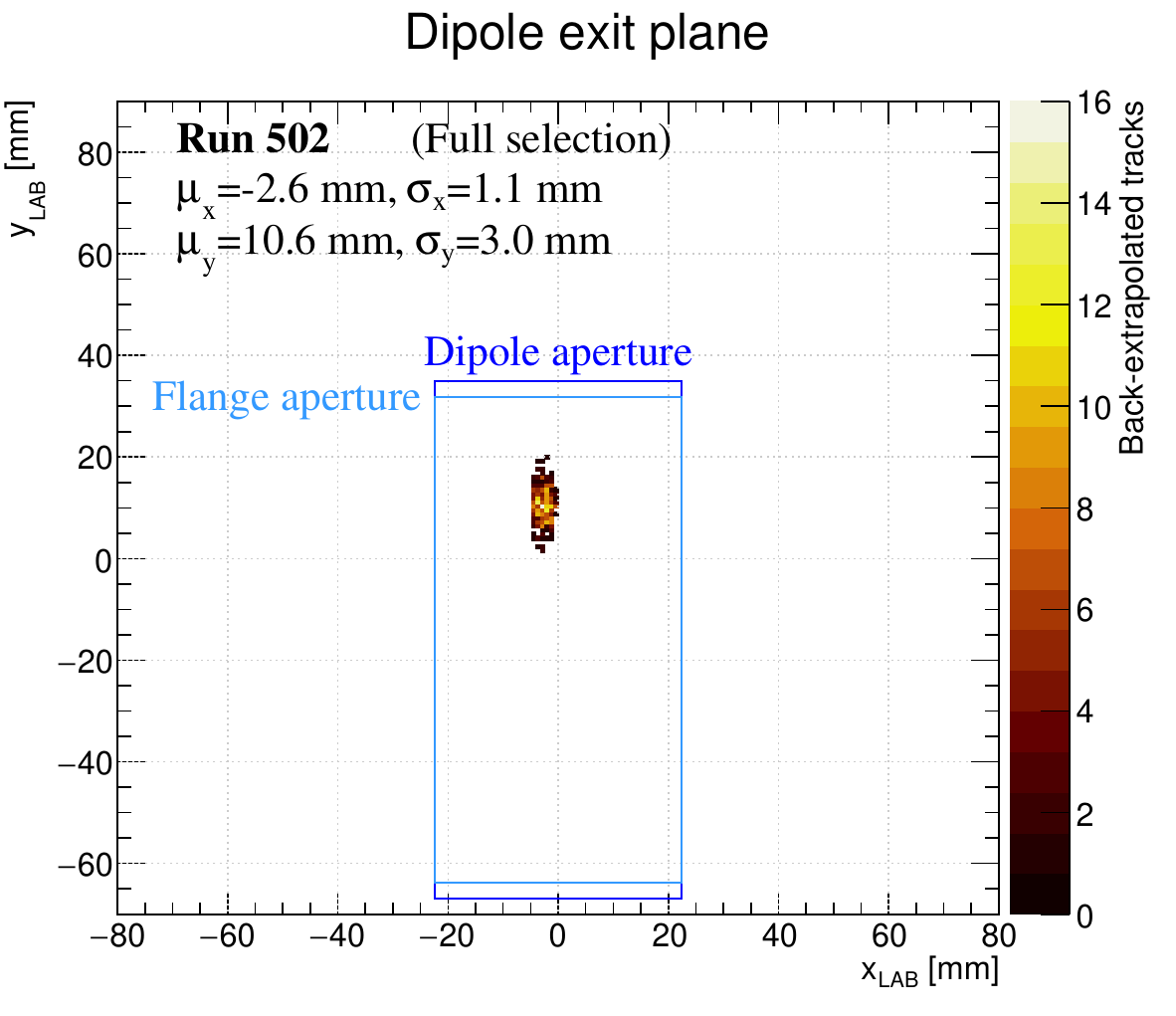}\end{overpic}
\begin{overpic}[width=0.49\textwidth]{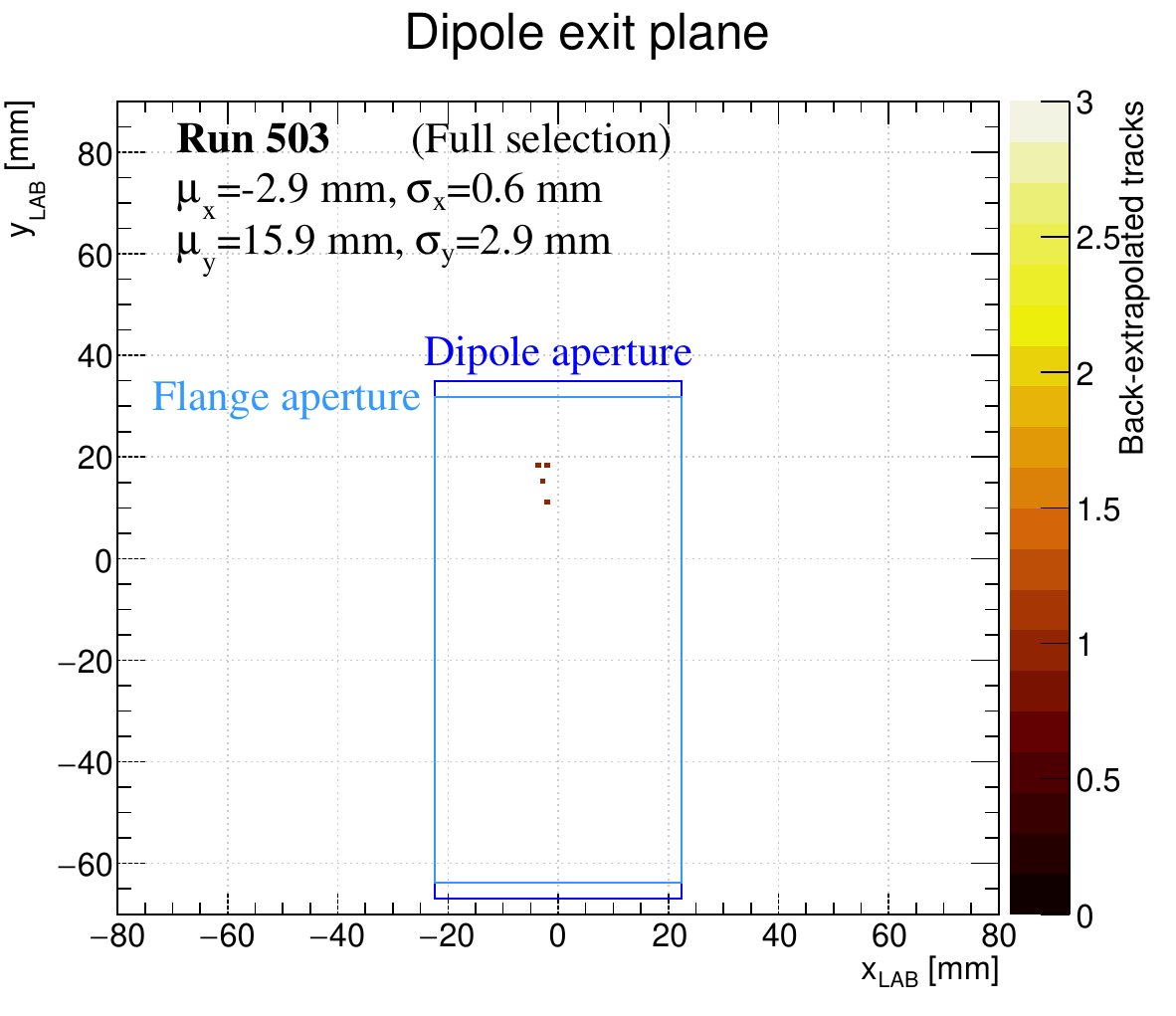}\end{overpic}
\caption{The transverse positions of the positron-like tracks at the dipole exit plane after full global alignment and after the full selection, including the spot cut (see Eq.~\ref{eq:spotcut}) for the Be window Run~502 (left) and the dump-only Run~503 (right).}
\label{fig:dipole2}
\end{figure}

The local and global alignment results discussed in Sec.~\ref{sec:localalignment} and~\ref{sec:globalignment} are used in the subsequent analysis.
We can now go back to tightening up the parts of the selection from Table~\ref{tab:selection} that were relaxed during the alignment process.
Particularly, the last two cuts of the ``Track level - tight'' section are restored, where the spot cut is defined as 
\begin{equation}
\left(\frac{x-\mu_x}{R_x}\right)^2
+
\left(\frac{y-\mu_y}{R_y}\right)^2
<1
\label{eq:spotcut}
\end{equation}
with the center point defined as $(\mu_x,\mu_y)=(-2.6,10.6)$~mm and the radii defined as $(R_x,R_y)=(2,10)$~mm.
These values are motivated by the Xsuite picture in Fig.~\ref{fig:dipole0} (right).
With the formulation of this final cut, we can now proceed to estimate the rates of signal and background as well as their $p_z$ spectra.
The final distributions at the dipole exit plane for the Be window Run~502 and the dump-only Run~503 are shown in Fig.~\ref{fig:dipole2} including all cuts from Table~\ref{tab:selection} with the final spot cut defined by Eq.~\ref{eq:spotcut}.

\subsection{Rates}
\label{sec:rates}
\begin{figure}[pos=!ht]
\centering
\begin{overpic}[width=0.8\textwidth]{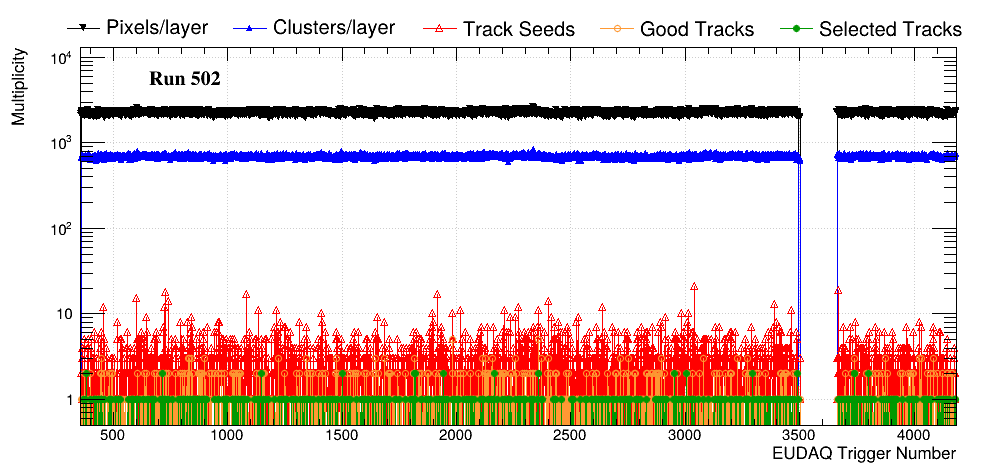}\end{overpic}
\begin{overpic}[width=0.8\textwidth]{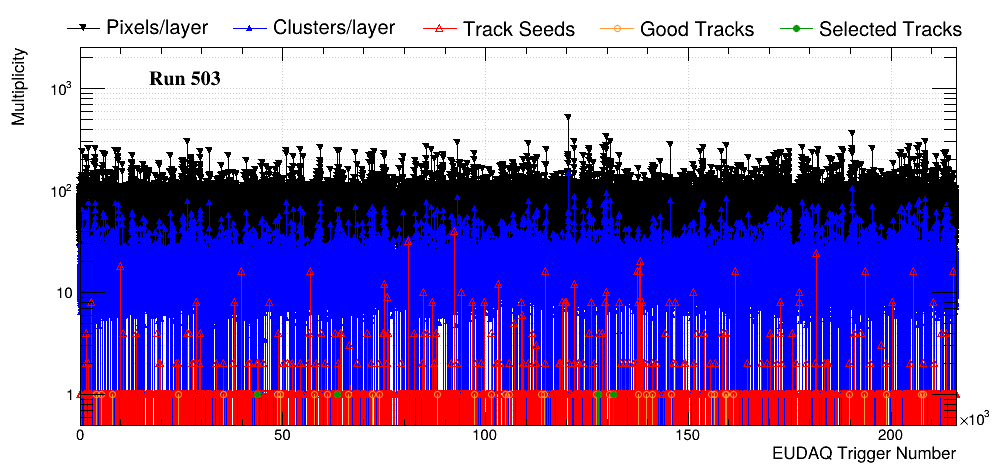}\end{overpic}
\caption{The tracking evolution after full global alignment and after the full selection, including the spot cut (see Eq.~\ref{eq:spotcut}) for the Be window Run~502 (top) and the dump-only Run~503 (bottom).}
\label{fig:rates}
\end{figure}

The rates (number of tracks per good BX) are summarized in Table~\ref{tab:rates}.
The dump-induced background rate in Run~503 is smaller by four orders of magnitude than the signal rate in Run~502.
We furthermore plot in Fig.~\ref{fig:rates} the evolution of the analysis.
This is done starting from the average number of pixels per layer per BX to the number of tracks per BX (going through the intermediate steps of clusters, track seeds and good tracks).
The BX is identified here as an EUDAQ trigger number (horizontal axis in the plots).
It can be seen in Fig.~\ref{fig:rates} that there are no local bursts throughout the two runs, where large numbers of positrons are produced in just one or a few BXs and the production is relatively even across data-taking.
The rate estimation is sensitive almost entirely to the parametrization of the spot cut defined by Eq.~\ref{eq:spotcut} and the tight cut on $\tilde\chi^2$ defined in Table~\ref{tab:selection}.
Systematically varying these two cuts in reasonably small ranges around the nominal values individually, leads to a variation of at most 46\% in the signal rate of Run~502 and 56\% in the background rate of Run~503.
The ranges we use are $(R_x,R_y)=([1,3],[7,15])$~mm (see right plot of Fig.~\ref{fig:dipole0}) and $\tilde\chi^2\leq [2,5]$ (see Fig.~\ref{fig:chi2}).
The $R_x$ variation is tested with the nominal $R_y$ value and vice versa.
The largest deviation in Run~502 corresponds to the $(R_x,R_y)=(1,10)$ variation with 234 surviving tracks (nominal case is 437).
The largest deviation in Run~503 corresponds to the $\tilde\chi^2<5$ variation with 9 surviving tracks (nominal case is 4).
In both runs, the resulting up/down uncertainties are symmetrized using the largest one.
The systematic uncertainty on the signal rate partly stems from the short run duration ($\sim 6$~minutes) and we estimate that longer run times would allow to constrain it further.\\

\begin{table}[pos=!ht]
\centering
{\footnotesize
\begin{tabular}{ccccccc}
\toprule
\midrule
Run & Duration & BXs & BXs  & Hit density         & Tracks & Positrons Rate \\
    & [Hours]  & All & Good & [Hits/mm$^2$]        &        & [Tacks/BX]\\
\midrule
Run~502 & $\sim 0.1$ & 3826 & 3655 & 1.7 & 437 & $(1.20\pm 0.06_{\rm stat.}\pm 0.56_{\rm syst.})\times 10^{-1}$\\
\midrule
Run~503 & $\sim 6$ & 216175 & 205907 & 0.04 & 4 &  $(1.9\pm1.0_{\rm stat.}\pm 2.4_{\rm syst.})\times 10^{-5}$\\
\midrule
\bottomrule
\end{tabular}
}
\caption{A summary of the different yields and respective rates in the two data runs.}
\label{tab:rates}
\end{table}

In E320 we anticipate multiple runs of $\mathcal{O}(1~{\rm hour})$ each, with 5~Hz collision shots and 5~Hz background (laser is mistimed).
The background is expected to be much smaller, at $\lesssim 500$~pixels/BX compared to the $\sim 2200$ in Run~502.
Furthermore, the selection in the E320 runs can be much tighter, since in those runs we expect to produce only a few $\sim 0.01-0.1$ NBW positrons per BX (see full discussion in~\cite{Borysov:2025ehq}) and hence the focusing can be much tighter (similar to the one seen in Fig.~\ref{fig:nov_occupancy} and ~\ref{fig:feb_scans} (top left)).
Therefore, we can (i) reduce the variations' ranges, and (ii) include a narrow region of interest (RoI) cut at the detector layers around the focused positrons pattern.
Crucially, while the mistimed shots only tell us what is the background when there is no collision, the sidebands of the narrow RoI can be used for an in situ estimate of the background during collisions (in-time shots).
However, this kind of cut is not applicable for the focus used in Run~502.\\

To qualitatively compare the measured signal rate with the equivalent extracted from a full GEANT4~\cite{ALLISON2016186,AGOSTINELLI2003250,1610988} simulation of the full experimental area, we have generated $2.6\times 10^{10}$ primary beam electrons, summing up to $\sim 4$~BXs.
The simulation includes the full material distribution description and its subsequent impacts like MPS, E-loss and the generation of secondary particles, as discussed in~\cite{Borysov:2025ehq}.
There is no attempt to digitize the response of the pixels as done in the full-tracker study~\cite{Borysov:2025ehq}.
Instead, we assume that every simulated particle passing through a given layer will lead to a creation of a pixels cluster.
The impact of this choice on the simulation outcome is negligible since the ALPIDEs efficiency for MIPs is nearly 100\%~\cite{AGLIERIRINELLA2017583,MAGER2016434,ALPIDEManual} and since we do not use the cluster size in the analysis of the simulated particles.
The same acceptance requirements are otherwise applied to the simulation as similarly as possible to those used in the data analysis.
That essentially reduces to a requirement that the simulated particles pass through all detector layers and all apertures as well as the dipole spot.
We find 43 such positrons in the simulation.
The scaling to the beam charge (1~nC) of one BX is 0.24.
The production rate is therefore $\sim 10.3\pm 1.6_{\rm stat}$ positrons per BX.
This is about two orders of magnitude larger than the experimentally measured rate from Table~\ref{tab:rates}.
However, we note that the GEANT4 simulation assumes a perfect alignment of the entire setup.
In this case, even with the Run~502 ``loose'' focus settings, the GEANT4 simulation appears to be ``tightly'' focused at the chips (e.g. similarly to the top-left plot of Fig.~\ref{fig:feb_scans} in Run~490).
Therefore, to scale the number obtained above to the scenario that resembles the focusing seen in data we use the Xsuite simulation to manually scan the beam approach configuration.
The scan is stopped at the point that yields roughly the same behavior as seen in Fig.~\ref{fig:feb_scans}.
We then simulate the tight and the loose focusing cases with 10,000 primary particles in each.
The number of particles with five ``hits'' in the detector is recorded in each.
The ratio of that number between the tight and loose focusing cases of the Xsuite MC is computed ($M_{34}=1$~m and $M_{34}=26$~m as seen in Fig.~\ref{fig:feb_scans}, respectively).
This ratio, $r_{\rm focusing}$, is multiplied by the GEANT4 rate mentioned above to get the realistic expected rate given the imperfections of the setup alignment.
The ratio is found to be $r_{\rm focusing}\sim 0.009$, which gives an expected rate of $\sim(0.93\pm0.01_{\rm stat.})\times 10^{-1}$ positrons per BX.
This qualitative estimate is compatible with the measured one from Table~\ref{tab:rates}, even without accounting for any systematic uncertainty on $r_{\rm focusing}$, which is arguably large.

\subsection{Momentum}
\label{sec:spectra}
\begin{figure}[pos=!ht]
\centering
\begin{overpic}[width=0.7\textwidth]{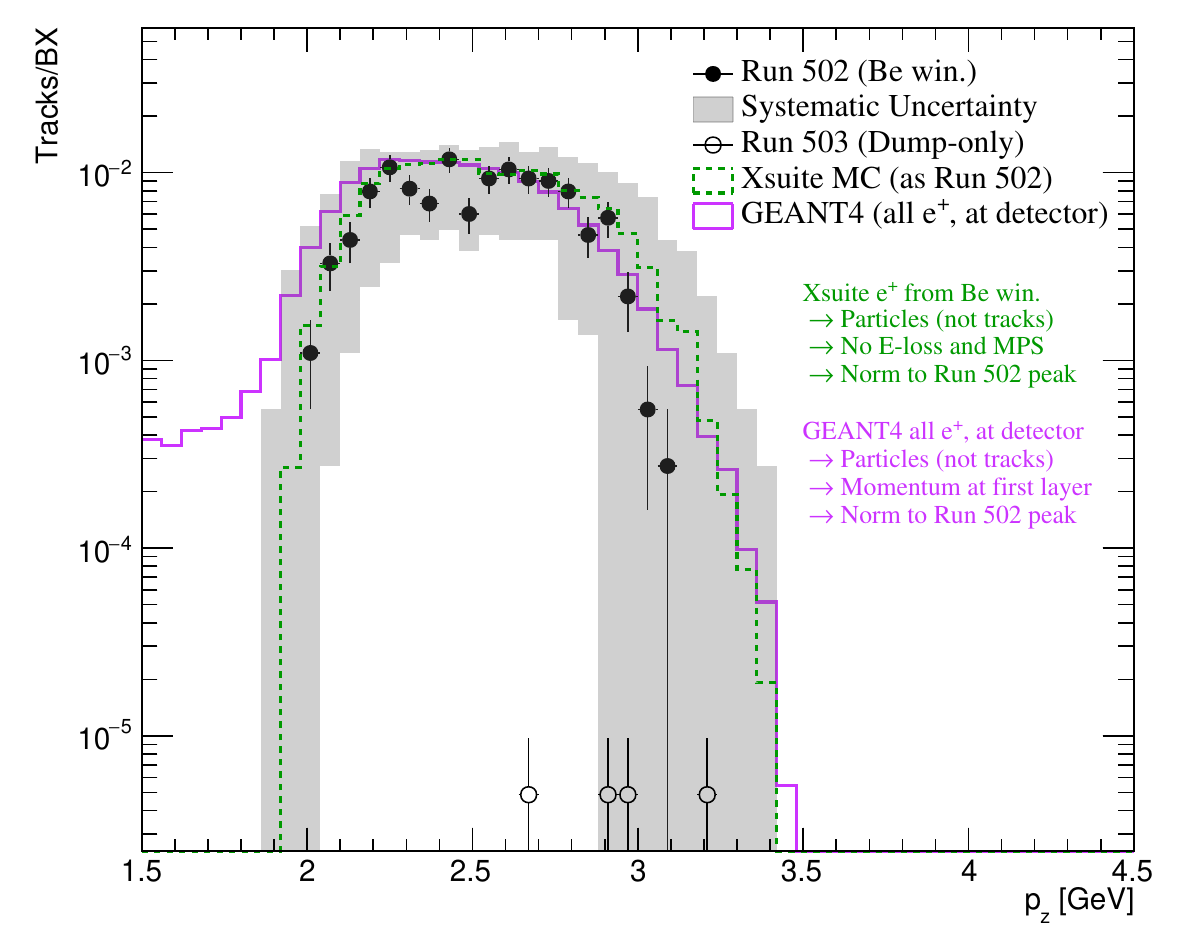}\end{overpic}
\caption{The $p_z$ spectra of positron-like tracks (per BX) after full global alignment and after the full selection, including the spot cut (see Eq.~\ref{eq:spotcut}) for the Be window Run~502 and the dump-only Run~503. The spectra are compared to both Xsuite toy MC (generated as discussed in Sec.~\ref{sec:globalignment}) and full GEANT4 simulation (with biased cross sections). The simulated shapes are normalized such that their maxima are equal to that of Run~502.}
\label{fig:energy_spectra}
\end{figure}

We now compare the momentum ($p_z$) spectra (shape-only) of the data with the simulation.
The spectra are plotted in Fig.~\ref{fig:energy_spectra}, compared with the equivalent from the Xsuite toy MC, as well as a full GEANT4 simulation.
The Xsuite particles' momenta are quoted at the production point since there is no difference between that and any other point along the beamline.
Contrarily, the GEANT4 particles' momenta are quoted at the first layer such that if the particles suffer from scattering and/or E-loss, their momentum will be different than the one at the production point.
This sample includes all positrons produced anywhere along the beamline, selected as discussed above.
We note that by using Eq.~\ref{eq:momentum_naive} we assume that to a good approximation $p_z\simeq p$.
This assumption holds even for the tail of the spectrum below $\sim 2$~GeV since the transverse momentum of the initial particles is negligible compared to their longitudinal momentum.
Unlike in Sec.~\ref{sec:rates}, the GEANT4 simulation used here, uses  arbitrarily biased (inflated) cross-sections to increase the statistics of the resulting positrons.
The biasing is done for both the Bremsstrahlung and the pair-production processes and it is introduced only at the volume.
For this reason, we focus on the shape comparison alone and normalize the maxima of the two simulations to the maximum of the data from Run~502.
The systematic uncertainty band on the shape of Run~502 is obtained from the uncertainties on the two  components, $\theta_{yz}$ (1~mrad, see Step 6 in Sec.~\ref{sec:localalignment}) and $\theta_{yz}^{\rm beam-det}$ (1~mrad, see  Sec.~\ref{sec:globalignment}), using Gaussian toys in the same way as discussed in Sec.~\ref{sec:localalignment}.\\

The spectra of the two simulations are matching reasonably well the bulk of the data spectrum, up to $\sim 2.9$~GeV.
However, two clear discrepancies are seen at the two tails between the data and the GEANT4 simulation, where the systematic uncertainty (gray band) cannot fully cover for these.
The discrepancy at the high-$p_z$ tail is attributed to high-$p_z$ positrons that have a positive $y_{\rm LAB}$ position component and a positive vertical momentum (positive $p_y$ in the LAB frame) component at the entrance to the dipole.
Positrons with $p_z\gtrsim 3.5$~GeV, with no displacement from the zero-dispersion axis and with no transverse momentum are out of acceptance since they pass below the bottom edge of at least one chip.
However, the displacement and transverse momentum configuration discussed above may give the high $p_z$ positrons a ``push'' up towards the detector.
This feature cannot, by construction, be reproduced with the simple field-angle-momentum relation given in Eq.~\ref{eq:momentum_naive}.
Moreover, it cannot be reproduced with tracking that is confined to the volume of the detector without back-propagating to the production point.
The discrepancy at the low-$p_z$ tail is attributed primarily to positrons that were produced at the Be window, but lost a substantial amount of energy mostly in the 500~$\mu{\rm m}$ stainless-steel vacuum exit window.
When examining the GEANT4 positrons at the production the distribution is effectively the same as the one from Xsuite.
These positrons do not change their trajectory in a significant way since the change in $p_z$ is tightly correlated with the one in $p_y$.
Hence, they will be reconstructed at their original angle (i.e. $p_z$) despite of arriving to detector at lower $p_z$ (after the interaction at the window).
The contribution from secondary positrons produced at different points along the beamline is negligible.

\section{Outlook}
\label{sec:outlook}
The E320 tracking detector prototype installed in Aug 2024 at the FACET\textsf{--}II accelerator, was commissioned in Nov 2024 and Feb 2025, and is already taking electron-laser collision data as of May 2025.
Using the commissioning campaigns' data, the detector is shown to work well for measuring single positrons per BX.
This rate corresponds to positrons that are produced in the Beryllium foil and pass through the detector fully.
Hence, it is tied to the magnets' settings, the detector position and other experimental conditions.
The measured signal rate is $(1.20\pm 0.06_{\rm stat.}\pm 0.56_{\rm syst.})\times 10^{-1}$ positrons per BX.
This rate is similar to the one expected in the E320 electron-laser collisions runs.
Crucially, this rate is measured in the presence of an extremely large background.
This illustrates that tracking can be operated reliably at an unprecedented hit density of $\sim 1.7$~hits/mm$^2$ due to background.
This density is already twice higher than what is expected by the HL-LHC experiments in their innermost tracking layers, roughly within a decade from now.

The systematic uncertainty size is attributed mostly to the limited statistics in the Run~502 sample and hence it can be constrained further with more statistics.
Thanks to the tight focusing used in the E320 runs, we expect this uncertainty to be even further constrained.
The tight focusing in the E320 runs enables a much tighter selection as well as a per-BX in-situ estimation of the background, which is expected to be much lower than that of Run~502 (at $\lesssim 500$~pixels/BX compared to $\sim 2200$ here).
We also show that when the foil is retracted, the false-positive background rate is smaller by four orders of magnitude compared to the signal rate.

Even with the lack of precise knowledge about the orbit of the electron beam, the $p_z$ spectrum shape of the data still shows a reasonable agreement with simulation, where the sources of the small discrepancies are well understood.
Finally, the spectrum normalization in data is only shown to qualitatively match that of the simulation, where again, the latter should be repeated when a better understanding of the orbit is obtained.

The initial analysis discussed here is fast and simple to execute.
Specifically, it provides a reliable measurement of the signal rate and its $p_z$ spectrum, almost independent from simulation.
Extracting finer and more precise information requires (i) more advanced tracking methods (e.g. the KF method from~\cite{Borysov:2025ehq}) and (ii) more pristine understanding of the actual experimental setup alignment (beamline elements positioning and orientation, beam axis, etc.).
These two points are currently being addressed and we leave that for a future work.

\section{Acknowledgments}
\label{sec:acknowledgments}
We thank the entire E320 collaboration as well as the FACET\textsf{--}II facility for their input and support.
FACET\textsf{--}II is supported by the U.S. Department of Energy under Contract No. DE-AC02-76SF00515.\\

We also thank the colleagues from the ALICE ITS2 project for useful discussions and help
related to the ALPIDE chips. We would like to especially thank Luciano Musa, Gianluca Aglieri Rinella, Antonello Di Mauro, Magnus Mager, Corrado Gargiulo, Felix Reidt, Ivan Ravasenga, Ruben Shahoyan and Walter Snoeys.\\

The work of Noam Tal Hod's group  at Weizmann is supported by a research grant from the Estate of Dr.\ Moshe Gl\"{u}ck, the Minerva foundation with funding from the Federal German Ministry for Education and Research, the ISRAEL SCIENCE FOUNDATION (grant No. 1235/24), the Anna and Maurice Boukstein Career Development Chair, the Benoziyo Endowment Fund for the Advancement of Science, the Estate of Emile Mimran, the Estate of Betty Weneser, a research grant from the Estate of Gerald Alexander, a research grant from the Potter's Wheel Foundation, a research grant from Adam Glickman and the Sassoon \& Marjorie Peress Legacy Fund, the Deloro Center for Space and Optics, and in part by the Krenter-Perinot center for High-Energy particle physics and by the Knell Family Institute for Artificial Intelligence.\\

S.\ Meuren is supported by the Agence Nationale de la Recherche (ANR) under the Chaire de Professeur Junior (CPJ) program. Furthermore, he is grateful for support from the visitor program of the Stanford PULSE institute.\\

D.\ A.\ Reis and T. Smorodnikova are supported by the U.S. Department of Energy Office of Science, Office of Fusion Energy Sciences under award DE-SC0020076.\\

S. Corde and S. Rego are supported by the Agence Nationale de la Recherche (ANR) with the g4QED project, Grant No. ANR-23-CE30-0011.

\bibliography{bibliography.bib}

\begin{appendices}
\section{Power, Trigger \& DAQ}
\label{app:daq}
\begin{figure}[pos=!ht]
\centering
\begin{overpic}[width=0.9\textwidth]{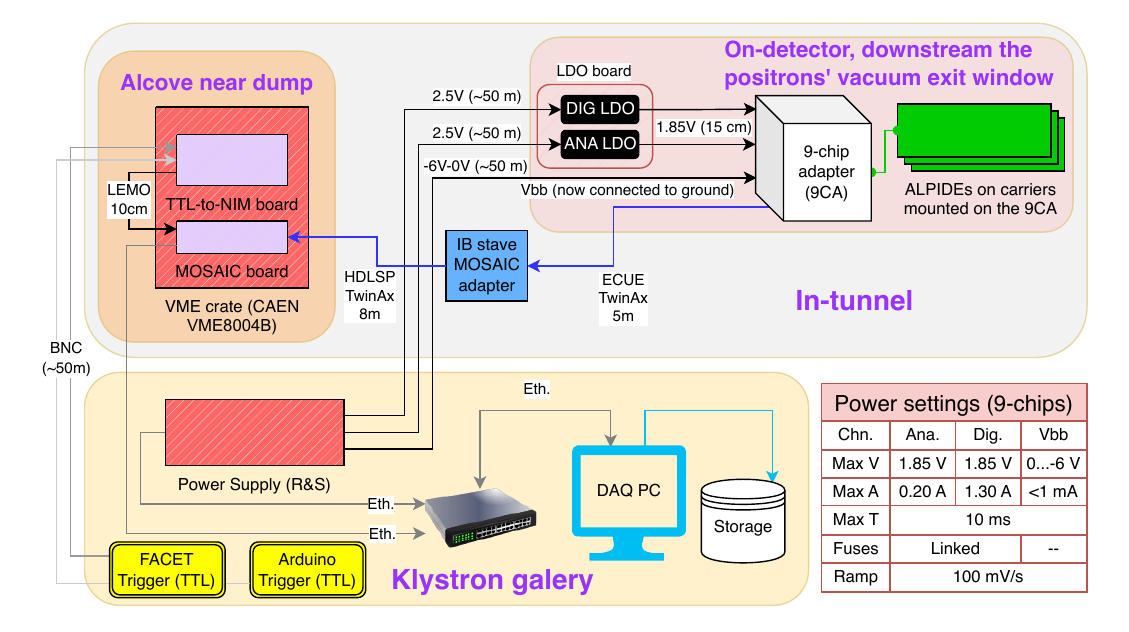}\end{overpic}
\caption{The schematic of the powering scheme, readout and trigger system for the prototype detector. 
The key elements are grouped by location, either in the
tunnel or in the Klystron gallery above. 
The different connection types and lengths are specified.}
\label{fig:ps_and_ro}
\end{figure}

A schematic overview of the powering scheme, trigger and readout system of the prototype detector is provided in Fig.~\ref{fig:ps_and_ro}, illustrating the key components, their locations, and connection lengths.

The 9CA is connected to an off-the-shelf power supply that provides channels for powering the digital and analog circuitry of the ALPIDE chips.
The bias voltage lines of the 9CA are connected to the ground, keeping the bias voltage applied to the chips' substrate at zero, as the built-in voltage in the pixel is sufficient to have a high detection prior to a significant irradiation of the chips. 
This also was, for example, the course of action for the Run~3 in the ALICE experiment.
We also note that a set of dedicated bias cables are already routed and can be connected to the 9CA if needed. 
The power supply is positioned approximately 50~m from the detector in the Klystron gallery above the accelerator tunnel. 
It delivers moderately low voltages of approximately 2.5~V per channel, with the bias channel used optionally. 
Voltage drops caused by current consumption, primarily in the digital channels, are manageable at this distance. 
To ensure stable operation, a custom low-dropout regulator (LDO) board stabilizes the power input to a fixed 1.85~V and also helps in filtering noise from the long supply lines.
The LDO chip used in the board has been validated as radiation-hard for this application\cite{Abovyan:2022qaz}.\\

\begin{figure}[pos=!ht]
\centering
\includegraphics[width=0.99\textwidth]{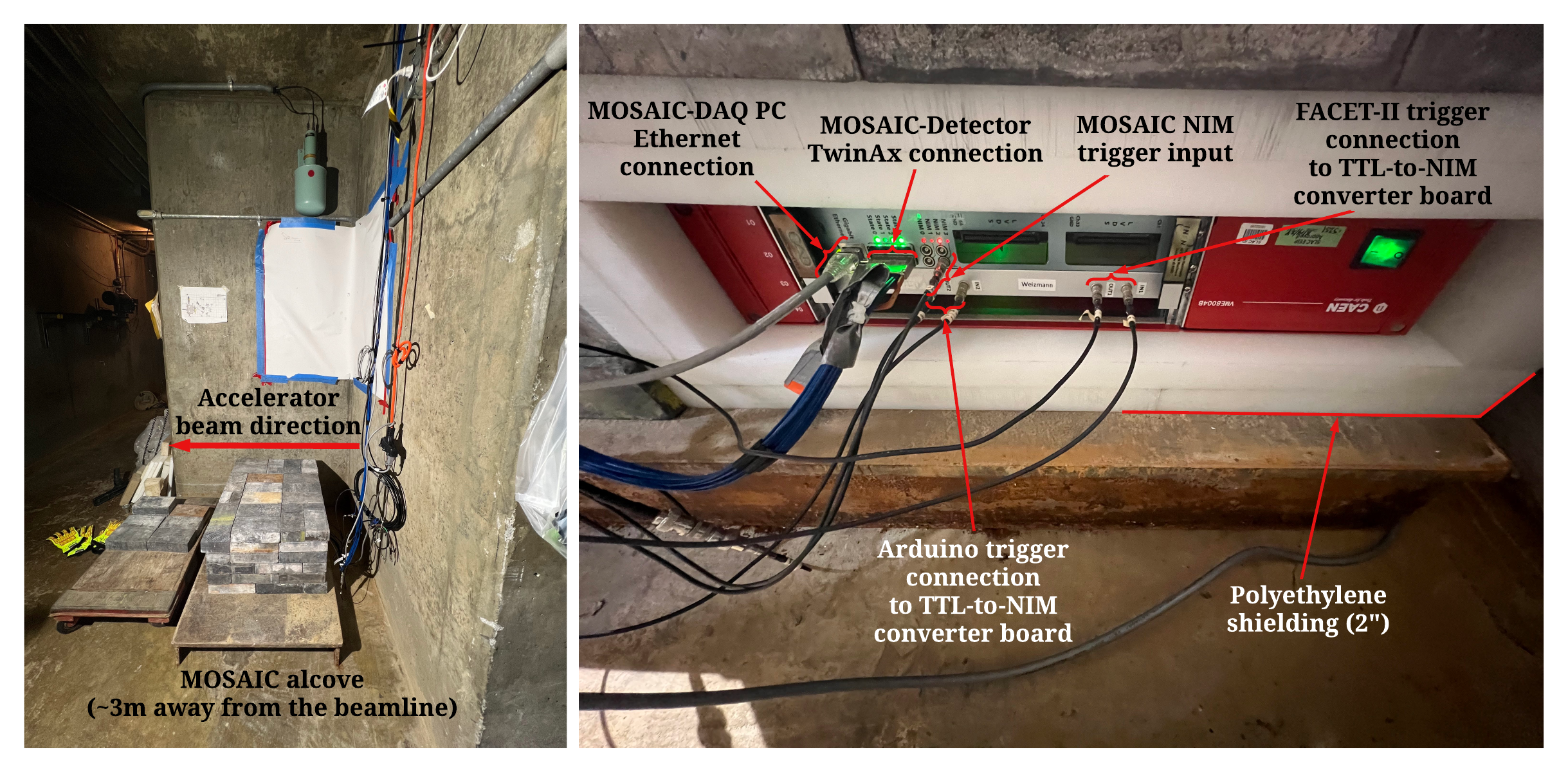}
\caption{MOSAIC board VME crate placement in the FACET\textsf{--}II accelerator tunnel. The setup is shielded by protective layers of lead and PE. Left: view of the setup from the accelerator aisle. Right: view of the setup from the alcove wall. Direction of the accelerator beam and the MOSAIC board connections are shown.}
\label{fig:mosaic_tunnel}
\end{figure}

The data and control signal transmission to and from the 9CA is managed by a MOSAIC readout board~\cite{refId0}. 
The connection utilizes two SAMTEC TwinAx cables with a combined length of 13~m, providing a Gigabit-capable channel to each ALPIDE chip. 
The MOSAIC board supports $\mathcal{O}(1~\text{Gb}/\text{s})$ communication, facilitating data transfer from the detector, temporary storage management, and connection to the control and data acquisition (DAQ) system. 
The board is housed inside a small 2-unit VME crate located in an alcove near the beam dump, approximately 10~m from the detector.
As the MOSAIC board is not radiation-hard, the VME crate is placed on an iron pallet on the tunnel floor, about 2~m upstream the dump, and shielded from the beam axis by a concrete wall.
All exposed sides of the crate are further protected by 2 inches of polyethylene (PE) and 4 inches of lead, with a small hatch left open for cable connections and airflow.
The shielded crate installed in the accelerator tunnel is shown in Fig.~\ref{fig:mosaic_tunnel}.
This is found to be sufficient to protect the MOSAIC boards from radiation and ensure their stable operation.

\begin{figure}[pos=!ht]
\centering
\includegraphics[width=0.49\linewidth]{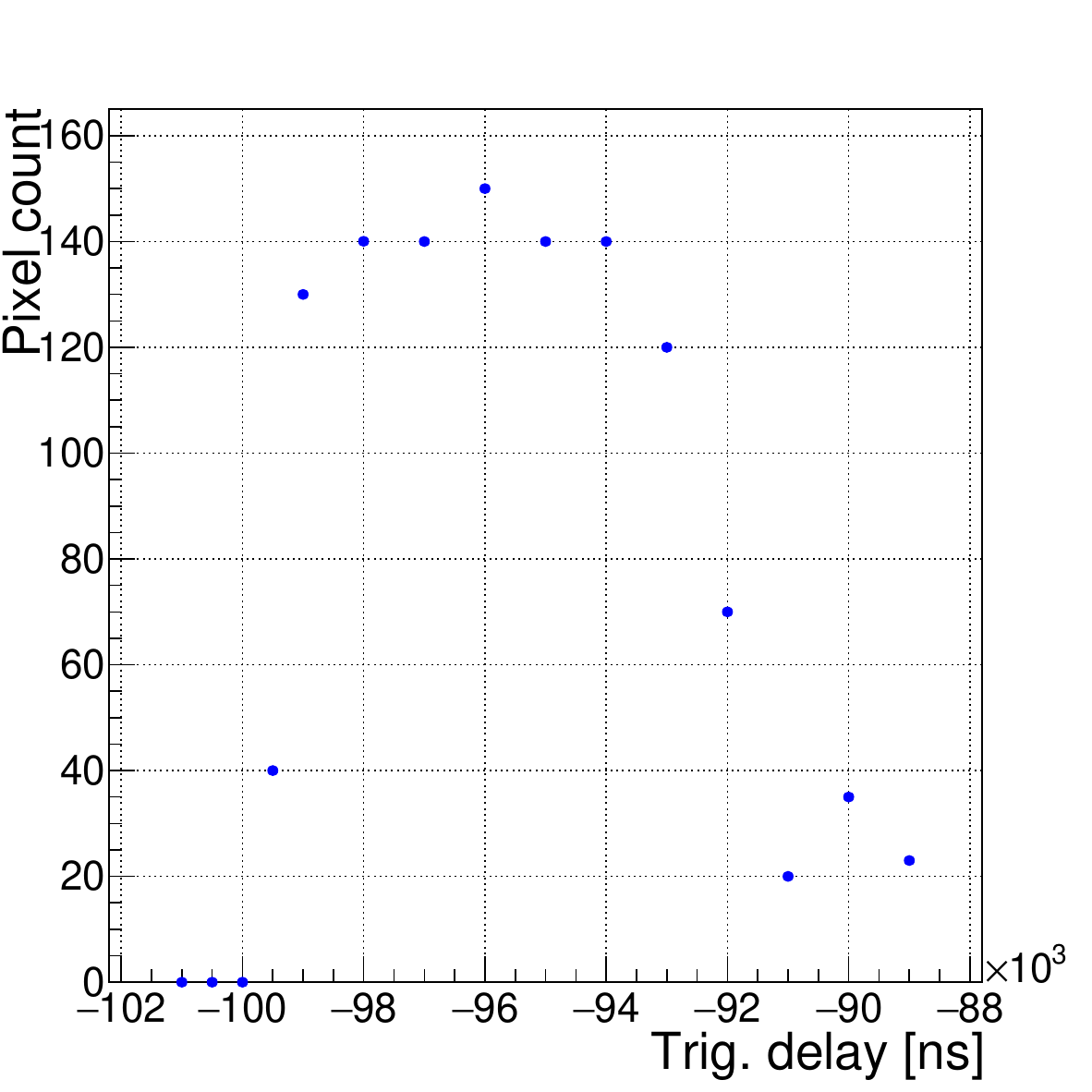}
\caption{The average occupancy in the chips as a function of the trigger delay. A trigger delay of $-96~\mu$s is adequate to ensure fully efficient data-taking.} 
\label{fig:delay_scan}
\end{figure}

The MOSAIC boards receive trigger signals synchronized with the accelerator clock. 
These signals are transmitted from the Klystron gallery in the TTL format using $\sim50$~m long BNC cables, converted to the NIM standard on a dedicated board within the same VME crate, and then delivered to the MOSAIC boards. 
The system includes an adjustable delay mechanism with a sub-nanosecond delay step and a range of $\pm 1.2$~ms relative to the accelerator's arrival time, enabling precise alignment of the detector's readout window. 
The trigger delay is adjusted to ensure that the readout coincides with the peak occupancy of the chips following the beam arrival.
Figure~\ref{fig:delay_scan} presents the average occupancy of the prototype chips as a function of the trigger delay.
The strobing time here and during all the data taking runs discussed was configured to 200~ns.
We scan the trigger delay in a wide range while the beam is present such that we expect to read at least a few pixels due to background.
We list the average occupancy in the chips versus the trigger delay and search for the plateau of maximum occupancy.
This plateau extends for approximately 
10~$\mu$s, which is consistent with the expected duration of the ALPIDE discriminator pulse~\cite{ALPIDEManual}.
The final delay value is fixed at $-96~\mu$s.

For redundancy and since the maximum trigger rate provided by FACET\textsf{--}II is 360~Hz, an independent trigger system, implemented using an Arduino board located in the Klystron gallery above the tunnel, is also deployed for the purposes of testing and performing the cosmic muons measurements for calibration and alignment of the detector.
The latter is performed in a semi-continuous readout mode, where the Arduino supplies a 1~kHz trigger, while the strobing time of the chips is set to 0.95~ms, thus allowing for almost continuous active readout.
To collect as many muons as possible in these runs, we can mechanically rotate the detector such that it faces the sky as discussed in Sec.~\ref{sec:tracker}.
A manual switch allows operators to alternate between the two triggers during accelerator shutdowns.

The detector services including power supply for digital, analog, and bias circuits, readout systems are routed from the top of the detector layers furthest from the beam axis, as illustrated in Fig.~\ref{fig:detector2}.
The MOSAIC boards connect to a network switch via Ethernet cables, which links them to the DAQ PC.
A few SSH tunnels are set up between the DAQ PC, FACET\textsf{--}II servers, and remote PCs at the Weizmann Institute of Science to meet DAQ requirements and enable remote control of the setup.

\begin{figure}[pos=!ht]
\centering
\includegraphics[width=0.99\textwidth]{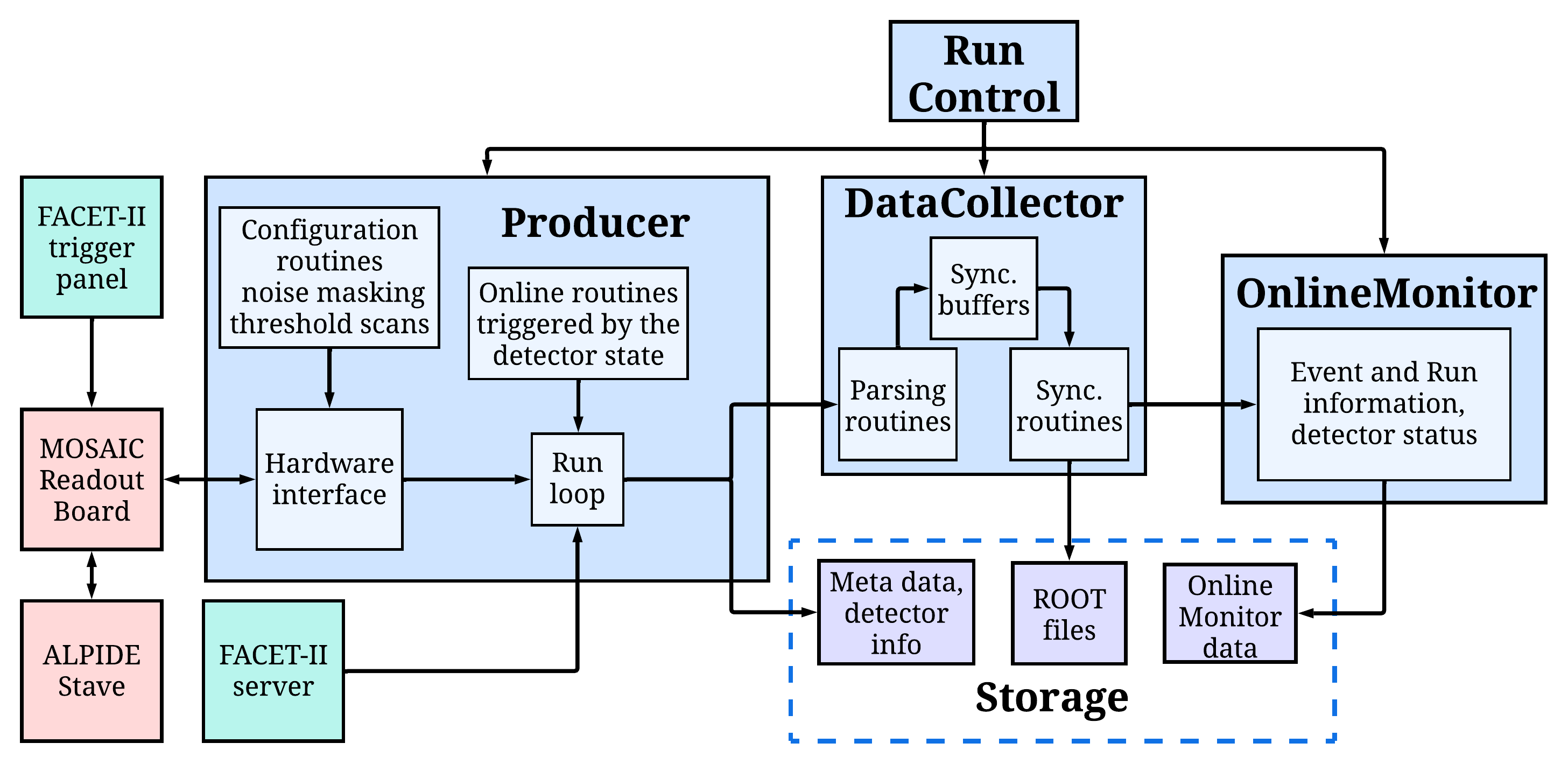}
\caption{A flow diagram of the control and data in E320 EUDAQ2 setup.}
\label{fig:eudaq_flow}
\end{figure}

The DAQ PC runs dedicated software, which controls and reads out the detector. 
The software is built within the EUDAQ2 framework~\cite{EUDAQ2020,Liu_2019}, which operates as a state machine managing communication between hardware and software components.
Fig.~\ref{fig:eudaq_flow} provides a schematic overview of the system’s control and data flow.

The \verb|RunControl| module serves as the user interface, enabling control of the system state and providing feedback on status of the connections and operations. 
The Producer acts as the hardware interface, managing configuration and readout of the ALPIDE chips and the MOSAIC board. 
It communicates with the ALPIDE chips via the MOSAIC board over Ethernet, sending commands and receiving data. 
In the case of E320, the \verb|Producer| is also responsible for interfacing with the FACET\textsf{--}II accelerator servers to ensure synchronization with its DAQ system, which operates within the EPICS framework~\cite{dalesio1991epics}.
The received data frames are preprocessed and forwarded to the \verb|DataCollector|, which decodes them and stores them in the DAQ PC memory. 
The decoded frames are then sent to the \verb|OnlineMonitor|, which provides real-time information about the detector status and data parameters.

The \verb|Producer| is responsible for configuring the detector, a critical component of the software's functionality. 
The configuration process begins with setting the in-pixel threshold for the chips, which determines the minimum charge required to activate a pixel. 
Proper threshold tuning is essential for accurate reconstruction analysis, as mis-configurations can lead to systematic underestimation or overestimation of the measurement errors. 
This tuning is achieved by adjusting dedicated DAC registers on the chips.
The in-pixel threshold during the data taking is set to approximately 130 elementary charges, while a typical minimum ionizing particle should yield at least 10 times more than that.

Once the threshold is set, a fake hit rate scan is performed to identify and mask pixels that falsely signal activity in the absence of external stimuli. 
These faulty pixels, typically caused by defects in the digital circuitry, are detected by repeatedly reading the pixel matrix under idle conditions and marking those that consistently return an activated state. 
To improve data quality and optimize bandwidth for physically meaningful signals, these pixels are masked using dedicated on-chip registers.
We typically find $\lesssim20$ pixels that need to be masked, per chip. This is a negligible fraction of the full matrix of $1024\times512$ pixels.

Further configuration includes defining operational parameters such as the strobing time window, which controls for how long the pixels can signal activity to be stored in on-chip memory, speed of the data-transfer links, and many more. 
Detailed descriptions of ALPIDE and MOSAIC configuration procedures, as well as their common operational modes, can be found in~\cite{ALPIDEManual, MOSAICManual}.

Beyond detector configuration, the \verb|Producer| establishes a connection to a server running within the EPICS framework on the FACET\textsf{--}II accelerator network. 
This connection enables the retrieval of real-time accelerator parameters, including magnetic field amplitudes, BX timestamps, beam profile monitor (BPM) readings, etc. 
These parameters are synchronized with detector data during runs, ensuring consistency at the reconstruction stage and facilitating cross-referencing with other detector subsystems.

\begin{figure}[pos=!ht]
\centering
\includegraphics[width=0.49\textwidth]{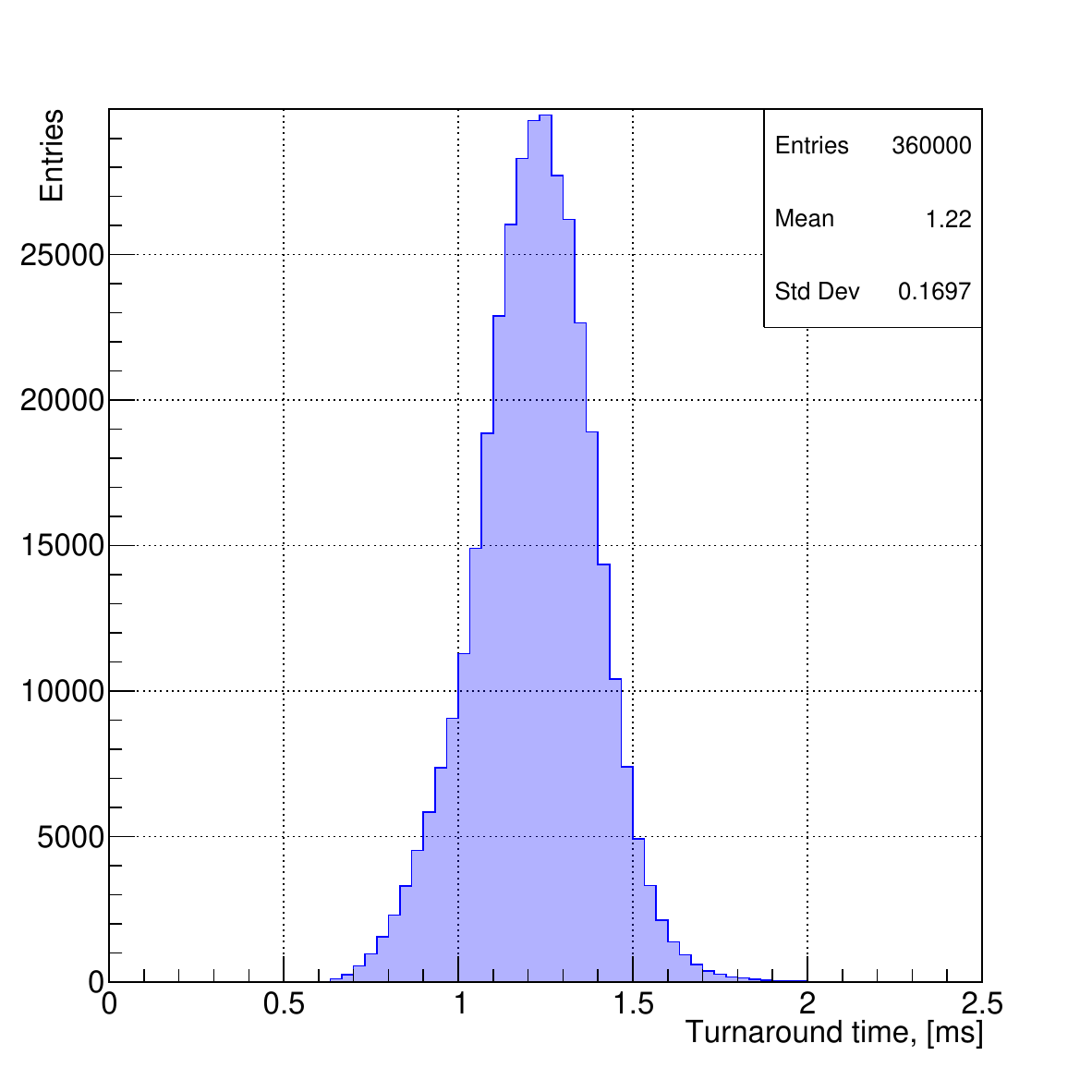}
\caption{Turnaround time distribution of the Producer-to-server transactions.}
\label{fig:turnaround_time}
\end{figure}

During data acquisition, upon receiving a trigger, the \verb|Producer| sends a request to the FACET\textsf{--}II server, which continuously monitors the accelerator state. 
The server then compiles a data frame and transmits it back to the \verb|Producer|. 
Given the event rate of 10~Hz, this transaction must be completed within 100~ms to maintain real-time synchronization and prevent data drifts or missed readouts. 
To ensure this requirement is met, the turnaround time of each transaction is continuously monitored and recorded as a part of the data frame.
The transaction time is defined as the duration between the \verb|Producer| sending a request and receiving the corresponding data frame.
The distribution of these transaction times is shown in Fig.~\ref{fig:turnaround_time}.
The data in this distribution were collected over a 10-hour period at a request rate of 10~Hz, under conditions where the FACET\textsf{--}II accelerator's servers and clients were operating at full capacity.
This ensured that the network load closely matched real data-taking conditions.
It has a narrow peaking shape with mean at 1.22~ms and standard deviation of 0.17~ms.
The mean transaction time is approximately two orders of magnitude faster than the required 100~ms limit, confirming that synchronization with the accelerator is not a limiting factor during data acquisition.

Once the detector is configured, data collection begins. 
The \verb|Producer| continuously monitors the MOSAIC board for arrival of new triggers. 
When the trigger arrives, it sends a readout requests after a delay of $\mathcal{O}(\mu{\rm s})$ to collect chip data from the on-board memory. 
Simultaneously, the \verb|Producer| queries the FACET\textsf{--}II server to log the accelerator state at the event time. 
The collected data is sent to the \verb|DataCollector|, where data frames are decoded and stored in a temporary buffer.

Frames from different chips are synchronized based on event timestamps and trigger IDs. 
Once synchronization is confirmed, the data is transferred to the DAQ PC memory, while a copy is sent to the \verb|OnlineMonitor|. 
Parameters such as chip occupancy and active pixel cluster size distributions are continuously monitored to ensure data quality. 

Extensive testing of the software demonstrated its capability to handle data rates of $\sim$45~MB/s while operating 9 ALPIDE chips.
This rate is several times lower than the tested bandwidth of the DAQ software, indicating its performance to be more than adequate for the anticipated workload.

\section{Data cleaning}
\label{app:cleaning}
We remove individual BXs (or groups of them), where the beam quality is poor.
To do that, we are retrieving independent information from different beam position monitors (BPMs), photo-multiplier tube counters (PMTs), Toroid charge monitors, Radiation monitors, etc. along the FACET\textsf{--}II beamline from the EPICS server per BX as discussed in Appendix~\ref{app:daq}.
All of these monitor readings should normally be stable around some baseline value throughout the data-taking.
However, the FACET\textsf{--}II beam conditions may change gradually and/or abruptly during a run. 
The abrupt type of changes is in principle not necessarily infrequent and even during a short (few minutes) run there could be multiple BXs, where the beam is lost somewhere upstream the IP chamber.
These losses are typically manifesting as distinct spikes (dips) above (below) the baseline value of the different monitors' readings.
In the tracking detector, these losses manifest as spikes in the pixel occupancies.
These BXs should be therefore removed from the data samples as they do not represent stable-conditions.
The veto algorithm to remove these BXs starts by putting all values in an array per monitor.
The array is sorted and the fraction of largest and smallest outliers (OL) in it, $f_{\rm OL}$ is trimmed.
We then calculate the unbiased (UB) mean $\mu_{\rm UB}$ and standard deviation $\sigma_{\rm UB}$ of the trimmed array, which effectively represent only the stable-conditions part of the run.
Finally, looping over the original array, we check if a given value is above $\mu_{\rm UB}+n\sigma_{\rm UB}$ (or below $\mu_{\rm UB}-n\sigma_{\rm UB}$) and reject the respective BX if so, where $n$ is some empirically found multiplier.\\

\begin{figure}[pos=!ht]
\centering
\begin{overpic}[width=0.99\textwidth]{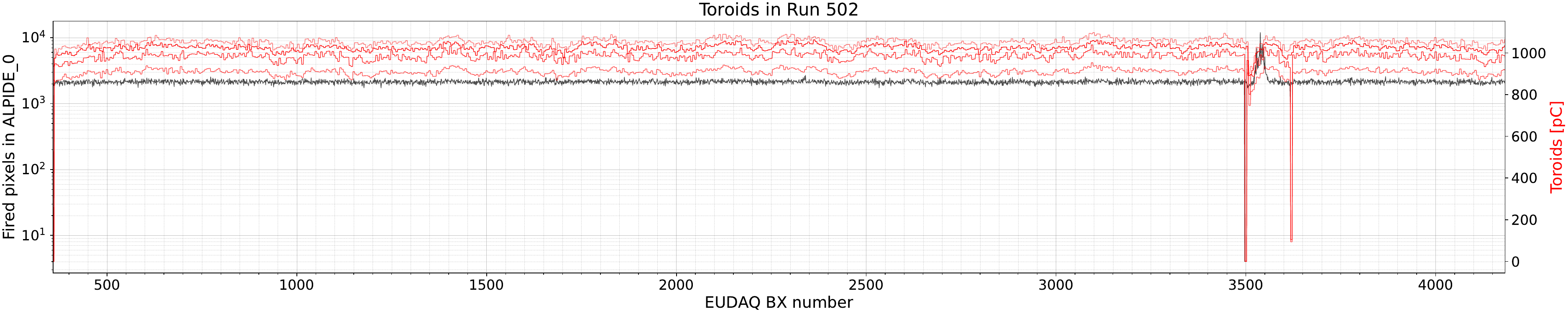}\end{overpic}\vspace{3mm}
\begin{overpic}[width=0.99\textwidth]{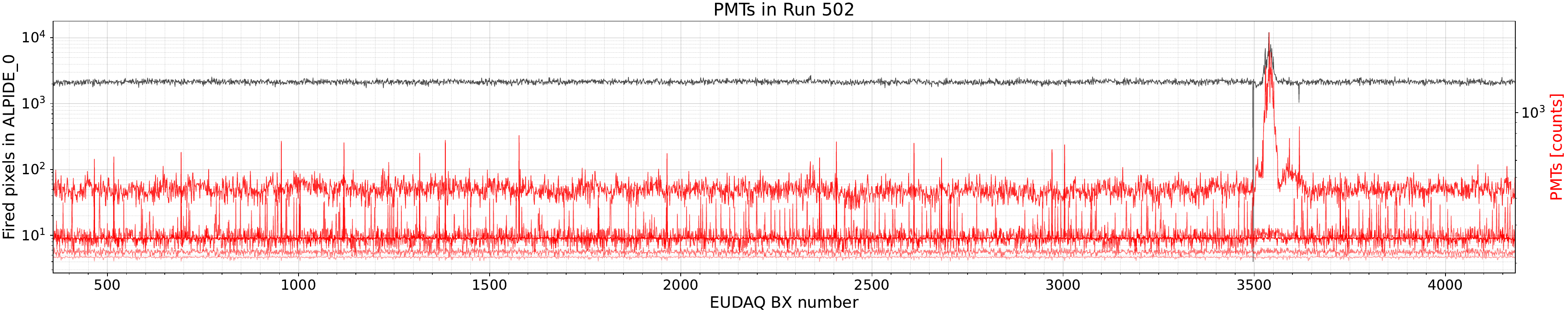}\end{overpic}\vspace{3mm}
\begin{overpic}[width=0.99\textwidth]{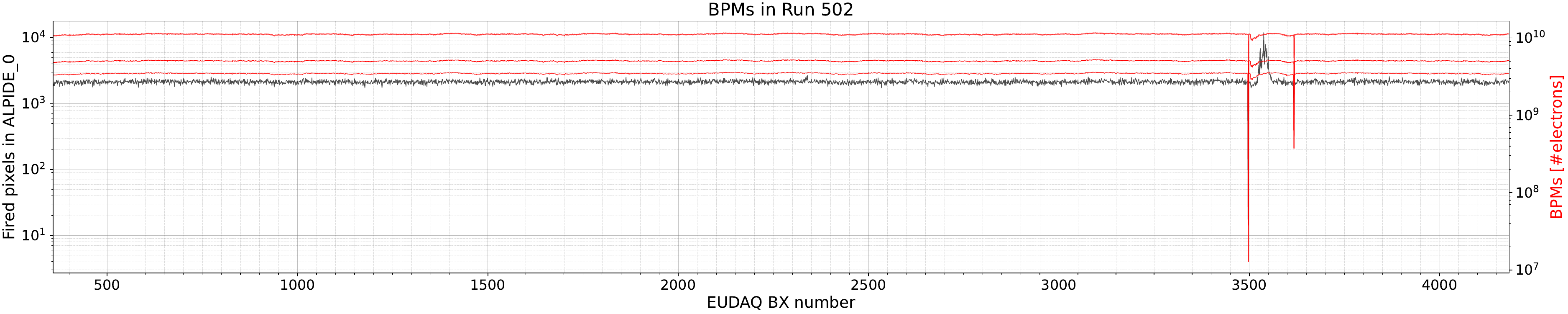}\end{overpic}\vspace{3mm}
\begin{overpic}[width=0.99\textwidth]{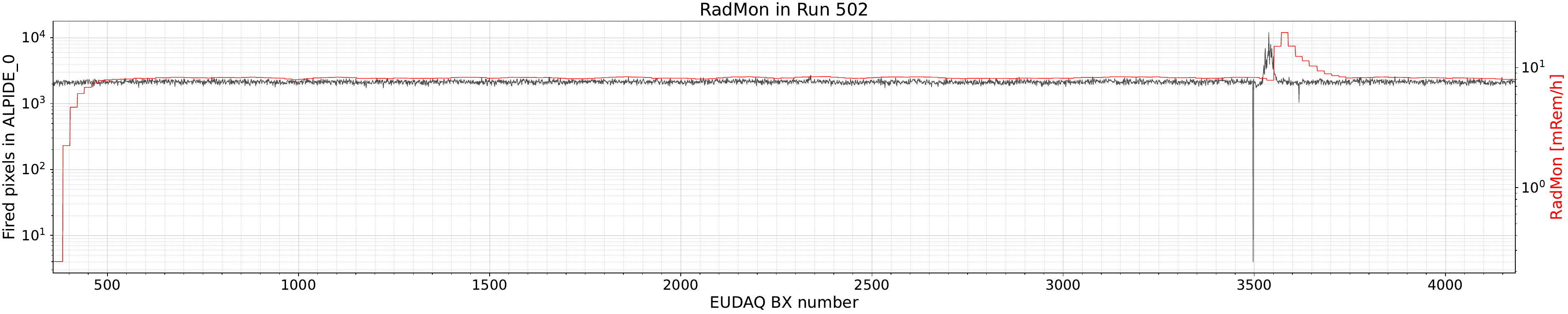}\end{overpic}
\caption{Examples of the monitors evolution throughout Run~502 before applying the cleaning algorithm. The pixel occupancy of the first tracking layer is plotted on the left vertical axis in black, while on the right axis the different red curves correspond to the different FACET\textsf{--}II monitors of the same kind, where applicable.}
\label{fig:cleaning_502}
\end{figure}

\begin{figure}[pos=!ht]
\centering
\begin{overpic}[width=0.99\textwidth]{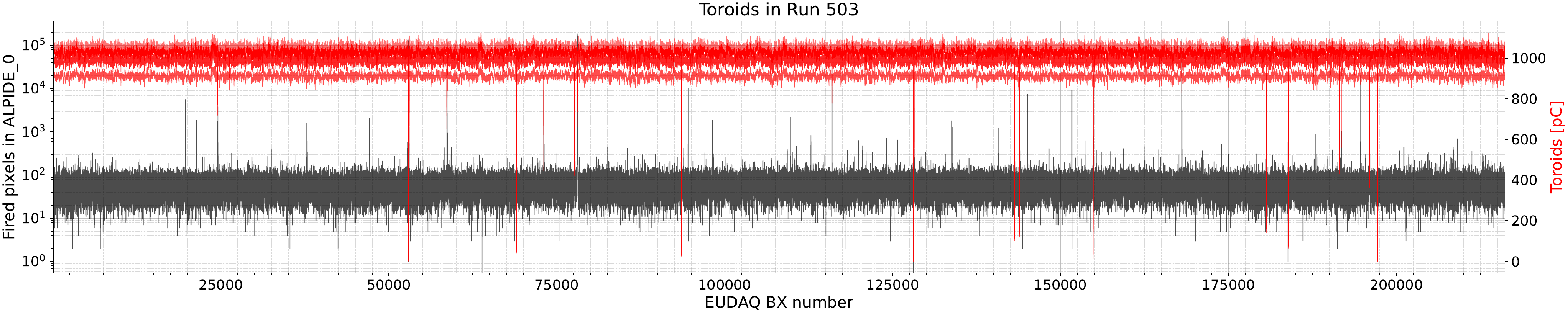}\end{overpic}\vspace{3mm}
\begin{overpic}[width=0.99\textwidth]{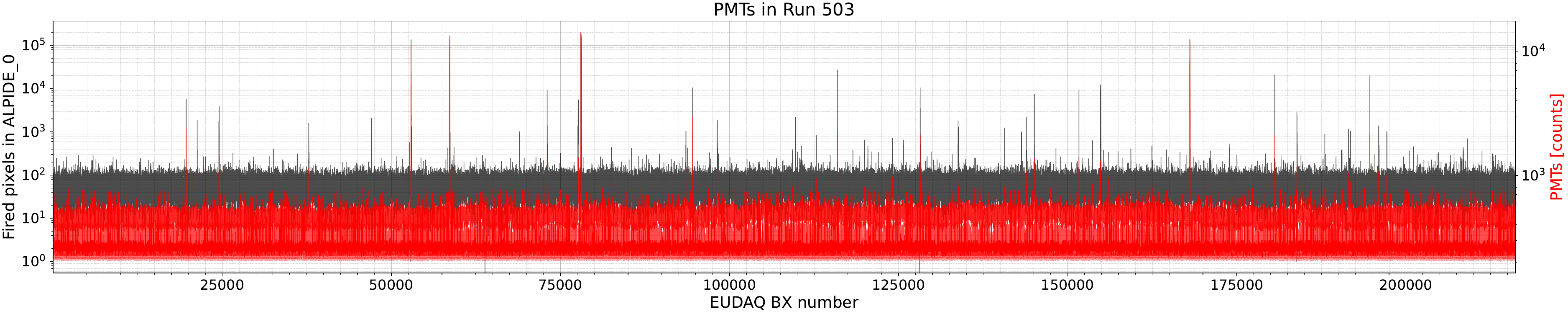}\end{overpic}\vspace{3mm}
\begin{overpic}[width=0.99\textwidth]{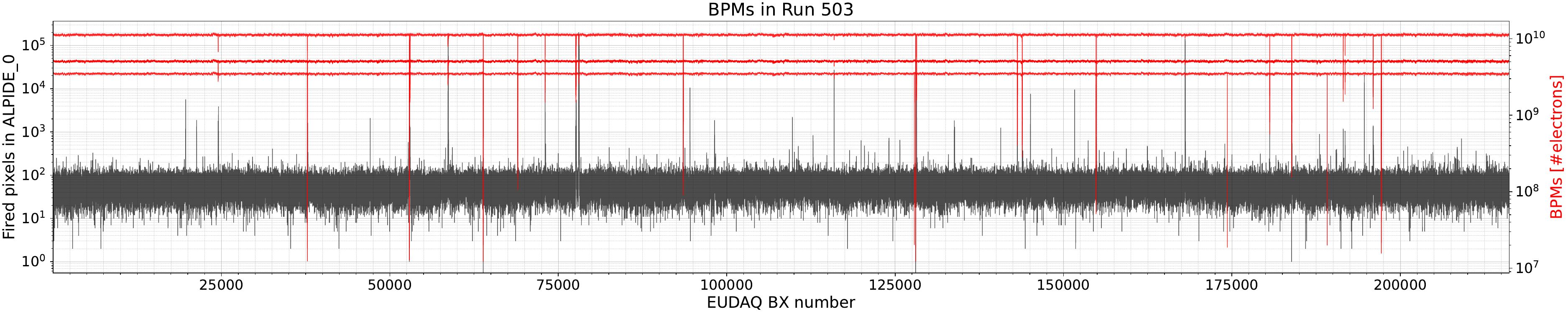}\end{overpic}\vspace{3mm}
\begin{overpic}[width=0.99\textwidth]{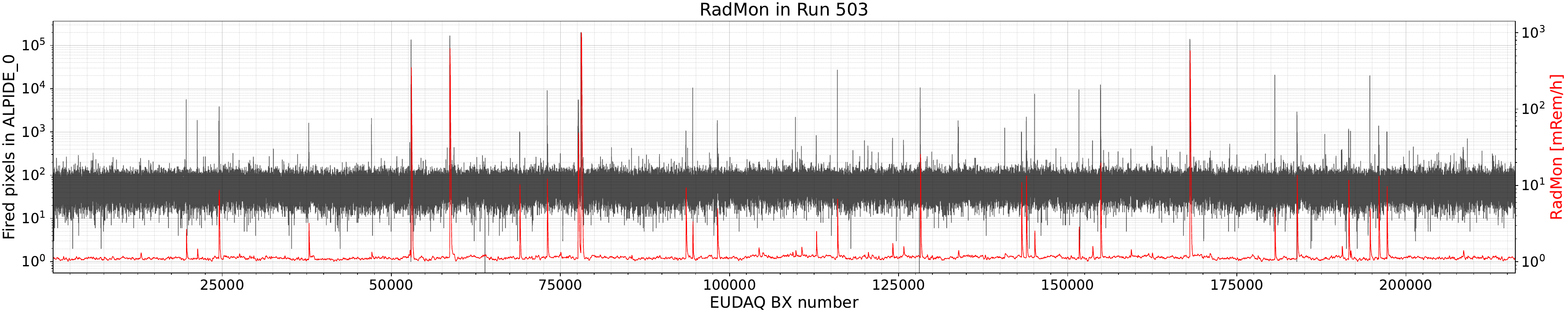}\end{overpic}
\caption{Examples of the monitors evolution throughout Run~503 before applying the cleaning algorithm. The pixel occupancy of the first tracking layer is plotted on the left vertical axis in black, while on the right axis the different red curves correspond to the different FACET\textsf{--}II monitors of the same kind, where applicable.}
\label{fig:cleaning_503}
\end{figure}

A few examples of the monitors evolution throughout the run are given in Fig.~\ref{fig:cleaning_502} for Run~502 and~\ref{fig:cleaning_503} for Run~503, both before applying the cleaning algorithm.
In both figures, the pixel occupancy in the entire chip of the first layer (\verb|ALPIDE_0|) is plotted on the left vertical axis in black, while the different independent monitors are plotted on the right axis in red.
In cases where there are several red curves plotted, they correspond to the different FACET\textsf{--}II monitors of the same kind.
It can be seen that there is a perfect correlation between the abnormalities seen in the different monitors and the spikes (or deficits) seen in the pixel occupancy along the run.\\

The algorithm is tuned per monitor per run in terms of $f_{\rm OL}$ and $n$.
After applying the algorithm, we effectively remove all spikes form the two black curves.
Specifically, in Run~502 there are 3,826 valid BXs, where 171 of those are removed ($\sim 4.4\%$ of the total).
In Run~503 there are 216,175 valid BXs, where 10,268, of those are removed ($\sim 4.7\%$ of the total).
All remaining BXs are used for the subsequent analysis.

\end{appendices}

\end{document}